\documentclass[reviewcopy]{elsart}
\usepackage{lineno}
\linenumbers

\usepackage[dvipsone]{graphicx}
\usepackage{textcomp}
\def\epem  {\ensuremath{e^+e^-}\xspace}
\def\cms   {\ensuremath{{\rm \,cm}^{-2} {\rm s}^{-1}}\xspace}
\def\invpb {\ensuremath{\mbox{\,pb}^{-1}}\xspace}
\def\ep    {\ensuremath{e^+}\xspace}
\def\en    {\ensuremath{e^-}\xspace}   

\RequirePackage{xspace}

\newcommand{\BHWIDE}     {BHWIDE}
\newcommand{\DAF}        {DA$\Phi$NE}
\newcommand{\GEANT}      {GEANT}
\newcommand{\GEANTTHREE} {GEANT3}

\newcommand{\SID}        {{\em SIDDHARTA}}
\newcommand{\MC}         {Monte-Carlo}
\newcommand{\TYVEK}      {Tyvek\texttrademark}

\begin{document}

\begin{frontmatter}

\title{
Measurement of the luminosity at the DA$\phi$NE collider upgraded with the crab waist scheme    
}

\author{M. Boscolo, F. Bossi, B. Buonomo, G. Mazzitelli, F. Murtas, P. Raimondi}
\author{G. Sensolini} 
\address{INFN/LNF, Via E. Fermi 40-00044, Frascati, RM, Italy}

\author{M.~Schioppa}
\address{Gruppo Collegato INFN, Via P.~Bucci - 87036, Rende, CS, Italy}

\author{F.~Iacoangeli, P.~Valente}
\address{INFN Roma, P.le Aldo~Moro 5, Roma, Italy}

\author{N.~Arnaud\corauthref{cor}}
\ead{narnaud@lal.in2p3.fr}
\corauth[cor]{Corresponding author.}
\author{D.~Breton, L.~Burmistrov, A.~Stocchi, A.~Variola, B.~Viaud}
\address{ Laboratoire de l'Acc\'{e}l\'{e}rateur Lin\'{e}aire, Universit\'e Paris-Sud, CNRS/IN2P3, 91898 Orsay, France}

\author{P.~Branchini}
\address{INFN Roma 3, Via della Vasca Navale, 84 - 00146, Roma, Italy}

\newpage

\begin{abstract}
The test of the crab waist collision scheme, undergoing at the \epem\ Frascati \DAF\ accelerator complex since February 2008, requires a fast and accurate measurement of the absolute luminosity, as well as a full characterization of the background conditions. Three different monitors, a Bhabha calorimeter, a Bhabha GEM tracker and a gamma bremsstrahlung proportional counter have been designed, tested and installed around the interaction point end of 2007-beginning of 2008. In this paper, we describe these detectors and present their performances in various operation conditions during the 2008 and 2009 \DAF\ runs. 
\end{abstract}

\begin{keyword}
Luminosity \sep \DAF\ \sep Crab Waist \sep Calorimeter \sep Bhabha scattering
\PACS 29.20.db \sep 29.40.Gx \sep 29.40.Vj
\end{keyword}

\end{frontmatter}

\tableofcontents

\clearpage

\section{Introduction}
\label{sec:intro}

Proposals of future flavor factories~\cite{ref:superb,ref:superkekb,ref:kloe2} emphasize the need of very high luminosities. For instance, the new generation of B-factories~\cite{ref:superb,ref:superkekb} requires improvements up to two orders of magnitude above the performances of the PEP-II~\cite{ref:pepii} and KEK-B~\cite{ref:kekb} \epem\ colliders. Among the ideas currently being developed to achieve this ambitious physics-driven goal, the crab waist compensation scheme associated with large Piwinski angle and low vertical beta function~\cite{ref:crabwaist} is very promising. Luminosities as high as $10^{36} \cms$ could be reached with beam currents similar to those operated routinely in today's accelerators, which would also help keeping the background under control. In addition to being based on existing technologies, this scheme would significantly limit the power (and hence the cost) needed to run such a new machine.

The \DAF\ accelerator, located in the National Laboratory of Frascati (INFN) and optimized for the production of $\phi$ mesons ($\sqrt {s}$ =1020 MeV) at a high rate, has been modified in 2007 to test the crab waist sextupole compensation scheme~\cite{ref:crabwaist} with the aim of reaching higher luminosity while controlling the background. After completion of this upgrade, operations restarted during winter 2007-2008.

Since 2000, \DAF\ has been delivering \epem\ collisions to three experiments KLOE~\cite{ref:kloe}, FINUDA~\cite{ref:finuda} and DEAR~\cite{ref:dear}, steadily improving performances in terms of luminosity, beam lifetimes and background. The best peak luminosity was $\sim 1.5 \times 10^{32} \cms$ with typical daily integrated luminosities of $\sim 8 \invpb$ during the KLOE run. According to calculations~\cite{ref:crabwaist}, the modified interaction region should increase the luminosity by a factor 3-5 with respect to the previous colliding scheme. To identify such a significant gain, a luminosity measurement precise at the $\sim 10\%$ level is enough. On the other hand, real time and accurate information is required regardless of the machine conditions to allow the \DAF\ operators to tune this new collider and to get relevant feedback for their optimization studies. Some redundancy between the various measurements is also important, in order to fight transient backgrounds which could impact strongly a particular detector.

Parallel to the upgrade of the \DAF\ interaction point 1 (IP), the \SID\ experiment (SIlicon Drift Detector for Hadronic Atom Research by Timing Application)~\cite{ref:siddharta} aiming at studying kaonic hydrogen and kaonic deuterium has been installed at the same location. The presence of this additional detector, whose operation requires a very good shielding against machine background, has consequences on the design and on the performances of the main luminometer. These are discussed in the following sections of this article. In principle, \SID\ can also provide a luminosity measurement based on the counting of charged Kaon pairs produced by the decay $\phi \rightarrow K^+ K^-$ whose branching fraction is well-known. However, this method suffers from a few limitations: a low rate (around 25~Hz at a luminosity of $10^{32} \cms$); the dependence on the exact center of mass (CM) energy because of the $\phi$ resonance lineshape; the need to separate efficiently true Kaons from minimum-ionizing particles, a background strongly machine-dependent; a limited duty cycle for technical reasons.

Therefore, various independent luminosity-oriented detectors have been built around the IP and put in operation beginning of February 2008 with a threefold goal: to guarantee an accurate measurement of the absolute luminosity, to monitor the background levels and to provide powerful and fast diagnosis tools to improve the machine. In the following, we describe the design, the construction and the operation of these detectors, such as the results achieved during the 2008 and 2009 \DAF\ runs.

The paper is organized as follows. Section~\ref{sec:overview} gives an overview of the processes used to measure the luminosity and presents the experimental setup. Then, Sections~\ref{sec:setup} and~\ref{sec:daq} describe in details the detectors and the trigger/data acquisition systems. The recorded Bhabha events are characterized in Section~\ref{sec:performances} which also explains the procedure used to estimate and subtract online the background. Section~\ref{sec:montecarlo} deals with the \MC\ simulation which is crucial to convert Bhabha event rate into absolute machine luminosity. Validation studies based on data-\MC\ comparisons are provided in Section~\ref{sec:valid} where the systematic error to the luminosity measurements is also computed. Experimental results are given in Section~\ref{sec:results} before we finally conclude in Section \ref{sec:conclusions}.

\section{Overview}
\label{sec:overview}

\subsection{Physical processes used for luminosity and background measurements} 
Since the early days of \epem\ colliders, well-known electromagnetic processes, such as Bhabha scattering~\cite{ref:bhabha} or single and double \epem\ bremsstrahlung~\cite{ref:brem}, have been used to monitor and measure collider luminosity. The Bhabha elastic scattering $\epem \rightarrow \epem$ has a very clean signature of two back-to-back tracks, of energy equal or close to the beam energy (510~MeV at \DAF\ ). The cross section $\sigma_{Bhabha}$ of this process has a very steep dependence on the polar angle $\theta$: $\sigma_{Bhabha} \propto 1/\theta^3$. Most Bhabha events are produced at small polar angle, an area where focusing quadrupoles or other machine elements are generally located, which limits the measured event rate. Nevertheless, Bhabha scattering at large angle still produces a sizable counting rate which can be exploited to measure luminosity; considering a polar angle range of 18$^\circ$-27$^\circ$, the Born cross-section of the process is as high as 5~$\mu$b.

The $\epem \rightarrow  \epem \gamma$  (single bremsstrahlung) process has a cross section of 169~mb for photon energies above 10~MeV, thus resulting in a very high counting rate. The photon undergoes a small angular deviation from the charged track original direction: 95$\%$ of the signal is contained in a cone of 1.7~mrad aperture. Moreover, its cross section depends only logarithmically on the CM energy and is therefore relatively independent of the actual machine parameters. On the other hand, it suffers heavily from background caused by particles interacting with the residual gas in the beam-pipe or lost by Touschek effect~\cite{ref:Touschek}. 

\subsection{Background from Touschek effect}

Achieving high luminosity is not enough: future uses of the crab waist scheme will depend on how well the backgrounds can be controlled in this configuration. For the \DAF\ collider, both the experimental machine-induced backgrounds and the beam lifetimes are dominated by the Touschek effect due to the use of dense beams at relatively low energy. The Coulomb scattering of charged particles in a stored bunch induces energy exchange between transverse and longitudinal motions. Small transverse momentum fluctuations lead to larger longitudinal variations as the effect is amplified by the Lorentz factor. Off-momentum particles can then exceed the momentum acceptance given by the radio-frequency (RF) bucket, or they may hit the aperture when displaced by dispersion. In both cases they get lost.

\begin{figure}
\begin{center}
\includegraphics[width=12cm]{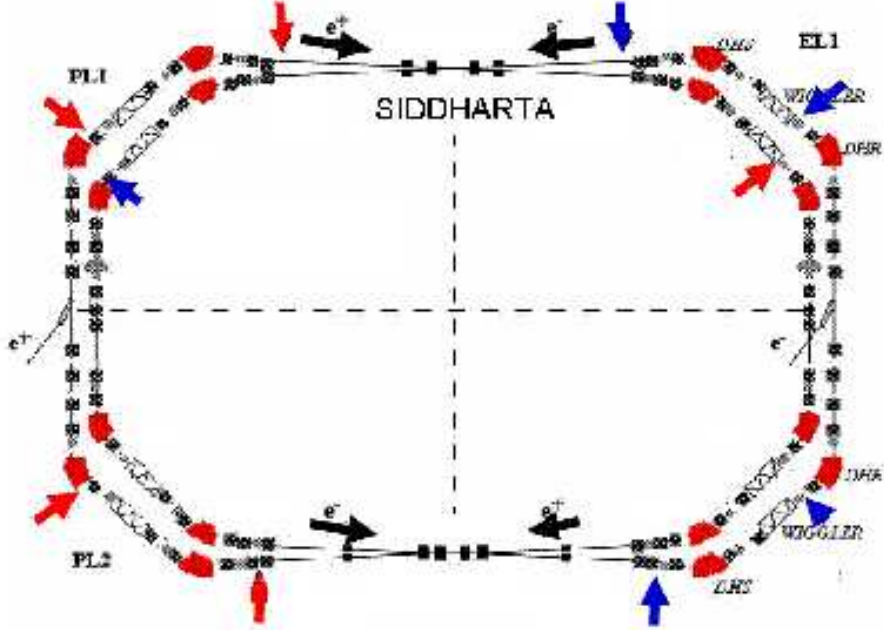}
\end{center}
\caption{\it { Layout of the \DAF\ main rings with the collimator locations shown by red and blue arrows for positrons and electrons, respectively. The IR collimators are located upstream of the IP while the other ones are in dispersive regions close to wigglers. } }
\label{fig:collimators}
\end{figure}

Several studies have been performed to control and reduce the background induced by the Touschek effect and to optimize the signal to noise ratio for all the experiments that have been running at \DAF. This has been achieved by adjusting optical parameters like the orbits at the IP or the strength of the machine sextupoles and by minimizing the radial beam size upstream from the interaction region (IR). Moreover, the insertion of a proper set of collimators (see Fig.~\ref{fig:collimators}), together with simulation-based tracking studies of Touschek scattered particles~\cite{ref:boscolo}, have also helped reducing the background significantly.

Particle losses due to Touschek effect at \DAF\ are expected to be higher for the crab waist configuration than for previous optics. However, the collimators are efficient even if longer jaws would have been more useful around the IR due to the low emittance. This hardware modification is planned in the near future for the next KLOE run. In the present setup, a very careful shielding has been designed to allow clean luminosity measurements and to increase the signal to noise ratio for the \SID\ detector. In addition, lead shieldings have been added behind the IR collimators to stop as many secondaries as possible.

\subsection{Detectors for luminosity and background measurements}

\begin{figure}[!h]
\centering
\includegraphics[width=14.0cm]{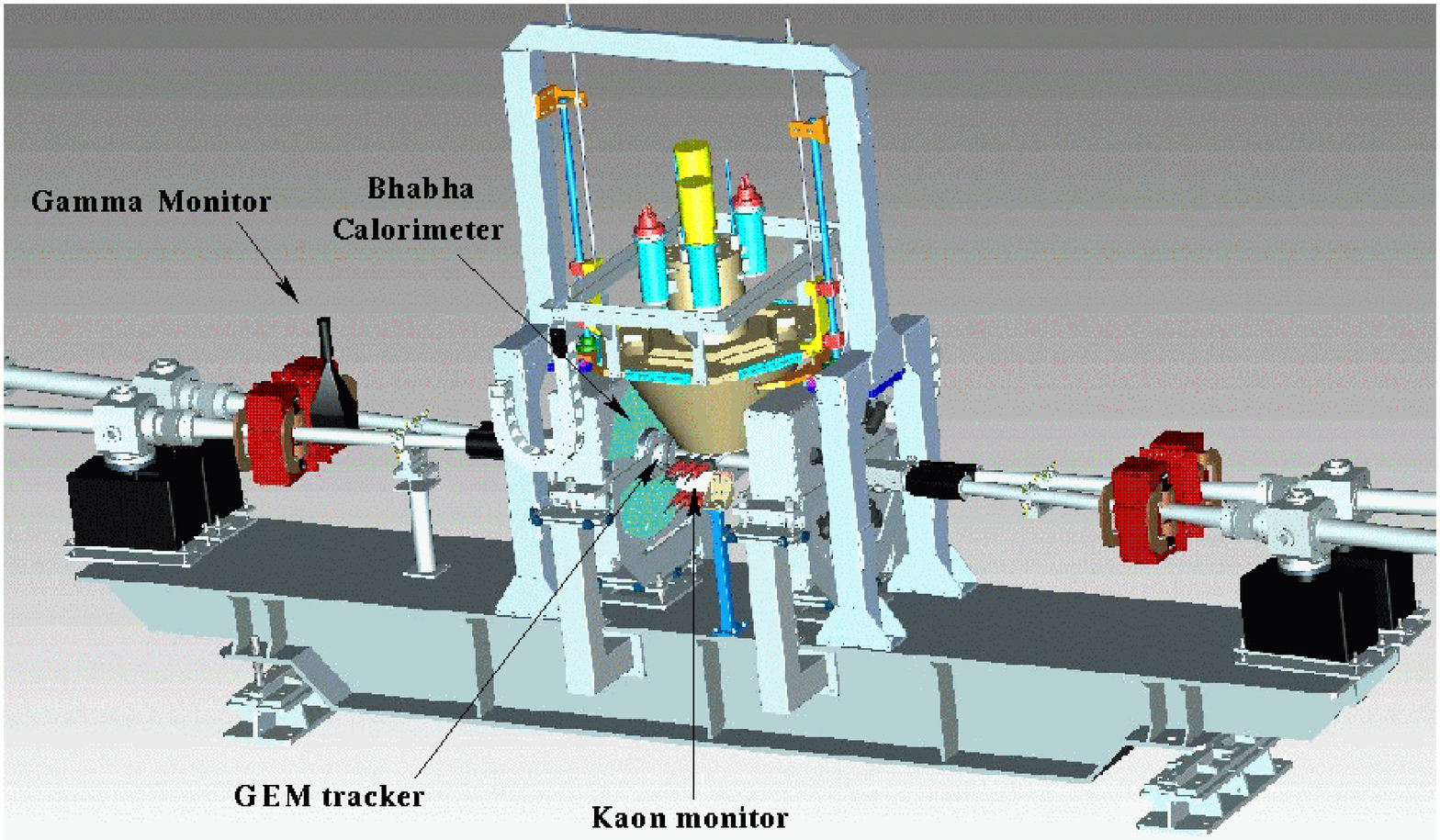}
\includegraphics[width=12.0cm]{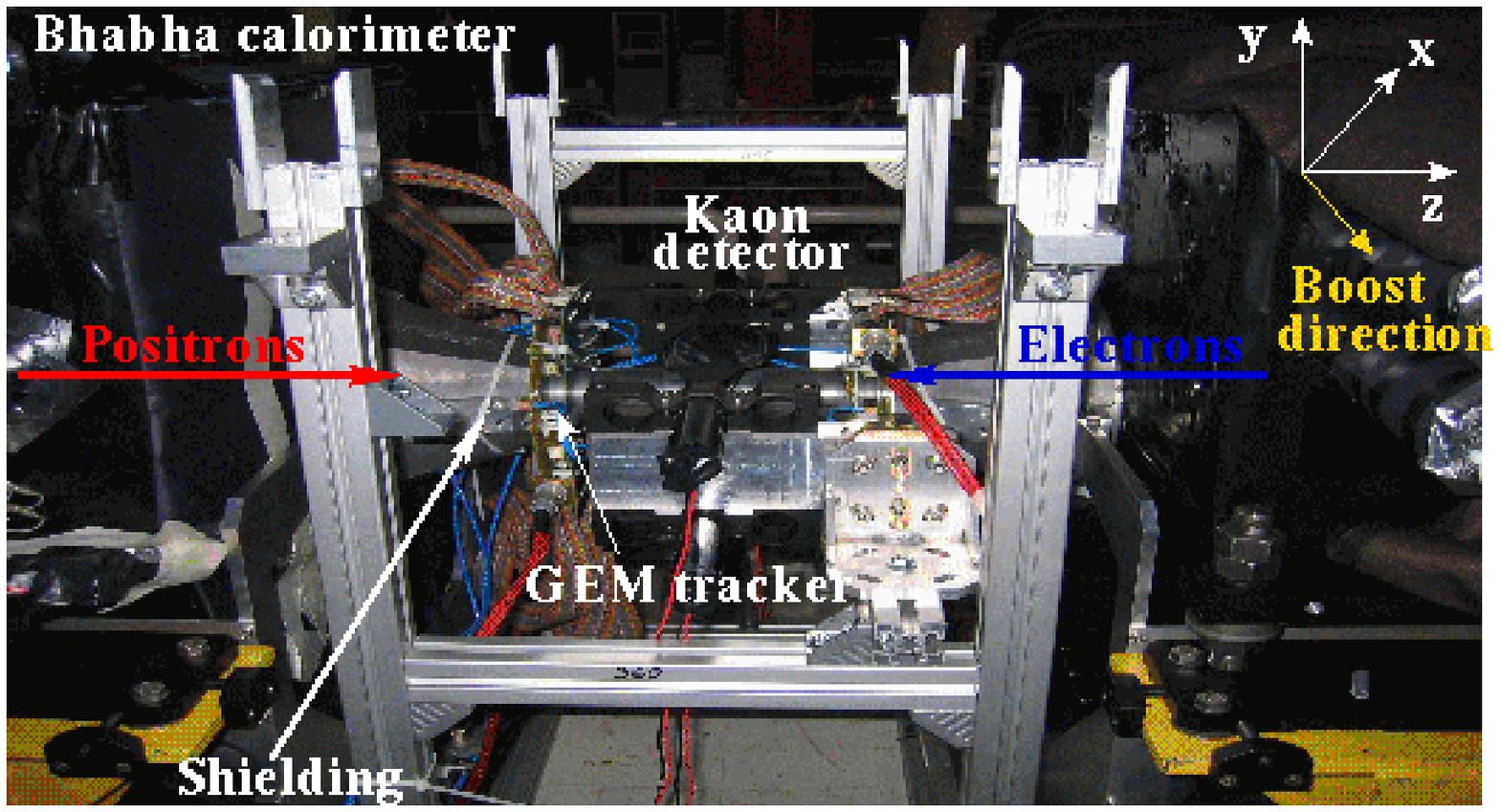}
\caption{\it { Overview of the upgraded  \DAF\ interaction region, showing  the Bhabha calorimeter, the GEM tracker and the gamma monitors. The bottom picture was taken in July 2008 during an access to the IP; on its top right corner one can see the system of coordinates which is used in the following. } }
\label{fig:setup}
\end{figure}

In the new \DAF\ interaction scheme, the two 510~MeV beams collide with a total crossing angle of 50~mrad (25~mrad per beam). At the IP, both beams are pointing toward the inner part of the ring; they are separated at 57~cm from the IP. Two low-$\beta$ permanent quadrupole magnets, the QD0s, are located on each side of the IP before the common beam pipe gets split into two separate sections. The QD0 magnets are approximately cylindrical, with a length of 24~cm and a 10~cm radius. Their front faces are located at 32.5~cm from the IP. The \SID\ detector is installed in the space between these two magnets; its shape is asymmetrical because of physics- and construction-related constraints -- see Ref.~\cite{ref:siddharta} for details. It mostly covers all the solid angle above a given polar angle (around 30~degrees) whose actual value depends on its shielding. An overview of the upgraded \DAF\ IR including the luminosity monitors (a Bhabha monitor and two forward gamma detectors) and the \SID\ detector is shown in Fig.~\ref{fig:setup}. 

Bhabha events are counted by a double-arm lead-scintillator calorimeter installed around the QD0 magnets. This detector can operate in coincidence with a GEM tracker located closer to the IP to help defining its polar acceptance. The luminosity number ${\cal L}_{meas}$ is extracted in real time by comparing the background-subtracted Bhabha event rate with the prediction from the \MC\ simulation~-- see Sections~\ref{sec:performances} and~\ref{sec:montecarlo} for details. One has

\begin{equation}
{\cal L}_{meas} = \frac{R_{meas}}{R^{MC}_{{\cal L}^{0}}}\times {\cal L}^0,
\label{eq:lumi}
\end{equation}

where $R_{meas}$ is the background-subtracted Bhabha rate measured at the \DAF\ IP and $R^{MC}_{{\cal L}^{0}}$ is the rate predicted by the \MC\ simulation assuming a luminosity of ${\cal L}^0=10^{32} \cms$. The reference \MC-based rate includes all acceptance effects: detector geometry, trigger, quality of the energy reconstruction, etc.

This luminosity measurement can be compared online with a ``geometrical'' luminosity estimate, ${\cal L}_{geom}$, which is based on beam sizes (as measured by the synchrotron light monitor, SLM) and beam currents.

\begin{equation}
{\cal L}_{geom} = \frac{ I^+ I^- }{ 4\pi N_b e^2 } \frac{ 120 }{ f_{RF} } \frac{ 1 }{ \sigma_x \sqrt{ \left[ \left( \sigma_{y,\, \mathrm{electron}}^{SLM} \right)^2 + \left( \sigma_{y,\, \mathrm{positron}}^{SLM} \right)^2 \right] \frac{ \beta_y^{IP} }{ 2 \beta_y^{SLM} } } } \times C_{\mathrm{CRAB}}
\end{equation}

with the $C_{\mathrm{CRAB}}$ reduction factor given by

\begin{eqnarray}
\nonumber
C_{\mathrm{CRAB}} &=& 0.55981 - 0.007474 \times I_{\mathrm{med}} - 8.121 \times 10^{-5} \times I^2_{\mathrm{med}} \\
\nonumber
I_{\mathrm{med}} &=& \frac{ 1 }{ 2 } \left( \frac{I^+}{N^+} + \frac{I^-}{N^-} \right)
\end{eqnarray}

$\beta_y^{SLM}$ is measured with the quadrupole oscillation method~\cite{ref:quadoscmethod}; $\beta_y^{IP}$ is estimated using the MAD machine model~\cite{ref:mad}; the $\sigma_{y}^{SLM}$ factors are computed from the SLM readouts. $N_b$ is the number of bunches in collision (i.e. $min(N^+,N^-)$) and $f_{RF}$ the RF system frequency, which is divided by 120, the harmonic number of the accelerator. The coefficients of the $C_{\mathrm{CRAB}}$ reduction factor have been computed numerically; they take into account various effects (hourglass, non-zero crossing angle, crab sextupoles) and the formula is adapted to the \DAF\ specific configuration of beams with different beta functions, emittances and bunch lengths~\cite{ref:zobov}. Data are updated every 15 seconds via the \DAF\ control system.

In addition to the Bhabha luminometer, two gamma monitors (PbW0$_4$ crystals) are located 170~cm away from the IP on either side and measure the rate of single bremsstrahlung events. These detectors provide a fast feedback when the machine conditions change; they have also been used to scan the beam profiles or to estimate the level of Touschek background. A detailed description of these three systems (the Bhabha calorimeter, the GEM trackers and the gamma monitors) can be found in the following section. 

\section{The experimental setup}
\label{sec:setup}

\subsection{Calorimeters}
\label{sec:calorimeters}

The Bhabha calorimeter is divided into four modules (two on each side of the IP) which surround the permanent QD0 magnets~-- see Fig.~\ref{fig:setup}. Its acceptance in polar angle starts at $18^{\circ}$ and goes up to about $27^{\circ}$; the angular upper bound depends on the exact \SID\ shielding configuration but the shower containment is degrading quickly above this angle. The modules are segmented in five azimuthal sectors, each $30^\circ$ wide.

The calorimeter design is driven by the need of having the main IR platform supporting its weight (about 400~kg in total) rather than the very fragile vacuum beam pipe. Each arm is made of two pieces (top and bottom halves) with a light aluminum skeleton sandwiching a stronger structure located in the machine mid-plane. A fan-shaped hole leaves a large and well-defined fraction of the $180^\circ$ acceptance free of any passive material. To keep the mechanics simple, the acceptance is covered by trapezoid (instead of ring-shaped) sectors. In the chosen design, the supporting structure is made of two parts connected by a longitudinal bar, one of each side of the fan-shaped hole.

\begin{figure}
\begin{center}
\includegraphics[width=8cm]{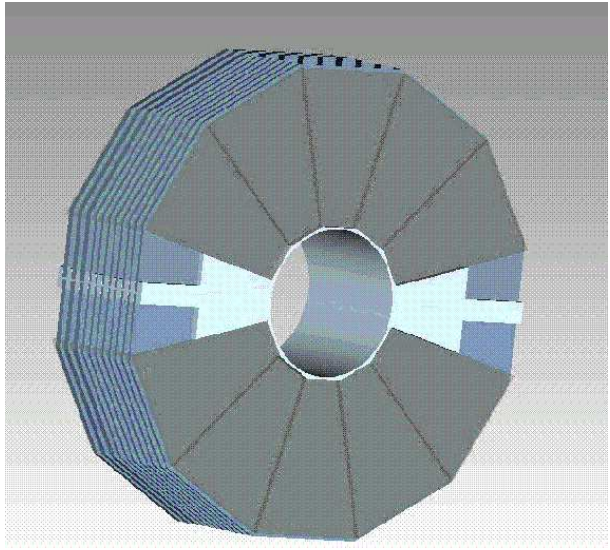}
\end{center}
\caption{\it { Drawing representing one arm of the Bhabha calorimeter. One can clearly see its structure and its granularity; among the twelve azimuthal sectors, ten are instrumented whereas the two around the beam plane are not. } }
\label{fig:ring}
\end{figure}

When the two halves are put together, the four sectors corresponding to the $\pm 15^\circ$ region around the beam plane cannot be instrumented~-- see Fig.~\ref{fig:ring}. This is not a flaw in the design as most of the machine background is located in this area: should they have been included, these most-lateral sectors would have been copiously hit by spurious particles. Having sectors $30^\circ$-wide in azimuth is a trade-off between the expected flexibility of the device (in particular one wants to be able to exclude temporarily a well-defined part of the acceptance in case of shielding modifications or changes in the background conditions) and the final number of channels to be instrumented.

Another major constraint of the calorimeter design is the presence around the IP of the \SID\ detector and of its shielding. Lead bricks are installed to stop as much radiation as possible before the particles enter the active part of the experiment. During the initial optimization phase, (February-August 2008) the shielding has been modified to improve the Kaon to background ratio. The vertical position of this detector has also been tuned (from 5 to 15~cm of the IP) in the meantime. The final shape has been installed in September 2008.

Each calorimeter sector is a sandwich of twelve trapezoidal 1~cm thick scintillator tiles, wrapped with \TYVEK~\cite{ref:tyvek} paper, alternated with eleven lead plates of variable thickness. Starting from the face close to the IP, the eight first plates are 0.5~cm thick while the last three are 1~cm thick. Hence the total thickness of the sector is 19~cm. This design is a compromise between the need of having a good longitudinal containment of the 510~MeV charged particle showers (the total depth corresponds to about 12.5 $X_0$) and the requirement of having a detector not exceeding the permanent quadrupole length~-- as some mechanical structures are located right behind these magnets. Moreover, the further the calorimeter from the IP, the lower the polar angle up to which it is sensitive. Finally, the free space between the calorimeter and \SID\ is needed to accommodate the GEM tracker which is described in the following.

\begin{figure}
\begin{center}
\includegraphics[width=8cm]{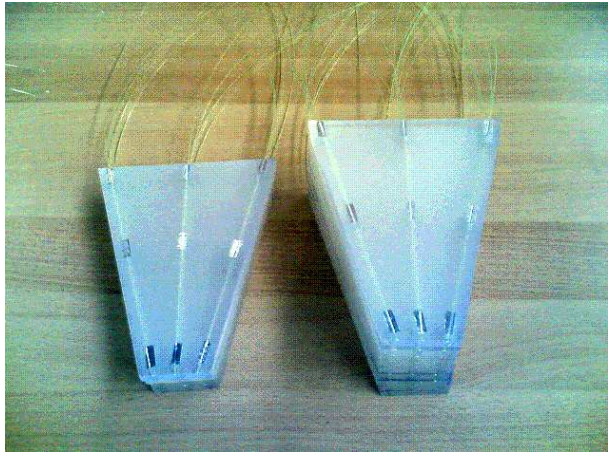}
\end{center}
\caption{\it { Picture of calorimeter tiles with the WLS fibers already glued on their radial grooves. } }
\label{fig:tiles-photo}
\end{figure}

The two-hundred and forty scintillator tiles have been produced with injection-molded technique in IHEP, Protvino~\cite{ref:tiles}.  Each tile has three, 2~mm deep, radial grooves on one face, (one in the middle and two 1~cm away from the edge of the tile) inside which wavelength shifting (WLS) fibers~\cite{ref:wls} of 1~mm diameter are placed~-- see Fig.~\ref{fig:tiles-photo} for details. The thirty-six WLS fibers of a given sector are plugged to an optical adapter to fit the photo-cathode of a Photonis-Philips XP 2262B photomultiplier (PMT)~\cite{ref:pmts} readout by a KLOE-based data acquisition system described in Section~\ref{sec:daq}.

\begin{figure}[!h]
\begin{center}
\includegraphics[width=6.0cm]{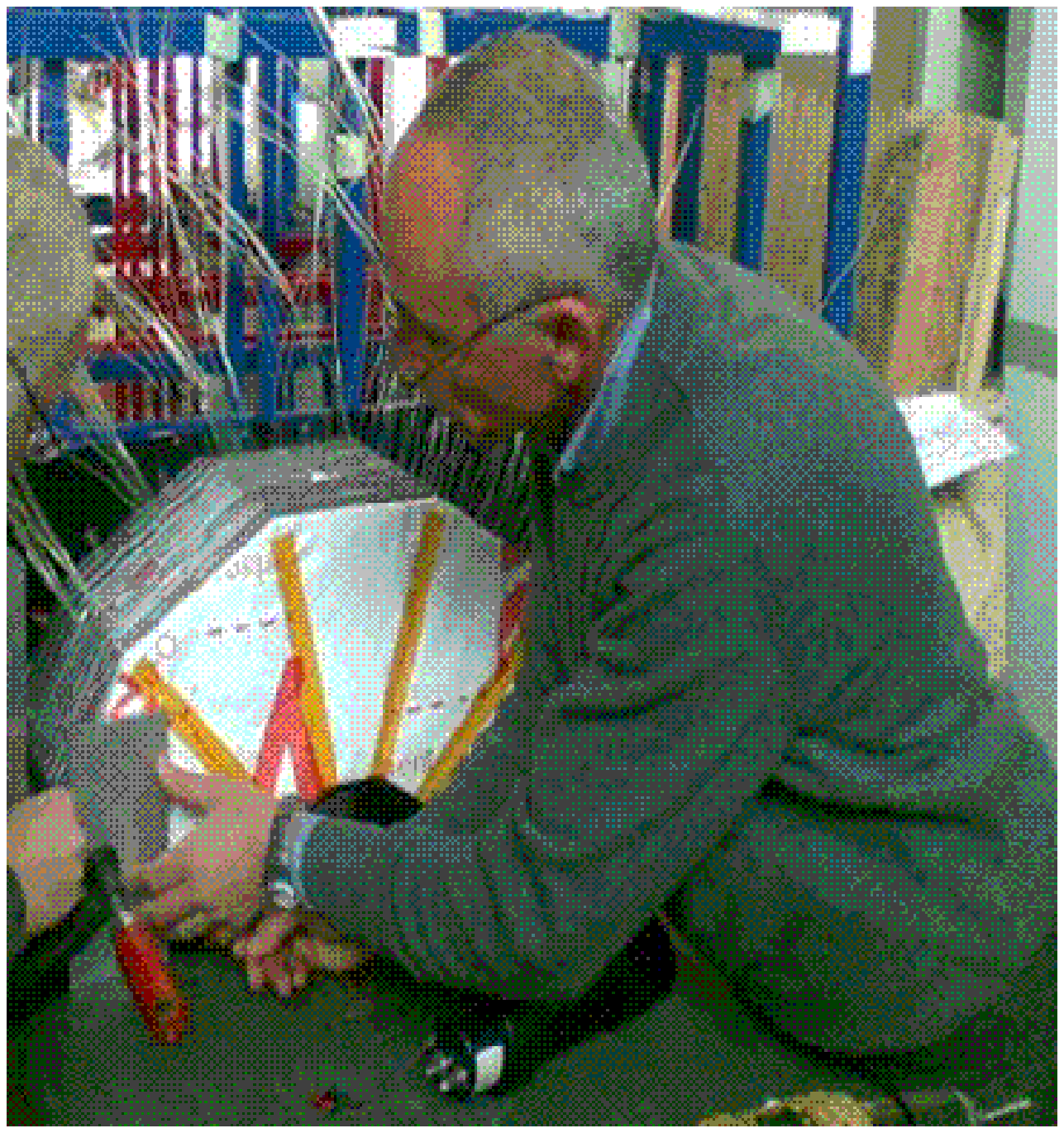}
\includegraphics[width=4.5cm]{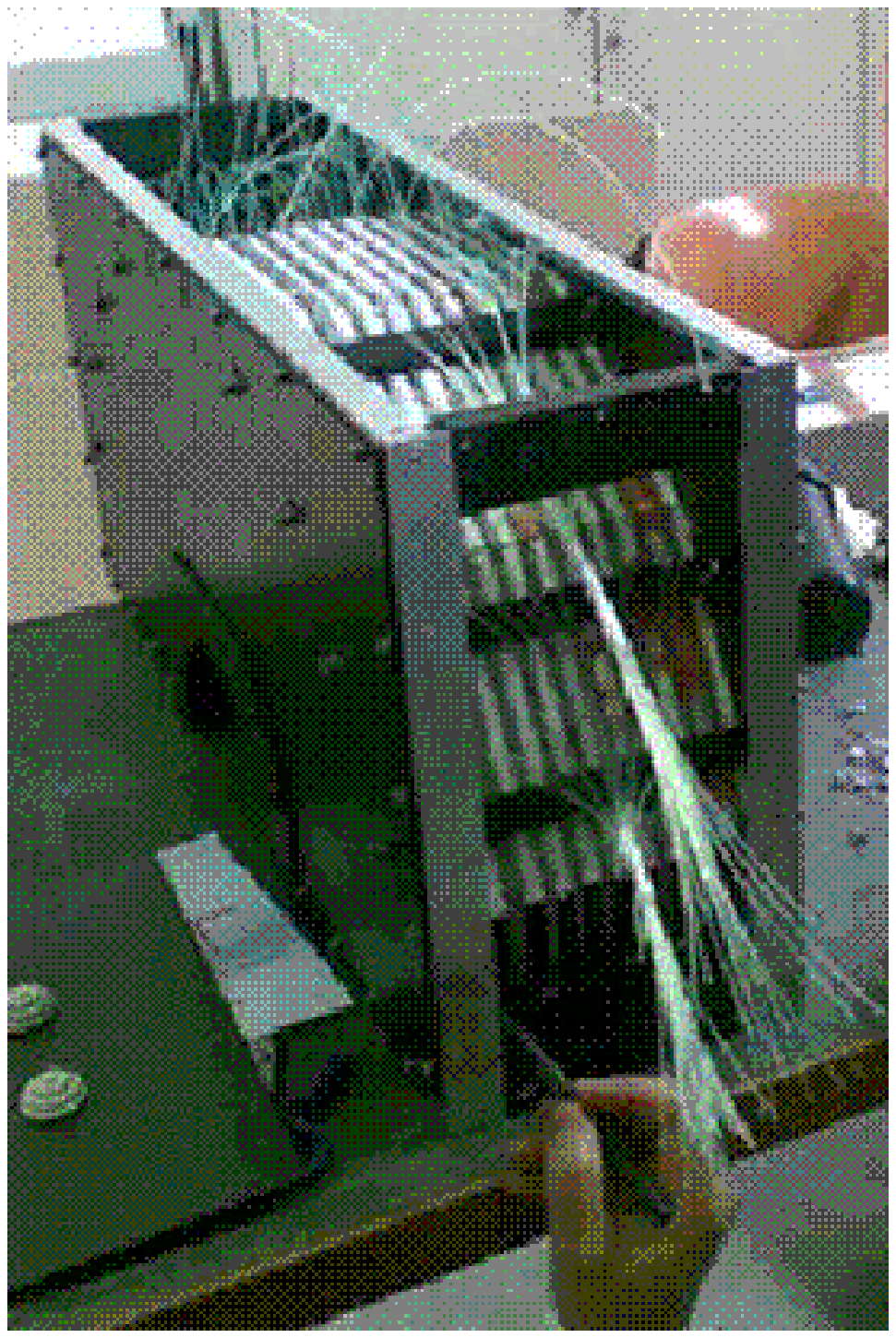}
\end{center}
\caption{ \it { Some pictures of a calorimeter module (five active sectors) assembly which took place in December 2007 and January 2008 at Frascati. The left picture shows the tiles wrapped in \TYVEK\ sheets to improve the light collection efficiency; the WLS fibers, already bundled together and waiting for being connected to the PMT, are also apparent. In the right picture, the module has been put in its holding structure and is in the final building steps before being moved to the \DAF\ IP. The side view allows one to see the twelve scintillator planes such as the eleven lead layers. } }
\label{fig:calo2}
\end{figure}

The design sector energy resolution of $~35\%$ at 510~MeV is accurate enough to rely on a threshold on the reconstructed energy to select candidate Bhabha events. After assembly, the four modules have been tested and calibrated at the \DAF\ Beam Test Facility~\cite{ref:btf}. Several position and energy (linearity and resolution) scans have been performed with beams of energy 141, 189, 283, 377 and 471~MeV respectively. Additional pictures of the calorimeter building process can be seen in Fig.~\ref{fig:calo2}.

\subsection{GEM trackers}
\label{sec:gems}

\begin{figure}[h!]
\centering
\includegraphics[width=10.0cm]{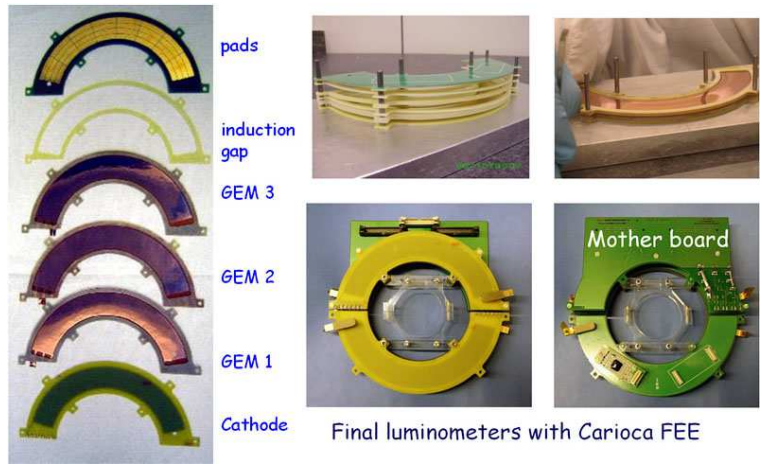}
\caption{ \it { Assembling a GEM tracker. } }
\label{fig:gemsview}
\end{figure}

Rings of GEM detectors~\cite{ref:gem} aiming at identifying charged particles from Bhabha events have been designed to be installed in front of each calorimeter, at a distance of 18.5~cm from the IP. Both trackers are divided into two vertical half moon-shaped units which surround the beam pipe. The top (bottom) half covers azimuthal angles between $14^\circ$ and $166^\circ$ ($194^\circ$ and $346^\circ$) respectively. Each of the four GEM trackers is segmented into thirty-two pads: four 6.5~mm-wide rings contain eight cells covering $19^\circ$ in azimuth each; the ring boundaries are located at 68, 74.5, 81, 87.5 and 94~mm respectively from the beam axis. The trackers are made of three GEMs, thin (50~$\mu$m) kapton foils sandwiched between two copper clads and perforated by a dense set of holes (70~$\mu$m diameter, 140~$\mu$m pitch). They are glued on a G10 frame and assembled between the cathode and the anode forming four gaps: from the IP to the calorimeter, the drift gap (3~mm), two transfer gaps (1 and 2~mm) and the induction gap of 1~mm. All the volume between the anode and the cathode is filled with a gas mixture of $Ar-CO_2-CF_4$ (45\%-15\%-40\%)~\cite{ref:lhcb}.

When a charged particle crosses the drift gap it generates 10-15 primary electrons which drift thanks to the 3 kV/cm electric field before being multiplied by the GEM foils. As a high potential difference (about 400~kV) is applied between the copper sides, the holes act as amplifiers and the gain of each layer is about 20~-- hence a factor 8~000 in total. The electron cluster coming out from the last GEM foil induces a signal on the anode which is amplified and discriminated by a Carioca Chip~\cite{ref:carioca} plugged on the back side of the pad PCB. The two front-end electronics cards designed and developed in Frascati are supplied through a mother board that houses two output connectors with 16 LVDS channels each and one input connector with low voltage supplies and two thresholds channels. High voltage is supplied to the four GEM trackers with a new system made of four active dividers~\cite{ref:hvgem}.

\begin{figure}[h!]
\centering
\includegraphics[width=6cm]{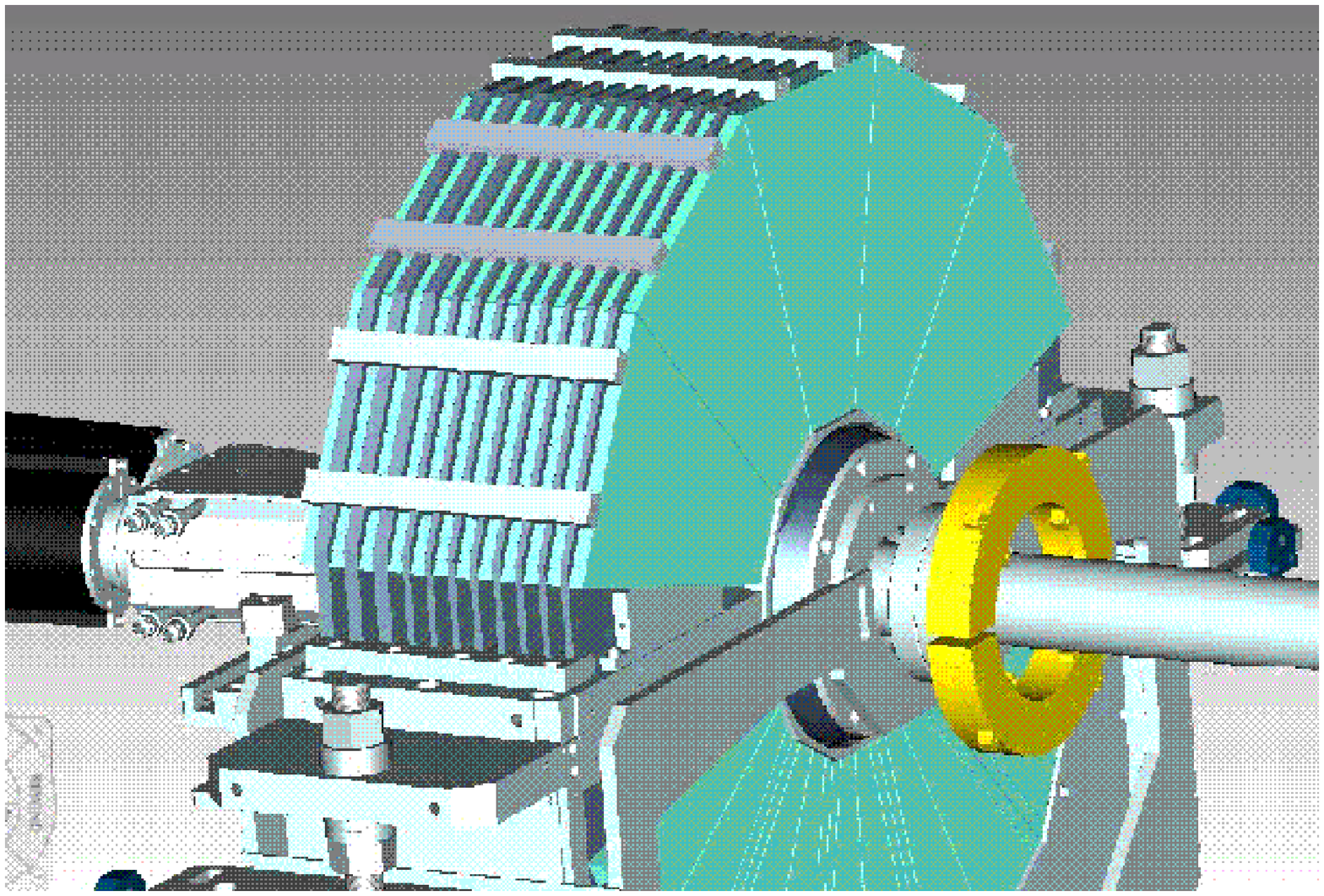}
\includegraphics[width=6cm]{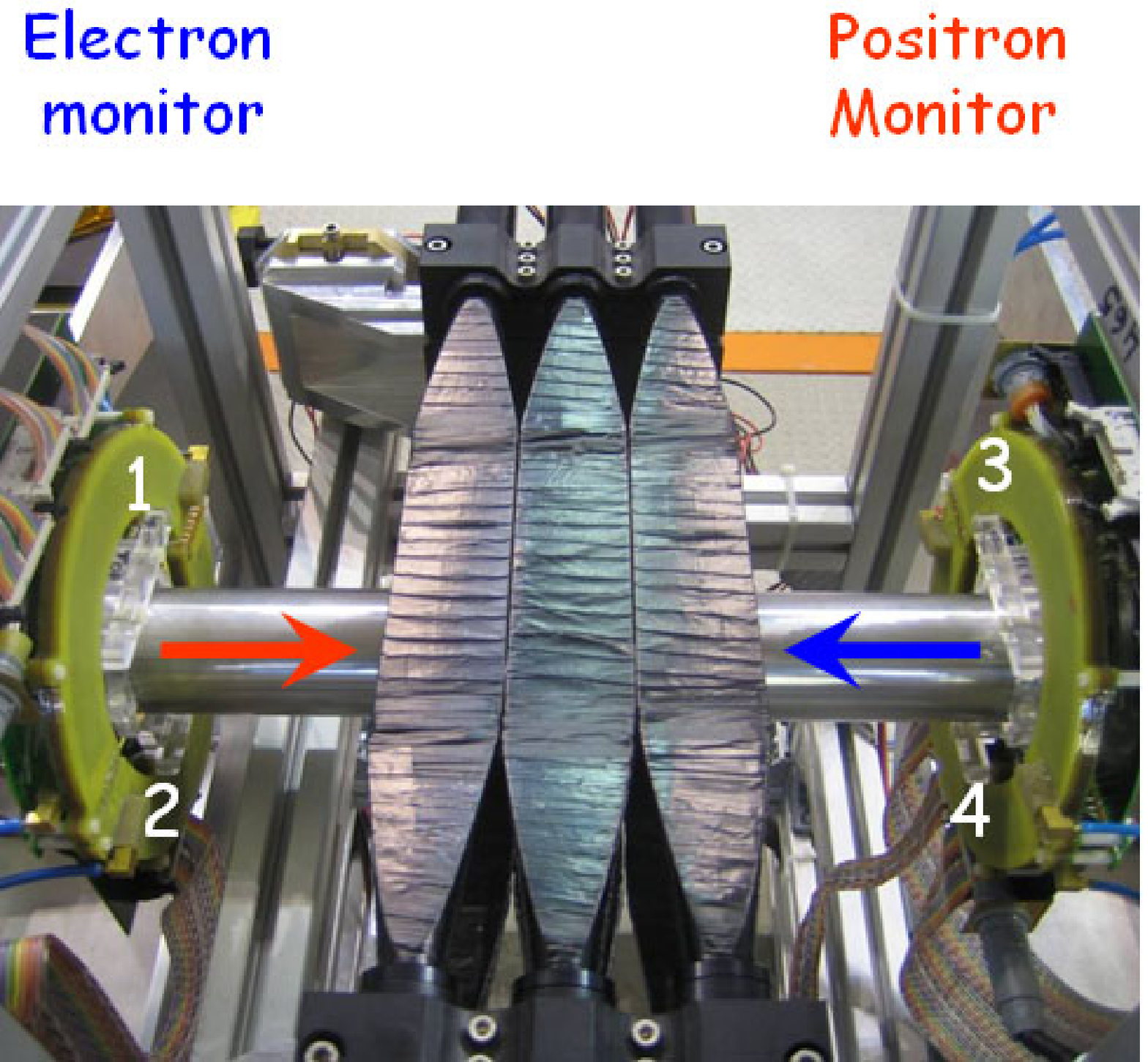}
\caption{ \it { Left: this drawing shows one side of the calorimeter with the GEM trackers visible~-- the rings around the beam pipe in front of the calorimeter structure. Right: top view of the \DAF\ IP with the \SID\ detector (the three oval-shaped structures in the center) and the GEM trackers (the circular shapes in front of the calorimeter visible on either side) installed in front of the Bhabha calorimeter. The numbering convention for the GEM trackers is described on this picture: trackers 1 and 2 monitor the electron beam while trackers 3 and 4 are sensitive to positrons; odd (even) modules are above (below) the beam line. } }
\label{fig:lumi-gem1}
\end{figure}

As described above, Fig.~\ref{fig:gemsview} shows the different layers of a GEM tracker (left picture) such as different steps of the assembly process (the four pictures on the right). Fig.~\ref{fig:lumi-gem1} presents one drawing of a calorimeter side with two GEM trackers in front of it (left plot) and an actual picture (right plot) of the IP taken at the beginning of the 2008 \DAF\ run, when the trackers were installed in their nominal location.

\subsection{Gamma monitors}
\label{sec:gammamon}

\begin{figure}[!h]
\centering
\includegraphics[width=10.0cm]{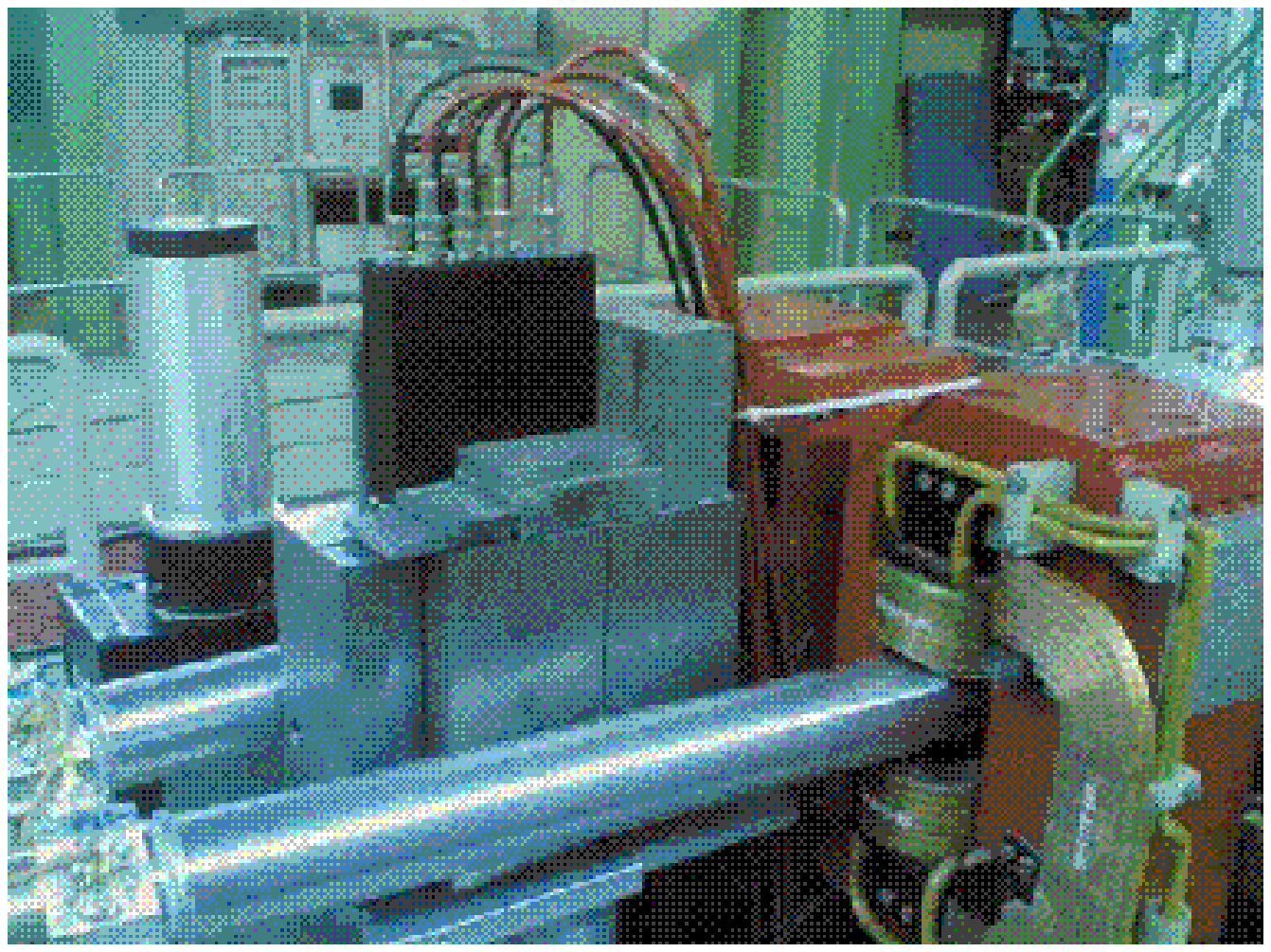}
\includegraphics[width=10.0cm]{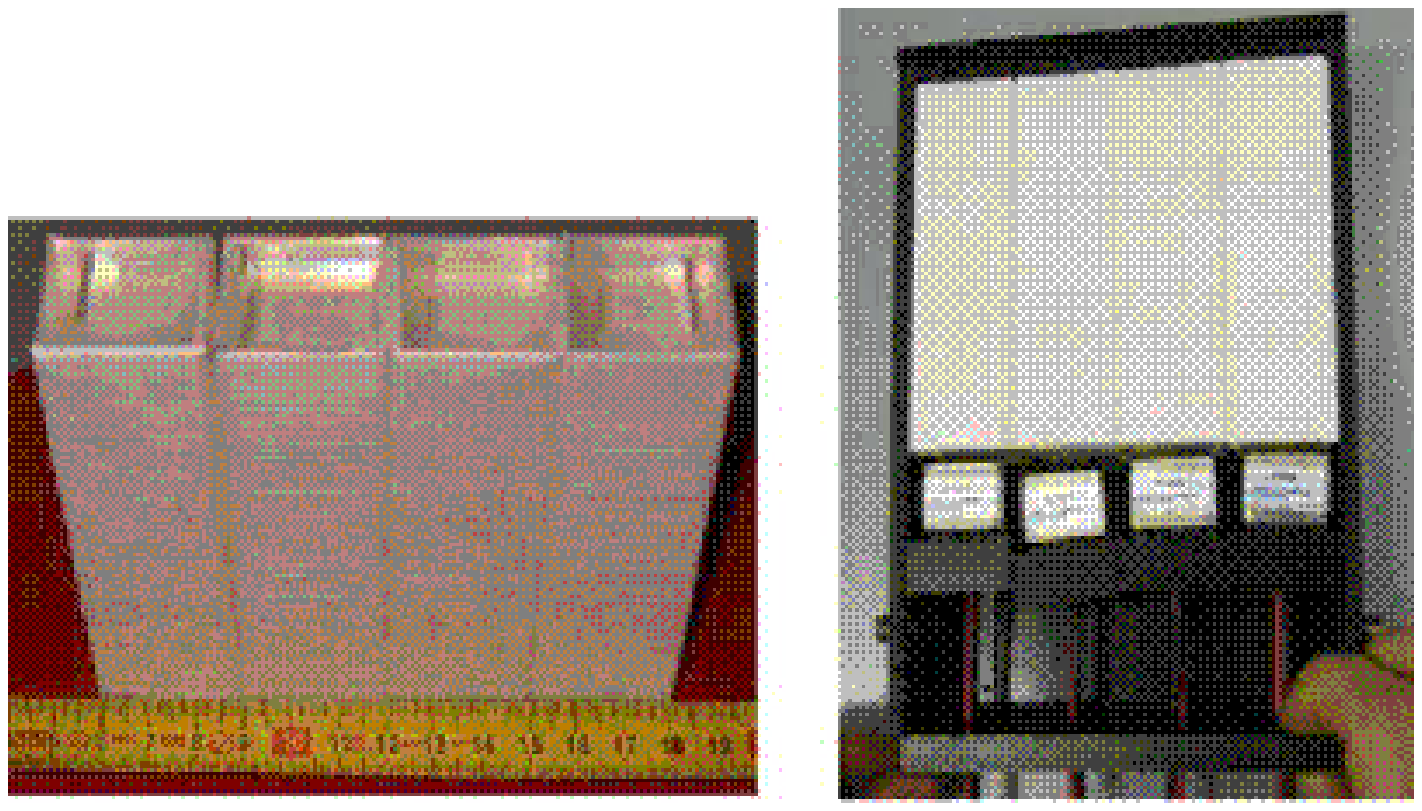}
\caption{\it { The gamma monitors are installed 170~cm away from the IP (top picture, taken in the \DAF\ hall). The window opened through the lead shielding is aligned with the trajectory of the photons produced in radiative Bhabha events. Each detector is made of four PbWO$_4$ crystals, wrapped in \TYVEK\ foils (bottom left picture). They are put together into a PVC box with the four PMTs whose entry windows are in direct contact with the top face of the crystals (bottom right picture). } }
\label{fig:gammon1}
\end{figure}

Two gamma monitors shown in Fig.~\ref{fig:gammon1} are located 170~cm away from the IP on either side and collect photons radiated by the electron or positron beams. These detectors are made of four PbW0$_4$ crystals which have a rectangular prism shape: a square section of $30 \times 30$~mm$^2$ and a height of 110~mm. The crystals are maintained together along their longer sides in order to form a structure 30~mm-wide in front of the photon beam and 120~mm-thick (about 13~$X_0$). Each crystal is readout by a Hamamatsu R7600 compact photomultiplier~\cite{ref:pmts2} which is in direct contact with the top face of the crystal. This design helps minimizing the background from particles hitting directly the PMT. Because of the boost introduced by the IP crossing angle, the photon trajectories are shifted toward the inner side (along the $-x$ direction) of the machine. To account for this deviation, the gamma monitors are located at $x=-5$~cm from the beam pipe and rotated by 4$^{\circ}$ in the horizontal plane with respect to the beam axis.

Thanks to their high logging rates, these detectors provide quick estimates of the luminosity and background variations. These quantities are then used for the machine real-time optimization. They cannot easily provide an absolute measurement of the luminosity as their counts would have to be corrected by a factor which is continuously changing as the beam currents and the machine conditions evolve. However, on a short time scale and as relative luminosity monitors, those counters have been found extremely useful.

\subsection{IP shielding}
\label{subsection:shielding}

\begin{figure}
\begin{center}
\includegraphics[width=8cm]{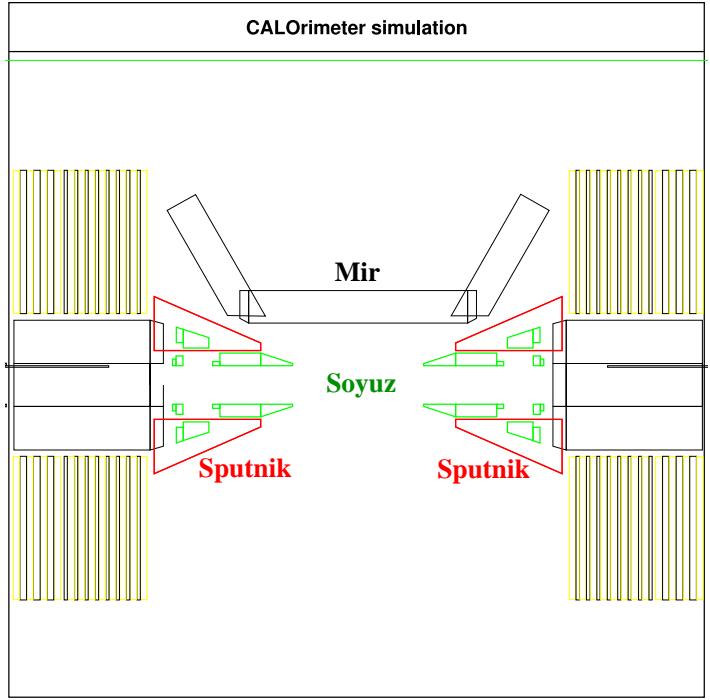}
\end{center}
\caption{\it { \GEANTTHREE\ picture showing a side view of the \DAF\ IP shielding surrounded by the Bhabha calorimeter. The small green components closer to the beam pipe form the Soyuz; the red triangles are the Sputnik while the black rectangular volume corresponds to the Mir shielding.} }
\label{fig:shieldings}
\end{figure}

During the course of the 2008-2009 \SID\ runs, ad-hoc lead shieldings have been constructed around the IP to protect this sensitive detector against low energy background photons. Three different types of shielding have been used, called ``Soyuz'', ``Sputnik'' and ``Mir'' respectively. The Soyuz defines the low angle edge of the  calorimeter acceptance by discarding tracks which are kicked back in the fiducial volume after showering in the QD0, a process potentially hard to simulate correctly. The Sputnik is meant to absorb the background generated by the beam bending inside the QD0 magnets; it surrounds the Soyuz. Finally, the Mir shield completes the shielding setup by closely surrounding the calorimeter. Fig.~\ref{fig:shieldings} shows these different components on a side view of the simulated IP~-- see Section~\ref{sec:montecarlo} for details.

The shape and the position of these elements are dictated by the existing machine optics and have been slightly changing over time. All in all, they result in a partial decrease of the luminometer acceptance. More importantly, they have a significant impact on the GEM detector location and thus on its performances, as described in Section~\ref{subsection:GEMperf}.

\section{Data acquisition and trigger}
\label{sec:daq}

Most of the front-end (FE) electronics and data acquisition (DAQ) components, as well as the PMT HV supplies, are taken from the KLOE experiment~\cite{ref:daq}. The signals coming from the calorimeter are amplified, inverted and delivered to a 3-stage splitter (SDS board). The first stage is used for timing measurements and consists of a constant fraction discriminator which delivers a current signal to the FE KLOE TDCs~\cite{ref:tdc} whose resolution is 1.04~ns. Then, the energy measurement stage includes a 3-pole Bessel filter followed by a KLOE charge ADC (0.25~pC resolution); the filter is needed to match the PMT signal bandwidth with the ADC input stage. Finally, the third stage sums all signals coming from the same calorimeter module and this information is used as input to the trigger logic. A candidate Bhabha event is defined as two high-enough energy deposits in back-to-back modules which are compatible in time~-- see Section~\ref{sec:performances} for details.

\begin{figure}
\begin{center}
\includegraphics[width=6in]{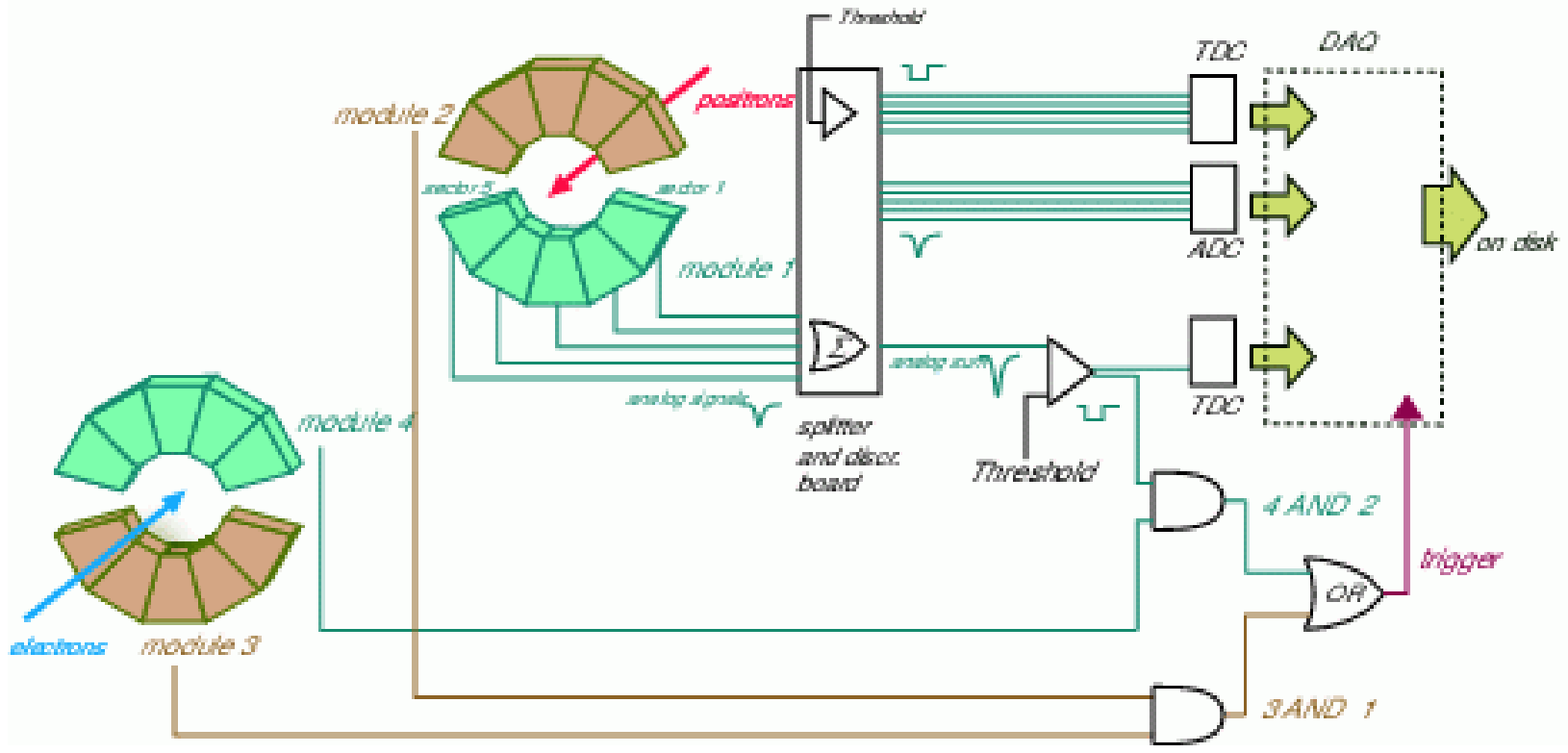}
\end{center}
\caption{ \it { Overview of the Bhabha calorimeter data acquisition and trigger systems. This detector is made of four modules (labeled M1 to M4) which are split into five azimuthal sectors each. A candidate Bhabha event is a time-compatible coincidence between two high-enough energy deposits in back-to-back modules: ``(M1 AND M4) OR (M2 AND M3)''. } }
\label{fig:daq}
\end{figure}

The KLOE FE modules are build around a custom bus, the AUX-bus~\cite{ref:daq2}, which allows a fast readout of the input data as it only uses the VME standard for initialization. The AUX interface builds sub-events crate by crate ensuring that the data are properly synchronized. A trigger-driven readout controller delivers these frames to the control manager board via another custom bus (the C-bus) which interconnects different crates. The event data are then stored into a FIFO memory where they are read using VME 32 bits block transfer mode~\cite{ref:branchini} by a MVME6100 processor running KLOE protocols. This CPU sends them to a PC computer via a gigabit Ethernet connection where they are finally written to disk. A simple graphical user interface (GUI) developed in JAVA performs the run control functions. The entire acquisition chain is installed in the collider hall, close to the \DAF\ IP. Fig.~\ref{fig:daq} summarizes the calorimeter data acquisition and trigger systems. 

The gamma monitor signals are split after the PMTs: one half is sent to the charge ADC of the KLOE data acquisition system (and not used further down for the measurements), while the other goes to an analog mixer. The analog sum of the four crystals is then discriminated and the counts are read by the \DAF\ control system via a VME scaler readout module. Finally, the GEM signal is triggered by the calorimeter; the data coming from the GEM detectors are discriminated, delivered to FE TDCs and readout via an AUX-bus. This scheme guarantees synchronization between calorimeter and GEMs.

\section{Performances}
\label{sec:performances}

\subsection{The Bhabha calorimeter}

\begin{figure}
\begin{center}
\includegraphics[width=2.5in]{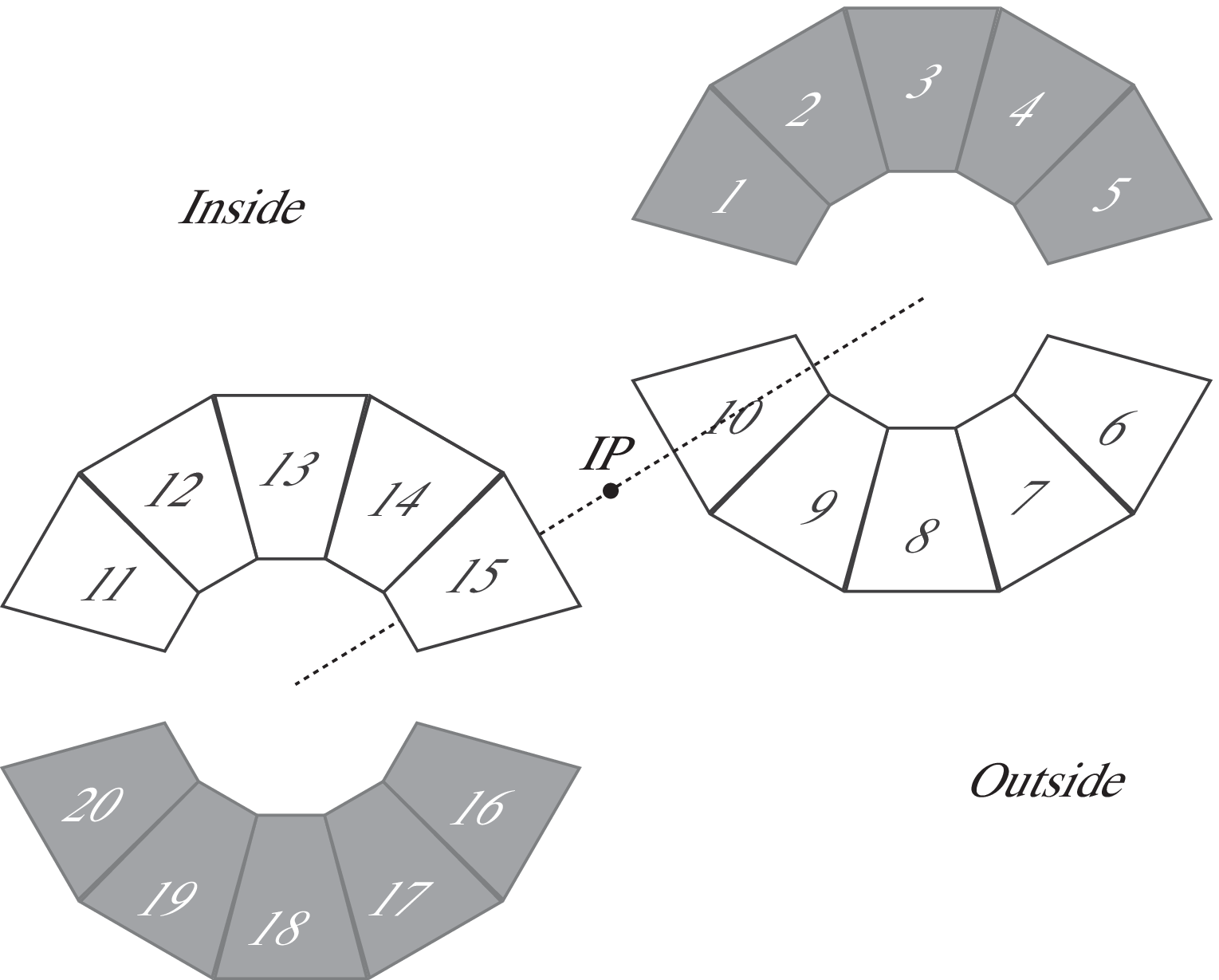}
\includegraphics[width=2.5in]{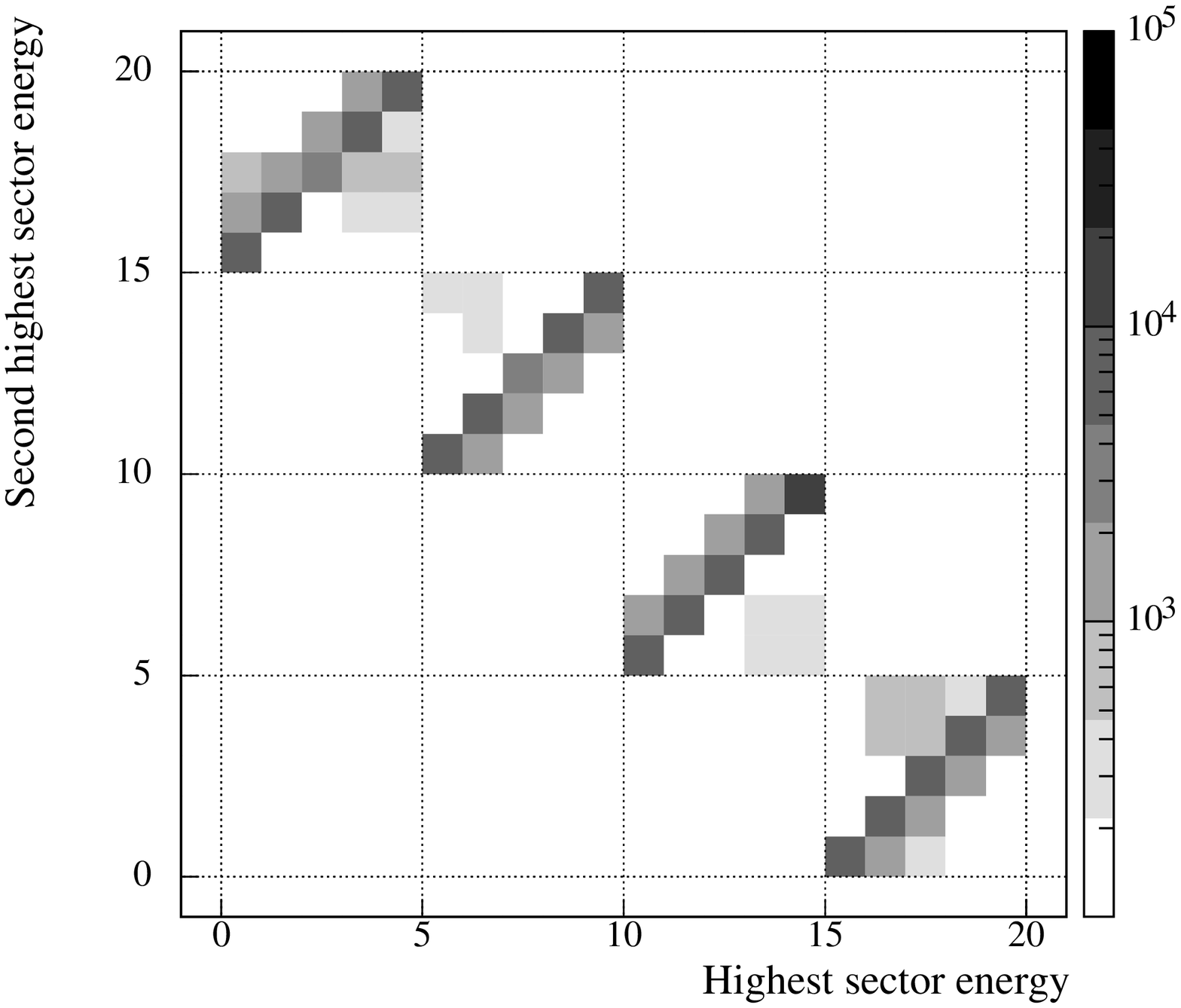}
\end{center}
\caption{\it { Given the numbering convention of the calorimeter sectors (see left plot), the position of the two calorimeter sectors with the highest energy deposits can be compared event by event. The result (2D-histogram on the right showing the sector number with the second highest energy deposit vs. the sector number with the highest deposit) shows that these two sectors are most of the time back-to-back, as expected from Bhabha events~-- note that the z-scale is logarithmic. }  }
\label{fig:bhabha}
\end{figure}

\begin{figure}
\begin{center}
\includegraphics[width=6in]{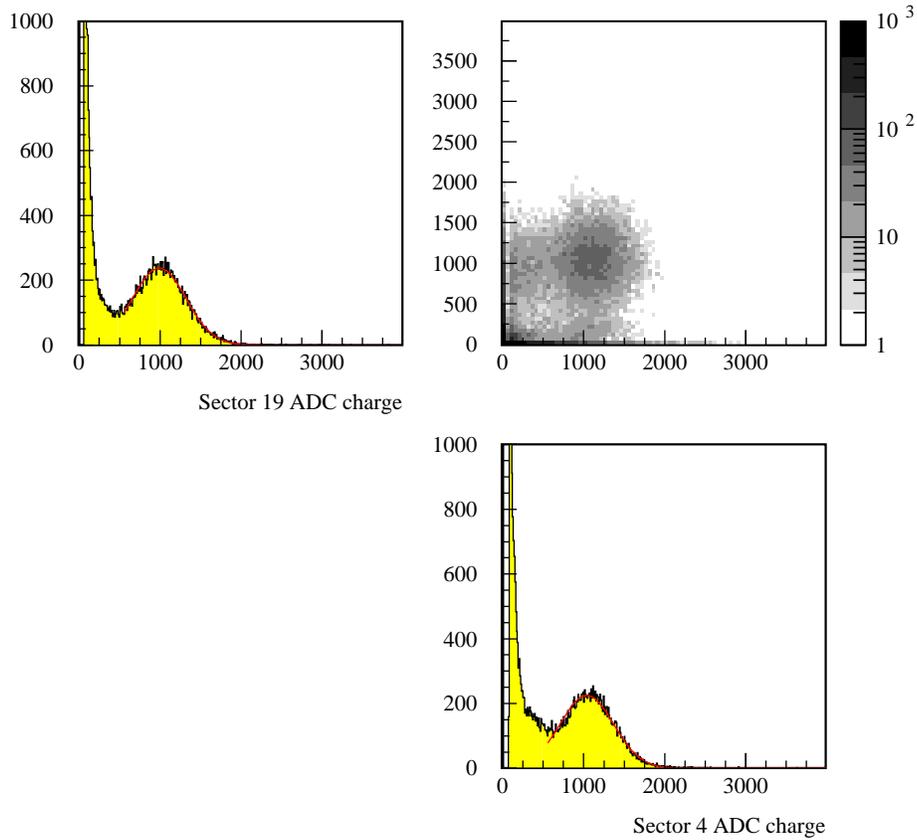}
\end{center}
\caption{\it { Top left and bottom right plots: ADC distributions of the back-to-back calorimeter sectors 19 and 4 respectively. The bumps around 1000 counts correspond to the true Bhabha events while events for which these sectors did not trigger are associated with smaller energy deposits. Top right plot: a 2D histogram showing the correlation between the two sector ADC charges; high energy deposits in one sector are associated with deposits of similar magnitude in the other sector, the expected signature of real Bhabha events. } }
\label{fig:bhabha2}
\end{figure}

Potential Bhabha elastic scattering events $\epem \to \epem$ are identified at the trigger level by requiring two coincident energy deposits above a given threshold in back-to-back modules. The locations of the two sectors with the highest energy deposits are indeed correlated, as shown on Fig.~\ref{fig:bhabha}. Taking as example the opposite sectors 19 and 4, Fig.~\ref{fig:bhabha2} shows the ADC distributions of each sector independently, such as the correlation between these two quantities. Energetic events corresponding to actual Bhabha events are common to both sectors.

\begin{figure}
\begin{center}
\includegraphics[width=14cm]{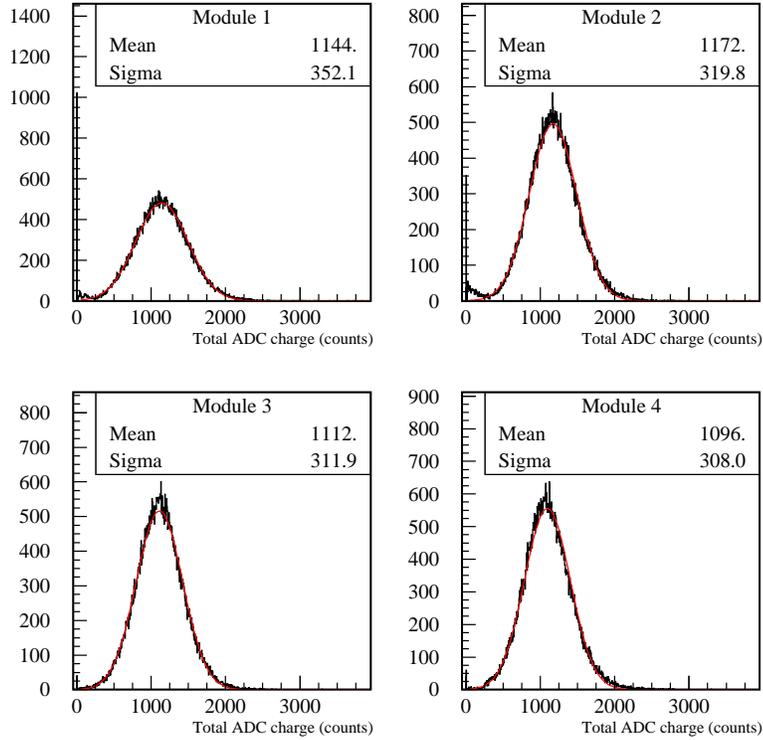} 
\end{center}
\caption{\it { Distribution of the ADC counts in the four calorimeter modules from events in which they are triggering. Fitting these curves provides the corresponding energy resolutions which are around 20\% at the \DAF\ energy (510~MeV); the calorimeter design performance is experimentally achieved in this representative high-statistics run.} }
\label{fig:bhabha3}
\end{figure}

Summing all the sector contributions from a given calorimeter module, Fig.~\ref{fig:bhabha3} shows the distribution of the ADC counts for triggered events in the four modules. Fitting these distributions by a Gaussian, one gets their energy resolutions.

\begin{equation}
\frac{ \sigma(E) }{ \sqrt{E} } \; \mathrm{ with} \; E=510~\mathrm{MeV}
\end{equation}

The results are 22\%, 19\%, 20\% and 20\% for the modules M1 to M4 respectively: the design performances are fulfilled in the whole calorimeter.

Accidental coincidences involving background particles hitting part of the calorimeter do contribute to the trigger. In order to convert the calorimeter rate into an absolute luminosity measurement, one needs to identify and remove this noise contamination, whose level is highly dependent on the actual machine running conditions and cannot be determined a priori. Therefore, a filtering procedure has been implemented (first offline and then directly in the DAQ farm) to improve the accuracy of the online luminosity provided to the \DAF\ control room. The raw trigger rate, $R_{raw}$, is corrected by a factor $0 \le CF \le 1$ to give the measured Bhabha rate $R_{meas}$ which is then scaled by the \MC\ prediction to compute the luminosity~-- see Eq.(\ref{eq:lumi}).

\begin{equation}
R_{meas} = R_{raw} \times CF
\end{equation}

\begin{figure}
\begin{center}
\includegraphics[width=6cm]{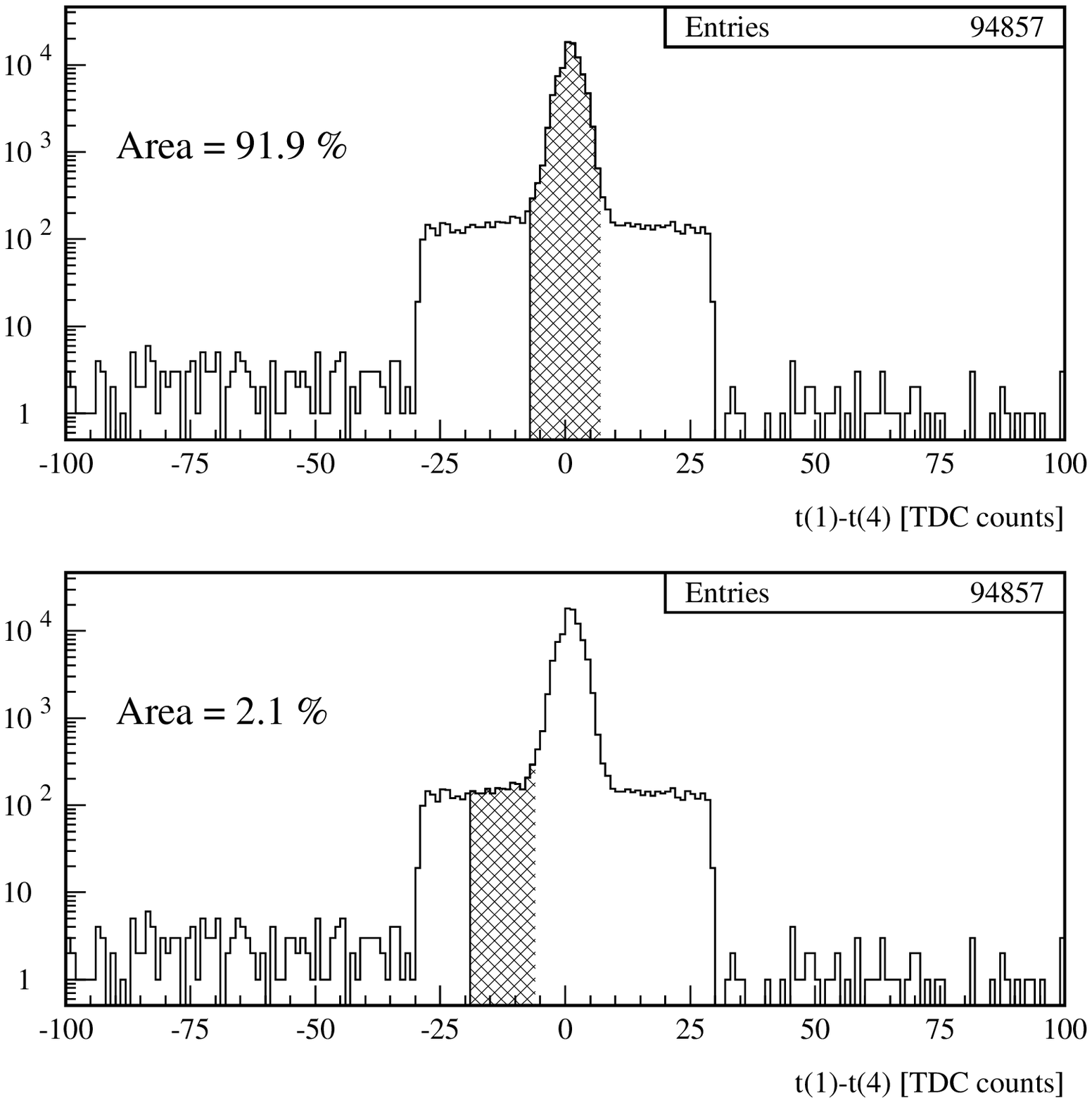}
\includegraphics[width=6cm]{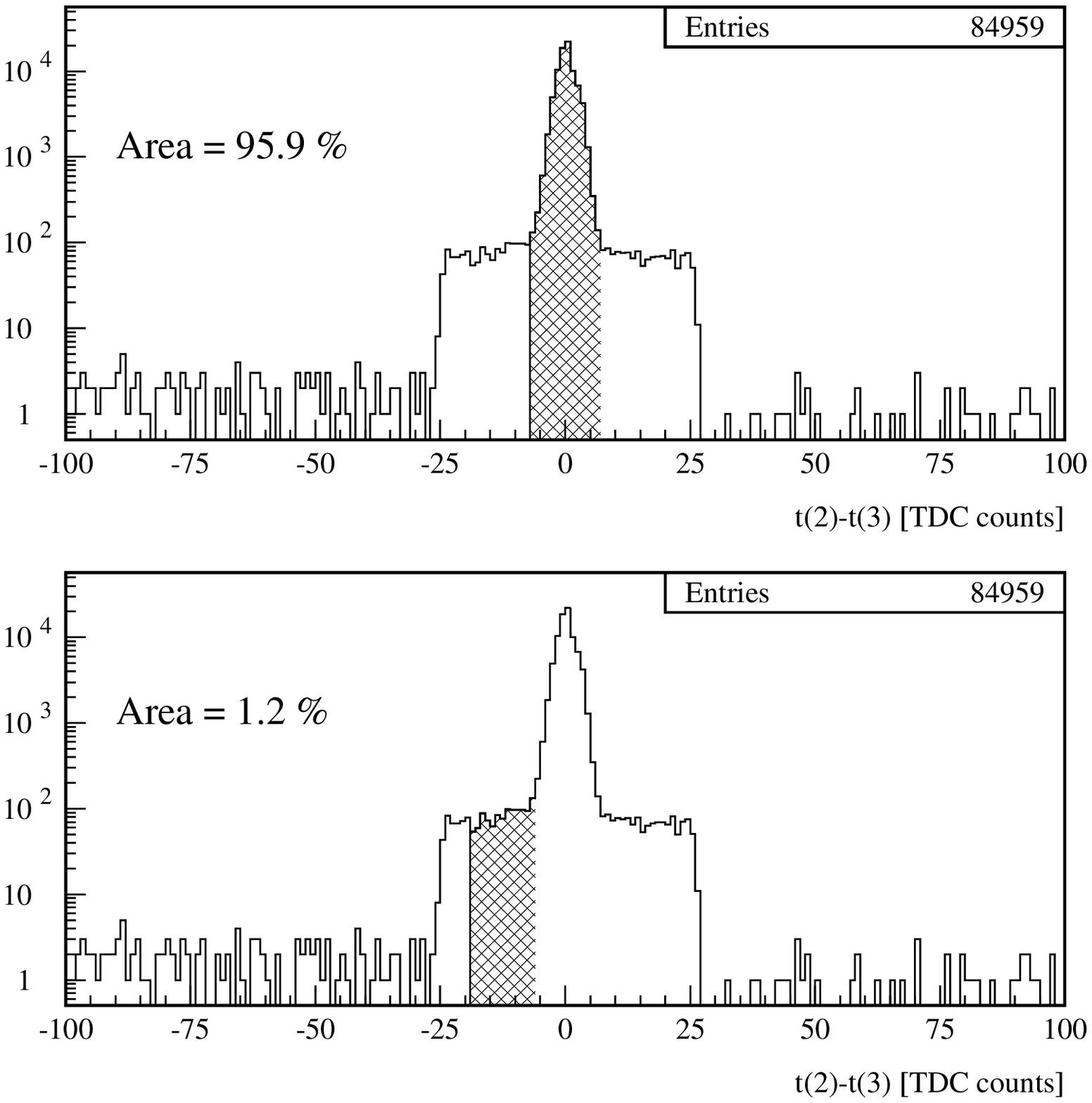}
\end{center}
\caption{ \it{ Distributions of the time difference $\Delta t$ between back-to-back modules triggering in coincidence (note the logarithmic $y$-scale on all histograms). Left column: modules M1 and M4; right column: modules M2 and M3. These histograms clearly show two components: a flat background from random triggers plus a Gaussian distribution centered at $\Delta t = 0$ which corresponds to genuine Bhabha events~-- see text for details. The width of this peak is a few ns (1 TDC count corresponds to 1.04~ns). The background level is about 10\% for the M1-M4 trigger and 5\% for the M2-M3 trigger in this example. } }
\label{fig:timediff-new}
\end{figure}

The value of the correction factor $CF$ is adjusted every 3000 events (at most every few seconds depending on the beam currents) by looking at the distribution of the time difference $\Delta t$ between the triggering modules. As shown in Fig.~\ref{fig:timediff-new} obtained from a representative high-statistics run, this distribution has two main components:
\begin{itemize}
\item a narrow Gaussian peak centered at $\Delta t = 0$ and corresponding to genuine Bhabha events (as demonstrated below),
\item sitting on top of a flat background coming from random coincidences between modules and whose width is determined by the duration of the digital signals building the coincidence ($\simeq$ 25 ns). 
\end{itemize}
The pattern is similar for both pairs of modules: M1-M4 on the left, M2-M3 on the right.

\begin{figure}
\begin{center}
\includegraphics[width=12cm]{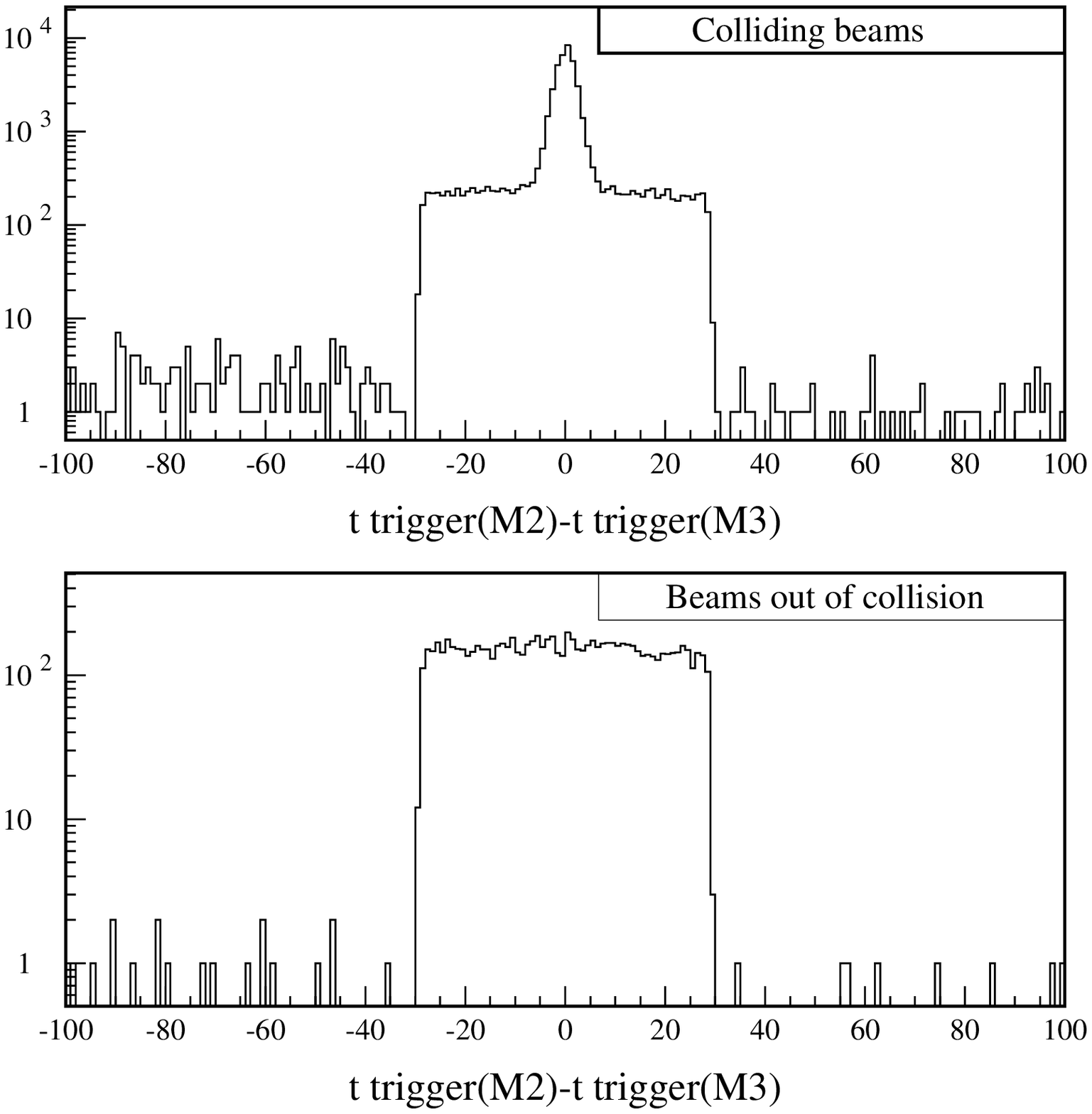}
\end{center}
\caption{ \it { Upper plot: histogram of the trigger time difference between the back-to-back modules M2 and M3 accumulated with beams in collisions; lower plot: same histogram, but this time for beams out of collisions (longitudinally separated). The in-time peak in the timing difference distribution does correspond to luminosity-induced Bhabha events.} }
\label{fig:r1520-r1522-ttr}
\end{figure}

\begin{figure}
\begin{center}
\includegraphics[width=15cm]{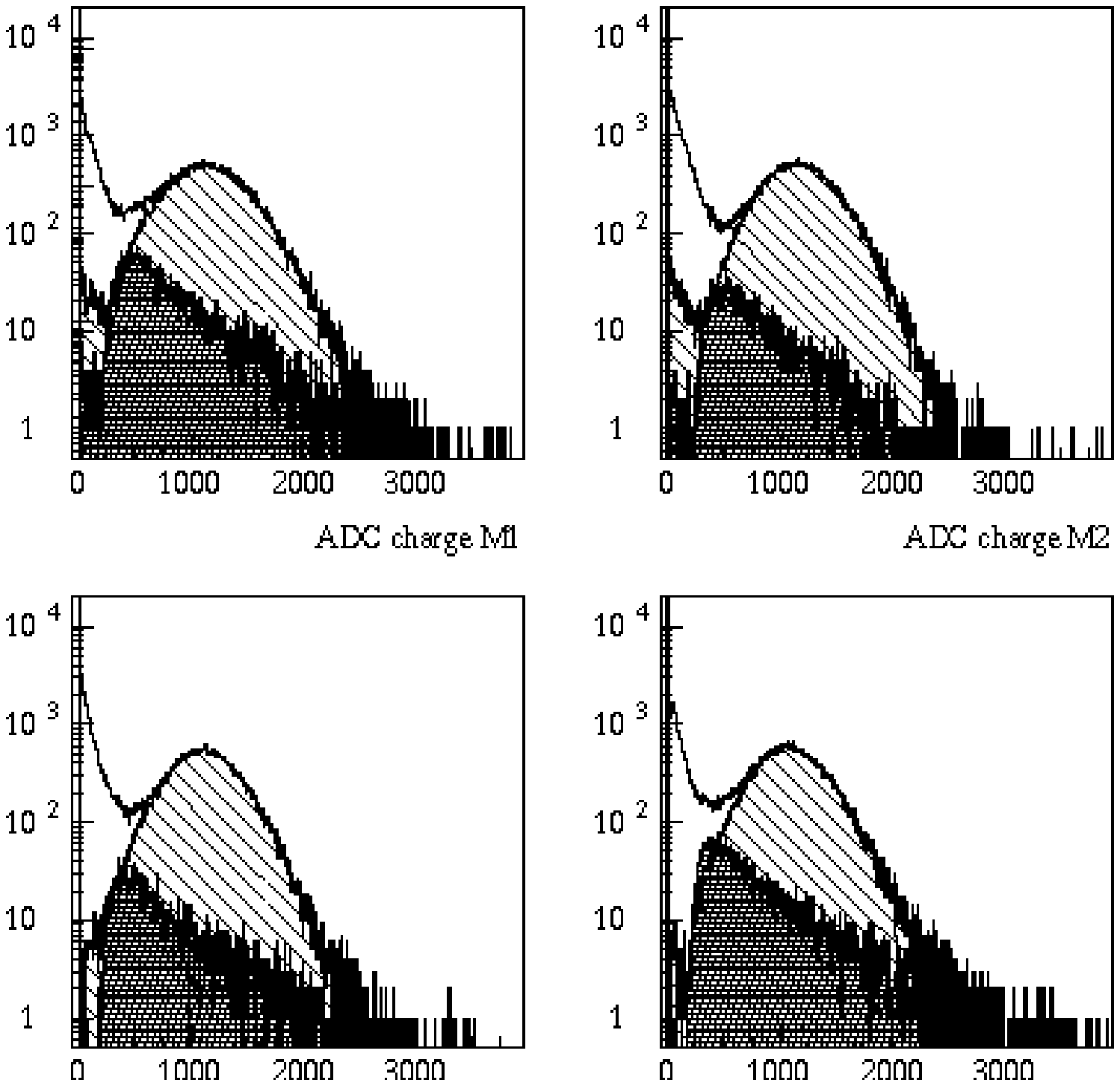}
\end{center}
\caption{ \it{ ADC distributions in the four calorimeter modules for the high-statistics run used to produce most of the plots in this section: all histograms show three components with similar patterns. Events for which that particular module did not trigger belong to the white area: little energy is deposited in these sectors, as expected from background fluctuations. Hence, these events are not real Bhabha decays for which one of the two particles is not detected. The hatched regions contain the other events for which the module triggered. Events for which the energy deposit is not coincident in time with the one measured in the opposite module (pure background case) are in the darker area. The deposited energy is lower than for the in-coincidence events (high purity Bhabha sample) which are in the light-hatched area. These plots and those in Fig.~\ref{fig:r1520-r1522-ttr} justify the timing criterion used to separate signal from background.} }
\label{fig:energydepositFour}
\end{figure}

Two observations allow us to justify the fact that the peak is produced by real Bhabha events. First, it disappears when the beams are longitudinally separated~-- see Fig.~\ref{fig:r1520-r1522-ttr}. Then, the energy deposited in the triggering modules is significantly larger for these events than for those which originate from the background. Using the same high-statistics run, Fig.~\ref{fig:energydepositFour} shows the ADC distributions in the four modules for all events. Three components are visible. 
\begin{itemize}
\item The white component peaking at 0 corresponds to events for which the other pair of modules (M1-M4 for the M2 or M3 histograms; M2-M3 for M1 and M4) triggered. As expected, there is none or little energy in the modules which did not 'see' a Bhabha decay.
\item The dark-hatched area contains events for which the module triggered while the trigger times were not coincident. These events correspond to the flat background shown in Fig.~\ref{fig:timediff-new}.
\item Finally, the light-hatched area shows the triggered events which are coincident in time. Still comparing this plot with Fig.~\ref{fig:timediff-new}, these are the genuine Bhabha events plus a small background component (note the logarithmic $y$-axis on both figures) one has to subtract to estimate correctly the luminosity.
\end{itemize}

\begin{figure}
\begin{center}
\includegraphics[width=15cm]{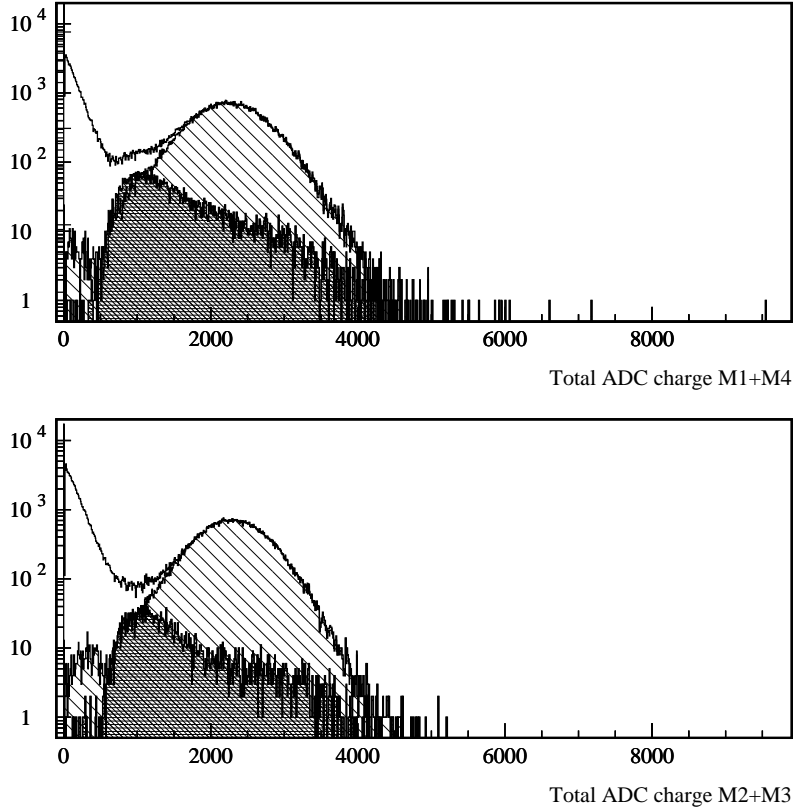}
\end{center}
\caption{ \it { Distributions of the sum of the energies deposited in opposite modules (M1-M4 and M2-M3) in the high-statistics run taken as example in this section. The three categories of events~-- not triggering (white area), triggering but coincidence out of time (dark hashed area) and triggering in-time (light hashed area)~-- are clearly visible and confirm that the Bhabha events fall in the third category. } }
\label{fig:energydepositTwo}
\end{figure}

Similar conclusions can be drawn from Fig.~\ref{fig:energydepositTwo} in which the charges measured by two opposite modules are summed. Triggering in-time events are mainly due to genuine Bhabha whereas the two other categories (triggering out-of-time or not triggering) are background-related.

To estimate the amount of background under the Bhabha peak, a sideband region containing only random coincidences is defined~-- see the hatched area in the bottom plots of Fig.~\ref{fig:timediff-new}. Counting the number $N_{in}$ of events in the $\Delta t$ range [-6;6] ADC count (corresponding to about $\pm 3\sigma$ of the Gaussian Bhabha peak, 1 count is 1.04~ns) and the number $N_{sb}$ of events in the sideband region (of equal width: 12 ADC counts), one has:

\begin{equation}
CF = 1.0 - \frac{ N_{sb} }{ N_{in} }
\end{equation}

\begin{figure}
\begin{center}
\includegraphics[width=10cm]{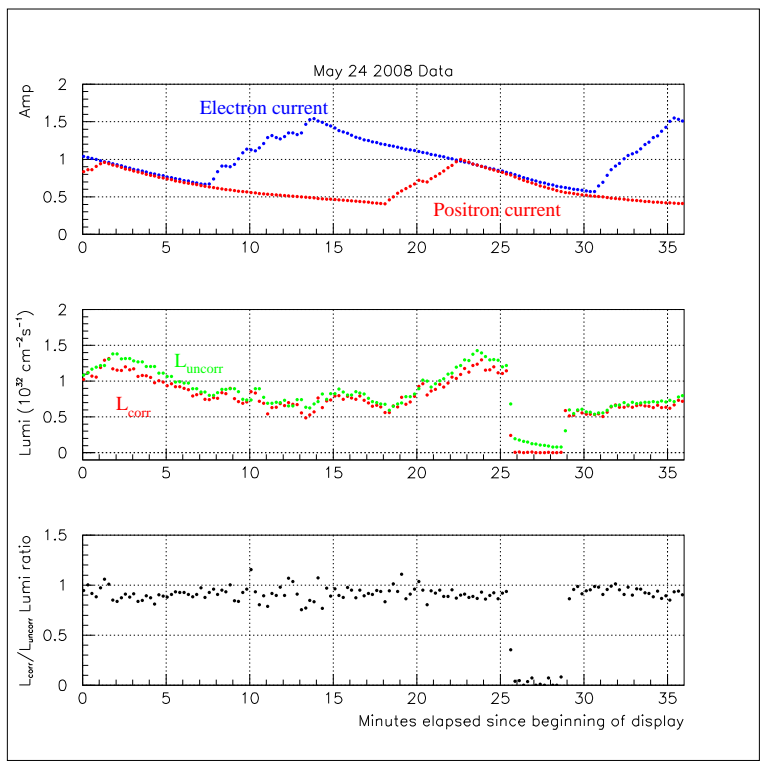}
\end{center}
\caption{ \it { Display of about 35 minutes of data taken on May 24$^{\mathrm th}$ 2008. From top to bottom: beam currents, uncorrected $L_{uncorr}$ and corrected $L_{corr}$ luminosities, ratio $L_{corr}/L_{uncorr}$. For this particular period of running, the background level is between 10-20\%. Moreover, when the beams are longitudinally separated (between minutes 26 and 29) $L_{corr}$ goes immediately to 0 (as expected) whereas $L_{uncorr}$ does not. } }
\label{fig:20080524}
\end{figure}

This simple procedure provides an accurate real-time measurement of the absolute luminosity which is among the main feedback signals used by the \DAF\ operators. Fig.~\ref{fig:20080524} displays about 35 minutes of data taken on May 24$^{\mathrm th}$ 2008 to show the improvement brought by this background-subtraction algorithm. The top plot displays the beam currents, the middle one the uncorrected $L_{uncorr}$ and corrected $L_{corr}$ (background-subtracted) luminosities and the bottom one the ratio $L_{corr}/L_{uncorr}$. Apart for a few marginal points, the uncorrected luminosity is always greater than the corrected one, by 10-20\% in this particular example. The difference is even more striking between minutes 26 and 29, a time period during which the beams are longitudinally separated. If $L_{corr}$ goes immediately to 0 as expected, $L_{uncorr}$ remains sensitive to a background component whose decay time is on the timescale of a minute.

\subsection{The GEM trackers}
\label{subsection:GEMperf}

\begin{figure}[!h]
\centering
\includegraphics[width=10cm]{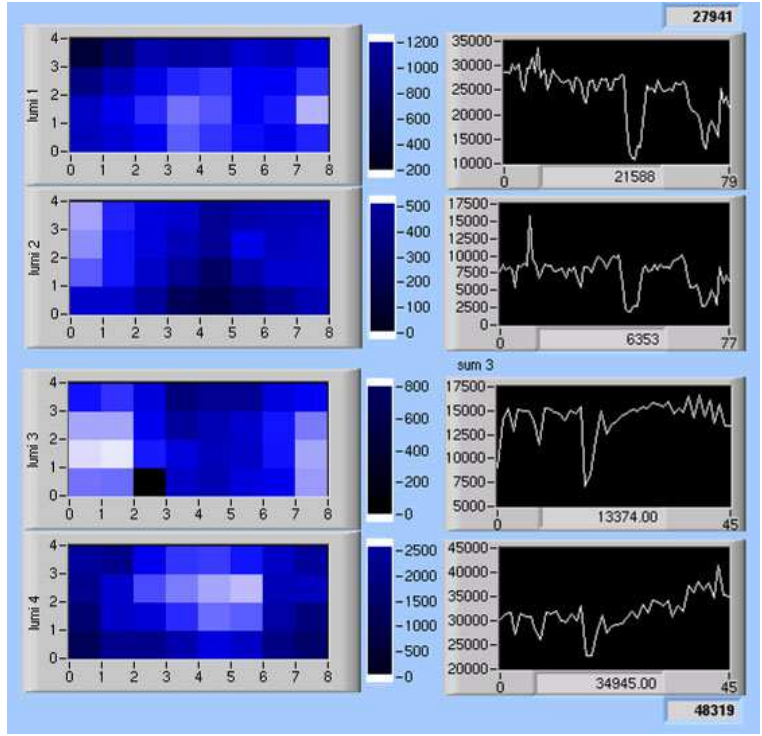}
\caption{ \it { These plots show how the GEM trackers were used as background monitors at the beginning of the upgraded \DAF\ commissioning~-- the rates presented in this example are typical of stable beam conditions. Each line of two plots corresponds to one of the four GEM trackers (recall that modules 1 and 2 are sensitive to the electron beam, 3 and 4 to positrons). On the left column, rates (in Hz) measured in each of the 32 tracker cells (a GEM module has an 8-fold segmentation in azimuth and a 4-fold segmentation in radius, see Section~\ref{sec:gems}) are shown: the lighter the color, the higher the rate. The background is typically around 1~kHz per pad. On the right column, the time evolutions of the tracker rates are displayed. Updated at 1~Hz, these plots (whose axis boundaries can be setup independently, which explains why the settings are slightly different on the four snapshots) have been used in real time by the \DAF\ operators to monitor the machine conditions. At the time this picture was taken, the electron beam background in the GEM was about 28~kHz while it was around 48~kHz for the positrons. More background was seen above (below) the beam line on the electron (positron) side. } }
\label{fig:gem_background}
\end{figure}

Unfortunately the GEM trackers could only be kept in their nominal position (18.5~cm from the IP) until April 2008 when the \SID\ lead shielding had to be extended to protect efficiently this detector from machine background. In that initial period, the GEM trackers were left out of the DAQ due to a conversion problem between the 128 LVDS Carioca channels and the ECL inputs of the KLOE TDCs. However, the four trackers were used as background monitors sending data every second to the \DAF\ control room. This information was available all the time, in particular during the injection phases. Fig.~\ref{fig:gem_background} shows a snapshot of the \DAF\ online stripcharts monitoring the background levels measured in the GEM trackers.
Pad-to-pad coincidences (64 opposite couples in OR) were also studied to get another estimate of the Bhabha rate. Yet, accidental coincidences were found to exceed significantly the geniune Bhabha rate and no significant variation was observed when the beams were put out of collision.

\begin{figure}[!h]
\centering
\includegraphics[width=6.0cm]{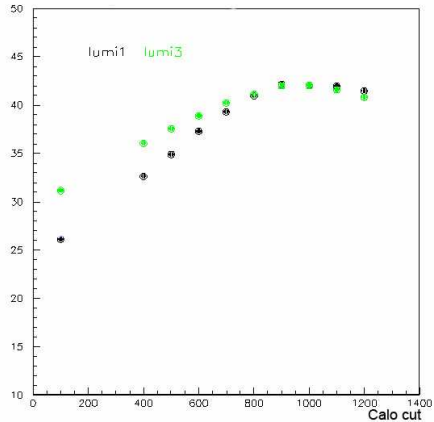}
\caption{ \it { Efficiencies (in $\%$) of the two GEM trackers located above the beam line (\#1 is sensitive to electrons, \#3 to positrons) versus the threshold applied offline on the calorimeter ADC output: at most 40-45\% of the calorimeter tracks are seen in the GEMs. This is explained by the fact that the trackers were closer to the IP than foreseen: a significant fraction of the tracks hitting the calorimeter passed through the GEM center hole and were not seen. The effect is enhanced by the strong polar angle dependance of the Bhabha cross-section. } }
\label{fig:gem_efficiency}
\end{figure}

\begin{figure}[!h]
\centering
\includegraphics[width=8cm]{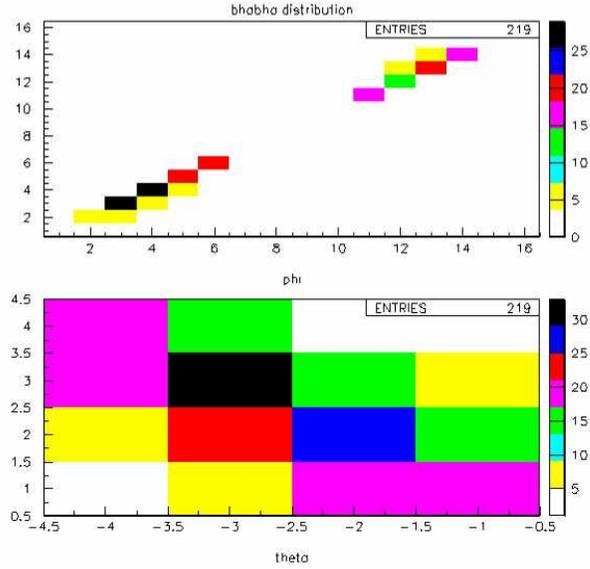}
\caption{ \it { Example of data acquired by the GEM trackers to check the collinearity of Bhabha events triggered by the calorimeter. The top plot shows the azimuthal correlation of the charged tracks~-- each pair of GEM trackers contains 16 $\phi$-sectors, 8 above the beamline and 8 below, whose numbering is chosen so that opposite sectors have the same label. The bottom plot shows the anticorrelation between the track polar angles. Combining both information shows that the Bhabha events are back to back, as expected. } }
\label{fig:gem_bhabha}
\end{figure}

In July 2008 the GEM trackers were installed again for a few days, but only 10~cm away from the IP in front of the lead shielding. As they were finally included in the DAQ, data were acquired when the calorimeter was triggering. Fig.~\ref{fig:gem_efficiency} shows the efficiency of two of the four GEM modules as a function of the offline calorimeter ADC threshold; the plateau at $40\%$ is due to the shift of the tracker $z$-positions: as the modules are closer to the IP, a large fraction of Bhabha tracks hitting the calorimeter pass through the GEM central hole and are thus not detected. The average hit cluster size was 1.2 pad; Fig.~\ref{fig:gem_bhabha} displays the $\theta$ and $\phi$ correlations of the hits in the four GEM detectors.

When the nominal \SID\ data taking run started, the GEM trackers were definitely removed from the IR.

\subsection{The gamma monitors}

\begin{figure}
\begin{center}
\includegraphics[width=4in]{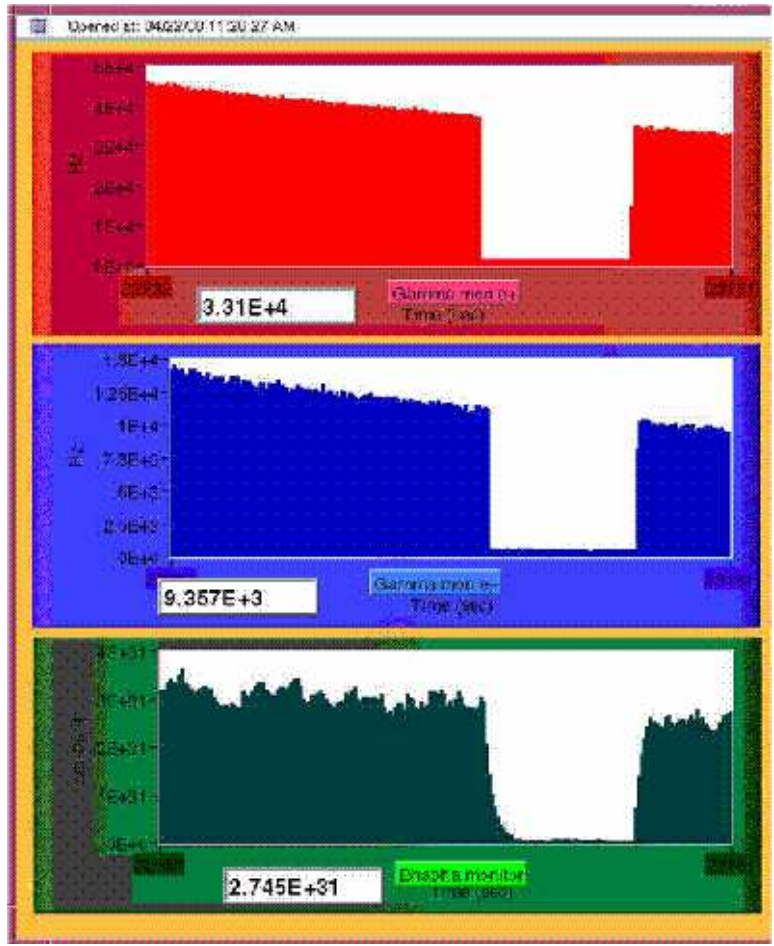}
\end{center}
\caption{\it { Time evolution of the gamma monitor (top and middle charts, $\ep$ side in red and $\en$ side in blue) and Bhabha background-subtracted (bottom chart in green) rates recorded during a machine test performed on April 22, 2008. All rates drop drastically when the two beams are separated by a 180$^{\circ}$ shift of the RF phase although the gamma monitors react quicker than the calorimeter. Similar conclusions can be drawn when the two beams collide again. } }
\label{fig:out-collision}
\end{figure}

The gamma monitors are mainly used to optimize the luminosity in real time as they are very sensitive to sudden changes in machine conditions. Thanks to the high radiative Bhabha cross section at low angle, the photon rate typically reaches several tens of kilohertz for a luminosity around $10^{32} \cms$. Therefore, these detectors can provide measurements with negligible statistical fluctuations every second. This is illustrated by Fig.~\ref{fig:out-collision} which shows the time evolution of the gamma monitor rates (top plot: $\ep$ side, red histogram; middle plot: $\en$ side, blue histogram) and of the background-subtracted Bhabha luminosity (lower histogram; in green) during a machine test in April 2008. At a certain time clearly visible on the charts, the beams are put out of collision by a $180^\circ$ RF phase shift; this new condition is immediately detected by the gamma monitors whereas there is some latency in the Bhabha rate. Sharp variations are also visible when the beams go back in collision. The smooth decrease of rates over the whole time range, visible in all three plots, corresponds to a period of coasting for both \DAF\ beams. One can also note that the residual background levels are very low in the absence of collision. 

\begin{figure}
\begin{center}
\includegraphics[width=5in]{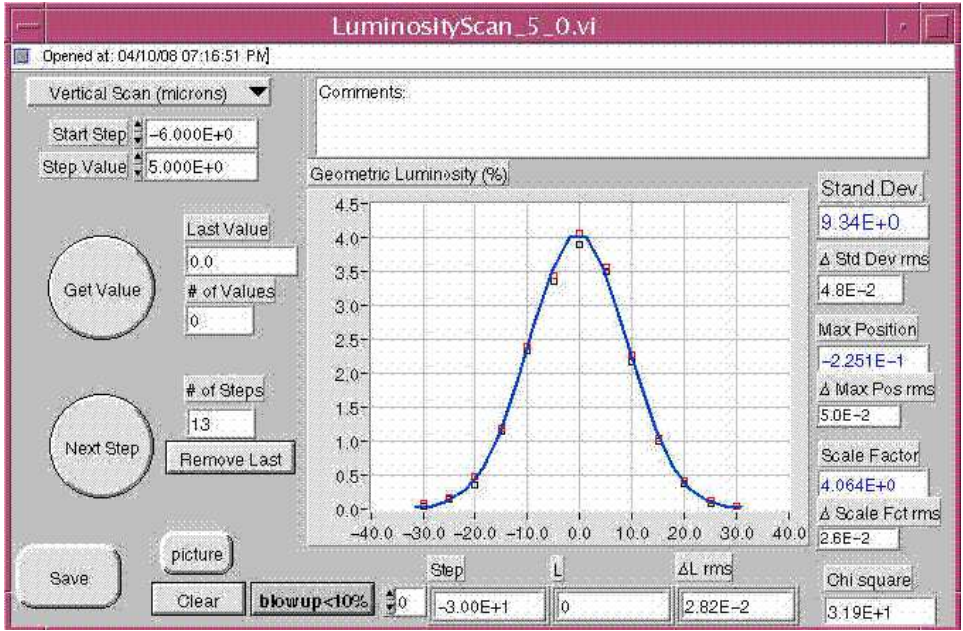}
\end{center}
\caption{ \it { \DAF\ control room monitor showing the result of a beam vertical scan (moving the electron beam while keeping the positron one fixed) performed in April 2008. Fitting the data coming from the gamma monitor allows one to estimate the quadratic sum of the two beam vertical sizes. The result, as displayed in the top box on the right column, is $\Sigma_y^{meas} = 9.34$~$\mu$m for this particular scan. } }
\label{fig:scan}
\end{figure}

The quick response of the gamma monitors and their low background contamination levels allow them to be used for precision measurements of the transverse beam size. The scan is carried out by shifting horizontally and vertically the beam trajectories. The rate variations measured during such test give access to the bunch profile. In order to be successful, this procedure must be very fast to avoid the result to be biased by the contribution of other effects (e.g. loss of luminosity when the beams are coasting, an effect which is relevant in the \DAF\ ring where typical beam lifetimes are lower than 1000~s). Fig.~\ref{fig:scan} shows a typical vertical scan performed in April 2008. In this example, the measured quadratic sum of the two beam vertical sizes is found to be $\Sigma_y^{meas} = 9.34$~$\mu$m. To match the upgraded \DAF\ running conditions, computations~\cite{ref:zobov_Sigmay} have shown that a corrective factor of $0.88$ needs to be applied to get the real value of $\Sigma_y$. Assuming that the two beams are Gaussian and of equal sizes, one finally gets $\sigma_y=5.8$~$\mu$m for this particular scan.
 
\section{\MC\ simulation}
\label{sec:montecarlo}

A \MC\ simulation of the luminometers is needed to validate the (radiative) Bhabha selection algorithms, to find the correspondence between the measured event rates and the actual machine luminosity and to study the impact of the background. For this purpose, a package has been developed to simulate the detector response in the \GEANTTHREE\ \cite{ref:geant3} framework. Input events are either Bhabha generated with the \BHWIDE\ \cite{ref:bhwide} \MC\ generator or simulated particles leaving the machine nominal orbit close to the IP~\cite{ref:boscolo}. After a detailed description of the simulation code (from the generation to the reconstruction of the events in the virtual detector), the main simulation results are presented.

\subsection{Event Generation}
\label{EvtGen}

The simulation is a two step-process. Events are first created at the generator level using the \BHWIDE\ code: both non-radiative and radiative Bhabha events (without limitation on the number of radiated photons) can be simulated. The BABAYAGA package~\cite{ref:babayaga} has provided cross-checks of the \BHWIDE\ performances as this software had been used in the past by the KLOE collaboration for a precise determination of the luminosity delivered by \DAF\ between 2000 and 2007~\cite{ref:KLOE_L_mes}. Both generators have been found in agreement. The four-momenta of the generated particles (assumed to be produced at the IP), as well as those of the potentially initial-state-radiated photons are stored in a PAW \cite{ref:paw} ntuple which is written to disk in order to be used as input to the \GEANT\ simulation. 

\begin{figure}
\begin{center}
\includegraphics[width=10cm]{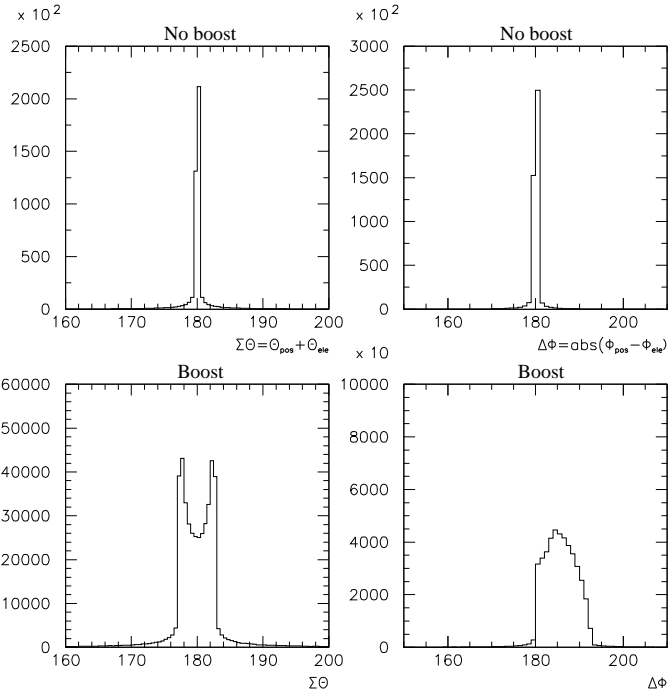}
\end{center}
\caption{ \it { Simulated Bhabha events without (top row) and with (bottom row) the \DAF\ horizontal crossing angle. CM generation is done with BHWIDE before the track parameters are converted to the laboratory frame described previously. The left column compares histograms of the sum of the two track polar angles: $\Sigma\Theta = \theta_{\mathrm{pos}}+\theta_{\mathrm{ele}}$; the right column shows the corresponding histograms for the variable $\Delta\Phi = |\phi_{\mathrm{pos}}-\phi_{\mathrm{ele}}|$ describing the azimuthal correlation between the electron and positron tracks. The $\Sigma\Theta$ and $\Delta\Phi$ distributions vary in shape with the boost and the mean value of the latter is also modified~-- beam trajectories are pointing toward the $-x$ direction. The non-zero widths of the top (no boost) distributions are due to the final state radiated photons.} }
\label{fig:boost-noboost}
\end{figure}

Like for the real machine (see Fig.~\ref{fig:setup}), the generated positrons are pointing toward the direction of the $+z$ axis while the electrons fly in the opposite direction. Convenient software switches allow one to select particular polar angle $\theta$ ranges for the charged tracks (whose distribution scales like $1/\theta^{3}$ in the CM frame). The horizontal boost due to the 25~mrad crossing angle between the two incoming beams and pointing toward the $-x$ direction is also simulated. As shown on Fig.~\ref{fig:boost-noboost}, its effect is quite significant: the polar and azimuthal correlations between the Bhabha positrons and electrons are modified by the non-zero crossing angle. Although it may look small, it impacts significantly the correlation between the particle trajectories which makes the understanding of its consequences mandatory to ensure accurate luminosity measurements. 

Particles leaving the nominal orbit close to the IP are expected to be the main source of background for the luminosity measurements. They are simulated with a code developed for the \DAF\ experiment \cite{ref:boscolo} which parameterizes the particle coordinates as a function of the path length ($s$) along the design trajectory. Touschek events are generated  continuously all over the ring, averaging the probability density function every three machine elements. Then, particles are tracked according to the optics over many turns or until they are lost. The actual beam pipe aperture is taken into account; nonlinear elements, sextupoles and octupoles, are also included in the tracking, allowing an intrinsic dynamical aperture calculation. As this simulation does not include the experimental setup, a dedicated procedure~-- see next section for details~-- needs to be applied prior to the \GEANT\ simulation, in order to go from the beam path frame to the \DAF\ coordinate system.

\subsection{Detector Simulation}
\label{DetSim}

The ntuple produced by \BHWIDE\ is used as input of a FORTRAN-based \GEANTTHREE\ code which simulates the region around the IP relevant for the luminosity measurements: from the $\ep$ gamma monitor to the $\en$ one ($z = \pm 170$~cm). The generated particles are positioned at the IP from where \GEANT\ propagates them until they leave the simulated mother volume, decay, or get absorbed. Along the way, \GEANT\ computes the effect of their interactions in the various materials they cross. Information relevant for the study of reconstruction algorithms or for the detector design optimization (such as the true impacts in the GEMs and colorimeter modules, or the number of photons detected in each calorimeter sector or in the gamma monitors) is stored in an output PAW ntuple for further analysis.

\begin{figure}
\begin{center}
\includegraphics[width=5.5in]{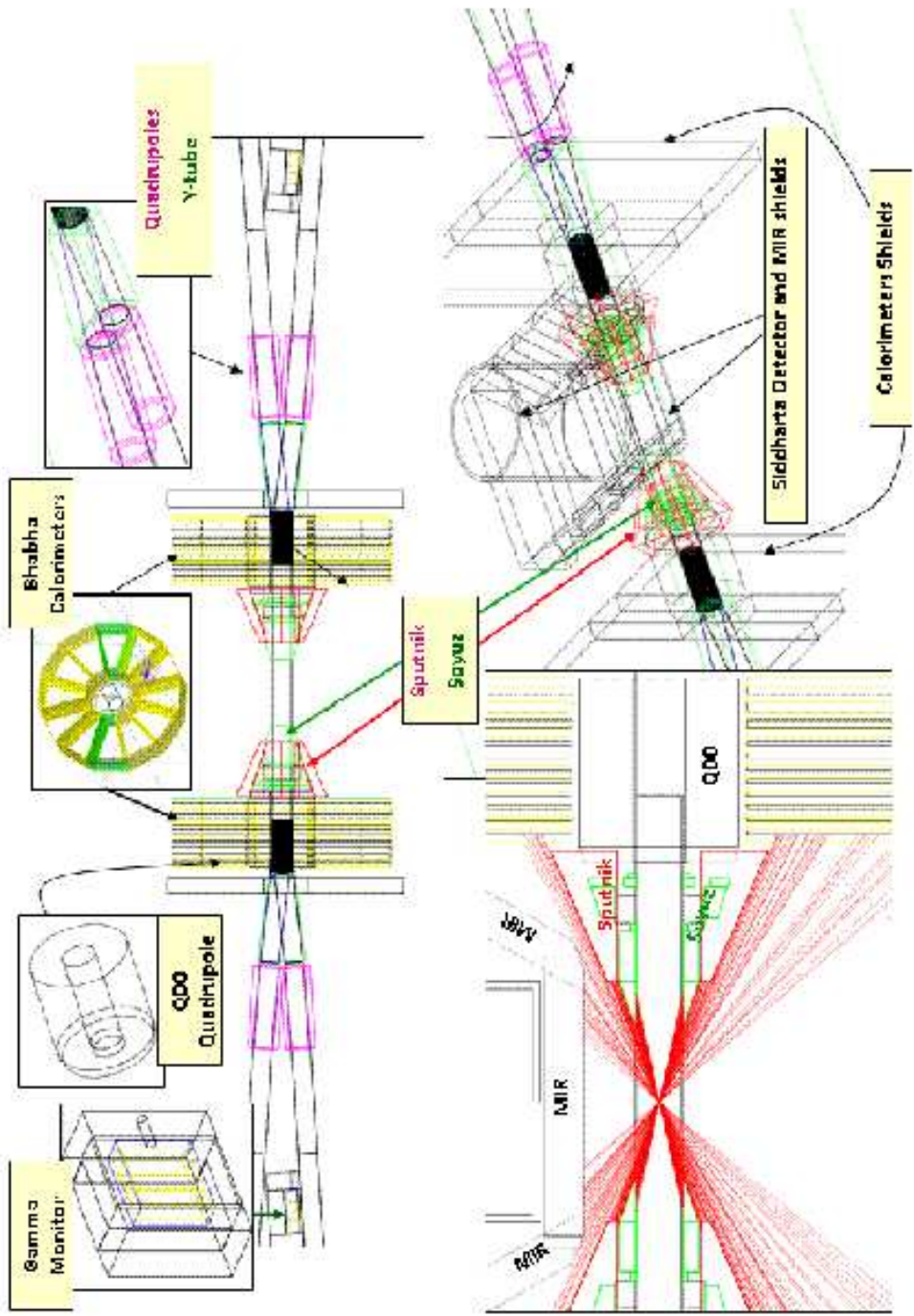}
\end{center}
\caption{\it { Overview of the experimental setup as implemented in the \GEANT\ simulation; see text for details. } }
\label{overall_fig}
\end{figure}

Fig.~\ref{overall_fig} gives an overview of the experimental setup as implemented in the \GEANT\ simulation. The left part shows the setup from above the beam plane~-- the \SID\ volume has been removed to make the drawing clearer. The main magnified elements are the Bhabha calorimeter modules, the gamma monitors, the QD0 permanent quadrupole magnets and the vacuum chambers whose geometry is particularly complicated in the region where the beam pipe, common to both beams at the IP, gets split (Y-tubes) to allow particles of opposite charges to be separated in the rest of the \DAF\ ring. The top right drawing shows details of the Soyuz, Sputnik and Mir lead shielding elements (see Section~\ref{subsection:shielding}); for a better readability, the Bhabha calorimeter has been removed from this picture. The big shieldings protecting the calorimeters from backward-orientated particle showers are also visible. Finally, the bottom right picture is a cut in the y-z plane which explains how crucial it is to simulate these shields accurately. The red tracks are Bhabha generated with the \BHWIDE\ package whose density reflects the $1/\theta^3$-dependence of the cross-section. Dramatic rate variations are visible around the shield edges, which means that a small error in their position would trigger large acceptance effects as the fiducial Bhabha volume would be incorrect.
 
The various luminometer elements, the machine components and shielding relevant for the luminosity measurements are simulated, with the proper dimensions and materials~-- see Fig.~\ref{overall_fig}. 
\begin{itemize}
\item The scintillator tiles, wrapped in their \TYVEK\  skin, and the lead absorber layers of the calorimeter. Losses of light in the tiles and in the optical fibers connected to the PMTs are also taken into account, such as the PMT quantum efficiencies.
\item The GEMs, with both their azimuthal and radial segmentations; the three kapton foils and their copper coating, the four longitudinal layers of gas (one drift and three transfer volumes) as well as the copper box containing the whole tracker.
\item The gamma monitor, made of four PbW0$_{4}$ crystals wrapped in a \TYVEK\ skin and put together in a PVC box.
\item The various sections of the aluminum beam-pipe, including the splitters (the green volume downstream of the calorimeter on the top-right plot) and the three cylindrical sections.
\item the QF1 (the cylinders upstream of the splitters) and QD0 quadrupoles (invisible as surrounded by the tile calorimeters), whose magnetic fields are also simulated.
\item The Soyuz, Sputnik and Mir shields.
\end{itemize}

Like for the Bhabha event generation, the input parameters are readout from a datacard and several software switches have been implemented to allow the user to customize the simulation: value of the horizontal beam crossing angle (defining the size of the boost), choice of the hardware pieces to be included in the IR and of the particles (electrons, positrons or photons) to be propagated in \GEANT, etc.

\begin{figure}
\begin{center}
\includegraphics[width=12cm]{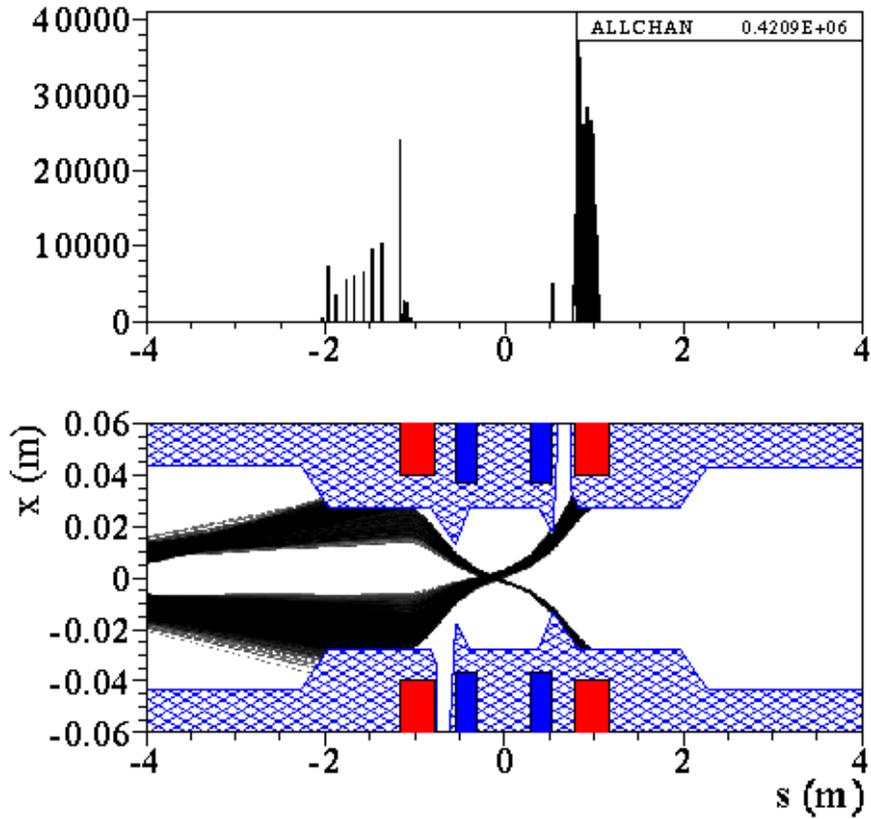}
\end{center}
\caption{\it { Distribution of particles lost in the IR versus path length $s$ (top plot) and the corresponding trajectories (bottom plot). The IP is at $s=0$ and the collimators are inserted; the blue and red rectangles show the position of the QD0 and QF1 quadrupoles respectively. } }
\label{fig:ptclosses}
\end{figure}

As mentioned in the previous section, the simulation of the particles lost from the beams~-- see Fig.~\ref{fig:ptclosses}~-- does not include the experimental setup around the IP. Hence, one needs to process the output of this code (positions and momenta of the particles when they leave the nominal orbit) before using it as input to the \GEANT\ simulation. The first step of this procedure consists in converting the particle parameters from a frame linked to the beam design trajectory to the IP frame defined in section~\ref{sec:overview}. Knowing the path length $s$, the particle transverse coordinates $(x_s,y_s)$, their first derivatives with respect to $s$ $(x'_s,y'_s)$ and the energy deviation with respect to the nominal value of 510~MeV, one can compute the particle position and momentum in the $(x,y,z)$ system of coordinates. 

Then, these particles are propagated backward in time until they are inside the beam pipe. For this purpose, one flips the true particle momentum and uses an 'inverted' \GEANT\ simulation of the IR where the vacuum beam pipe is now lead and all other materials vacuum. With this setup, the first \GEANT\ hit associated with a non-zero energy loss can only occur inside the beam pipe which provides the true position of the particle before it would leave it. Its true momentum is then defined as the opposite of the value computed by \GEANT\ at the last hit with null energy loss. These parameters are finally stored on disk in another ntuple which can then be used as input of the 'regular' \GEANT\ simulation which is finally used to simulate the Touschek background rates in the gamma monitors. As can be seen on Figs.~\ref{fig:gammon1} and~\ref{overall_fig}, these detectors are surrounded by lead bricks. This shield allows them to fulfill their primary task: a clean measurement of the photon emitted by radiative Bhabha events, in spite of the large beam backgrounds. Consequently, the acceptance of the gamma monitors is limited to a circular window, dug in the lead and aligned with the typical line of flight of a radiative Bhabha photon, whose diameter does not exceed 1~cm. Because of this reduced acceptance, Touschek particles  are not  expected to directly hit the gamma monitor. They have to be measured indirectly, via the secondary particles produced in showers caused by their passage through the QF1 magnet or the beam wall. A significant fraction of these secondaries have their energy lower than 10 keV. This is the limit below which \GEANT\ is no longer reliable in its description of the particle-material interactions. This is why the simulation of the Touschek background measurement is based on the GEANT4~\cite{ref:geant4} package, which has been improved w.r.t \GEANTTHREE\ to better deal with the low energy regime.

\subsection{Reconstruction}
\label{section:Reco}

\begin{figure}
\begin{center}
\includegraphics[width=6cm]{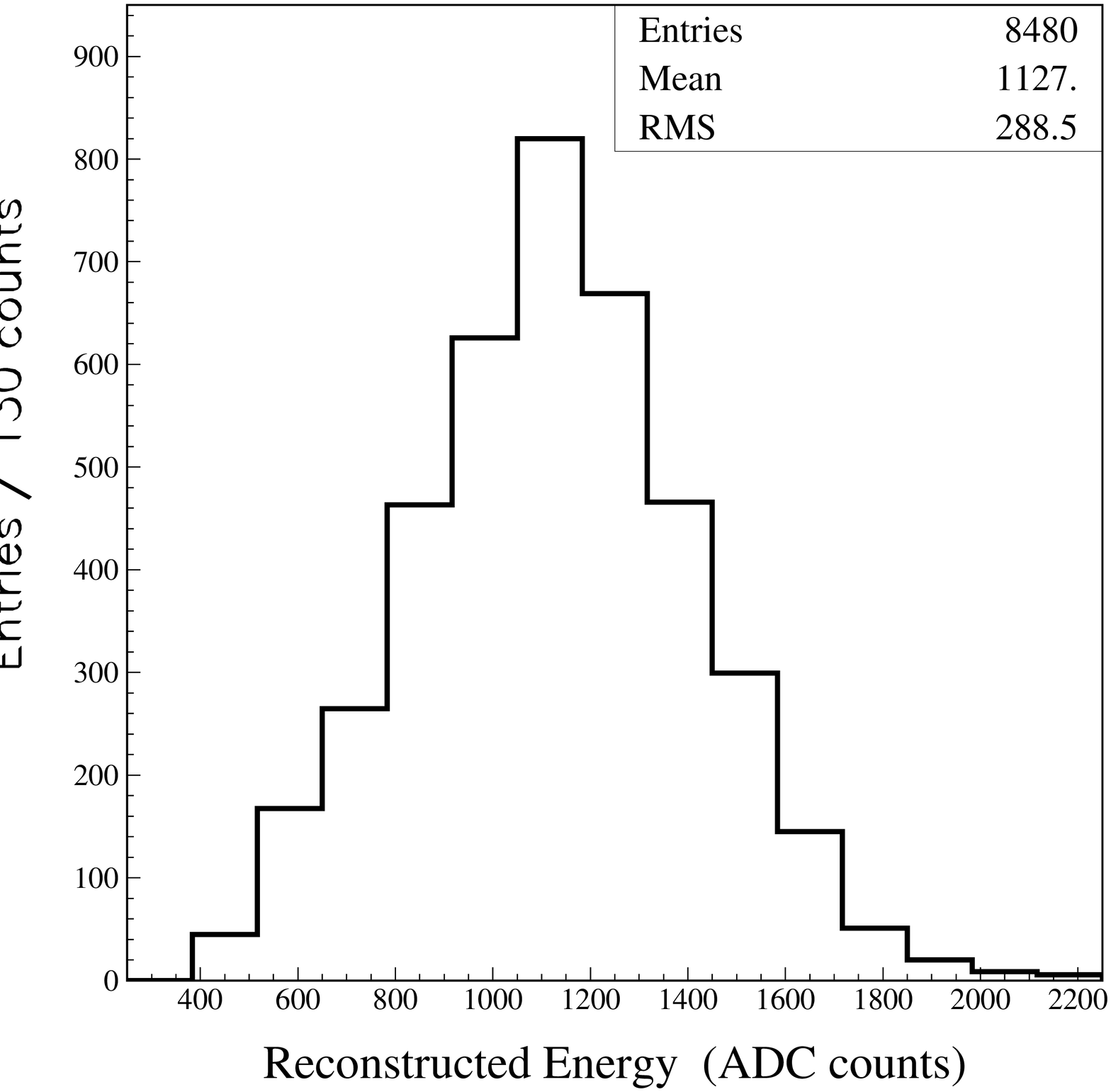} \\
\includegraphics[width=6cm]{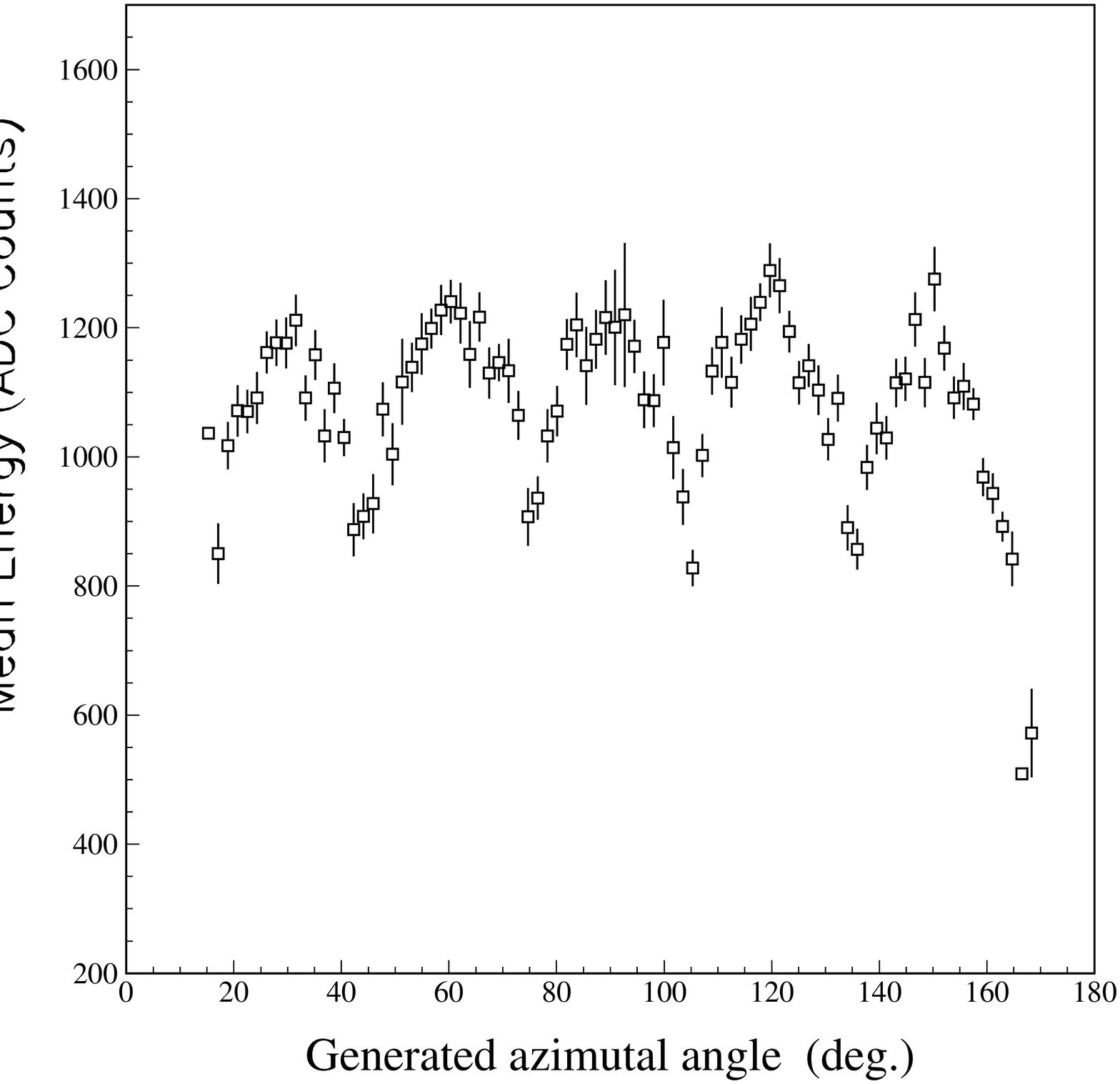}
\includegraphics[width=6cm]{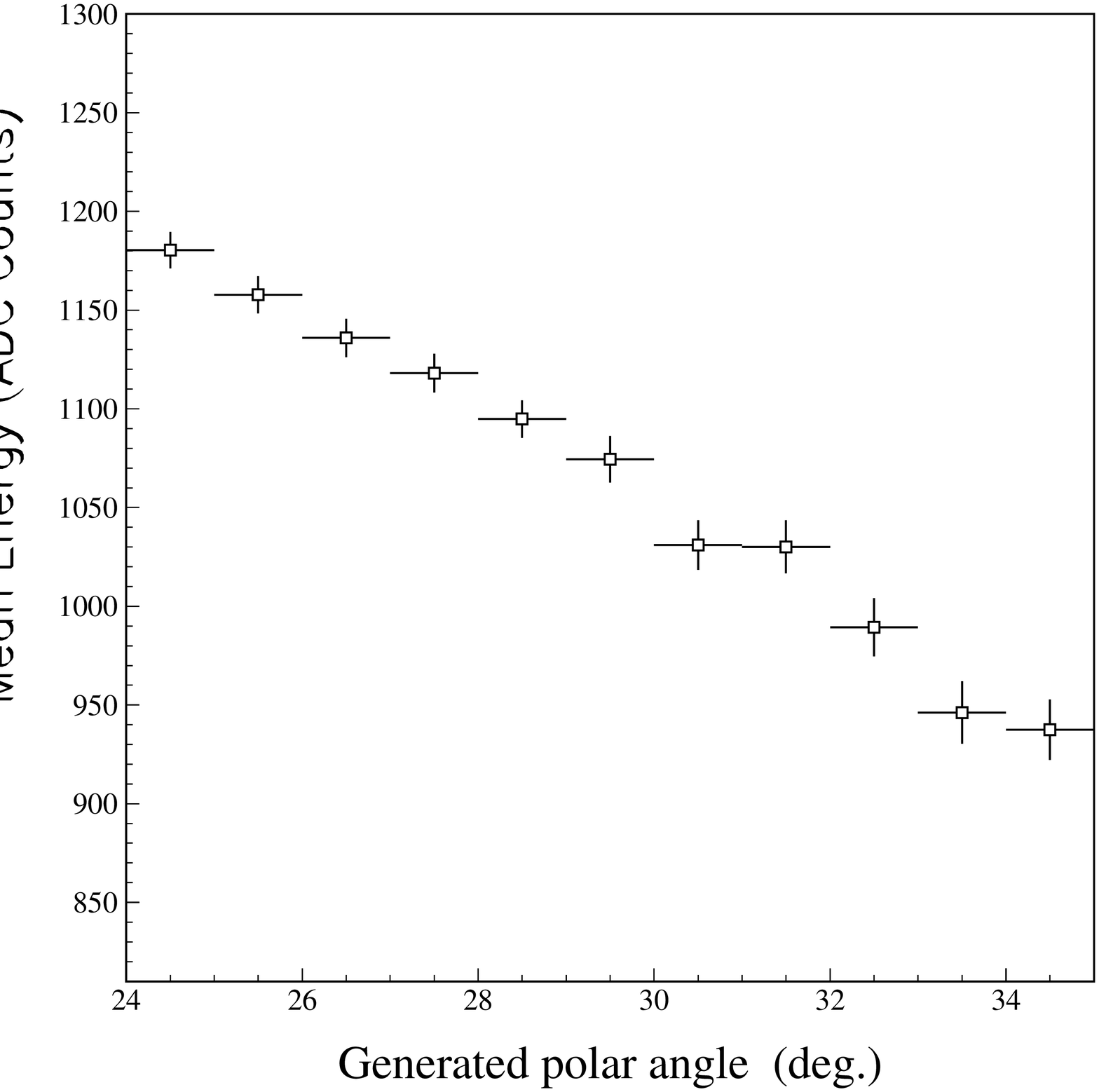}
\end{center}
\caption{\it { The top plot shows the distribution of the total number of photons collected by the PMTs in the ``most hit'' calorimeter sector and its two closest neighbors. The histogram has a Gaussian-like shape whose resolution is about 18\%, in agreement with the calorimeter design and the actual data. The two bottom plots show how this number depends on the true azimuthal $\phi$ (left side) and polar $\theta$ (right side) track angles. On the left plot, the calorimeter segmentation is clearly visible: the tracks which deposit the highest energy in average are those which hit the middle of the azimuthal sectors. In addition, the right profile histogram shows that the deposited energy decreases with the polar angle as the tracks get less contained. } }
\label{npho_fig}
\end{figure}

Let's now focus on the reconstruction of events in the Bhabha calorimeter. \GEANTTHREE\ simulates the showers initiated by the charged particles and computes in particular the amount of energy lost at each step of the tracking while the particles are crossing the scintillating tiles. The number of recorded photons is deduced from this quantity using a set of realistic parameters which are provided as input to the simulation: the scintillator photon yield per unit energy; the attenuation depth in this material and in the WLS optical fibers; the PMT quantum gain. The top plot in Fig.~\ref{npho_fig} shows the distribution of the number of reconstructed photons per event in a given module, summing up the 'most hit' sector and its two closest neighbors. As the average number of photons correspond to the nominal beam energy of 510~MeV, the energy resolution appears to be $18\%/\sqrt{E}$.  

Fig.~\ref{npho_fig} also shows the dependence of the number of collected photons on the true track angular coordinates ($\theta$ and $\phi$) at the calorimeter. The $\phi$ dependence (bottom left plot) comes from the calorimeter segmentation and from the correlation between the azimuthal track position and the distance between the hits in the tile and the fibers collecting the light in the material. This correlation also explains the $\theta$ dependence visible on the bottom right plot. The amplitude of this effect is obviously function of the properties of various components: scintillator, WLS fiber and PMT. These topics are discussed in details in Section~\ref{sec:valid} which is dedicated to the validation studies of the simulation and where a good agreement between data and \MC\ distributions is shown.


\subsection{Calorimeter thresholds}
\label{subsection:thresholds}

Another challenging part of the \MC\ simulation consists in choosing a proper set of thresholds to control the individual module triggers. As the reconstructed energy is directly proportional to the number of photoelectrons at the photomultiplier, the trigger thresholds are applied at this level. This is not an accurate description of the calorimeter trigger: in the real system (see Section~\ref{sec:daq}), each module threshold is applied on a pulse coming from the KLOE SDS board summing up the currents collected by the PMTs in the five sectors. This electronics chain adds some noise that smears the signal; thus, the pulses are not directly proportional to the numbers of photoelectrons. A systematic uncertainty is estimated to account for this discrepancy between data and \MC. The effect of the smearing is evaluated by varying the thresholds used in the simulations within a range centered on the value determined from the data (see below), and twice as large as the error on this determination.

The \MC\ thresholds are adjusted to those found in data which are computed using the following procedure. Taking the distribution of the total energy in a given module (defined as the sum of the pedestal-subtracted ADC values of the five associated sectors) and a cut value, one can count the number of events for which that energy exceeds the cut. Then, one can compute the fraction of those events for which the module actually triggered. The curve showing the variation of this fraction versus the cut value has a step-like shape which allows one to extract the threshold. For low (high) cut values, the fraction is close to 0 (1); the transition between the two states is sharp and occurs around the threshold. For the reason explained in the previous paragraph, the curve is not a pure step function but exhibits some Gaussian smearing. Therefore, it is fit with a 3-parameter model:

\begin{equation}
\phi( c, \rho, \tau, \sigma ) = \rho \times E\left( \frac{c-\tau}{\sigma} \right)
\end{equation}

where
\begin{itemize}
\item $c$ is the cut value;
\item $\rho$ is the {\it ratio} plateau (in the high cut region) which should be very close to 1;
\item $\tau$ is the fitted {\it threshold};
\item $\sigma$ is the {\it smearing} parameter, i.e. the width of the Gaussian;
\item finally, $E$ is the Gaussian error function.
\end{itemize}

\begin{figure}
\begin{center}
\includegraphics[width=6cm]{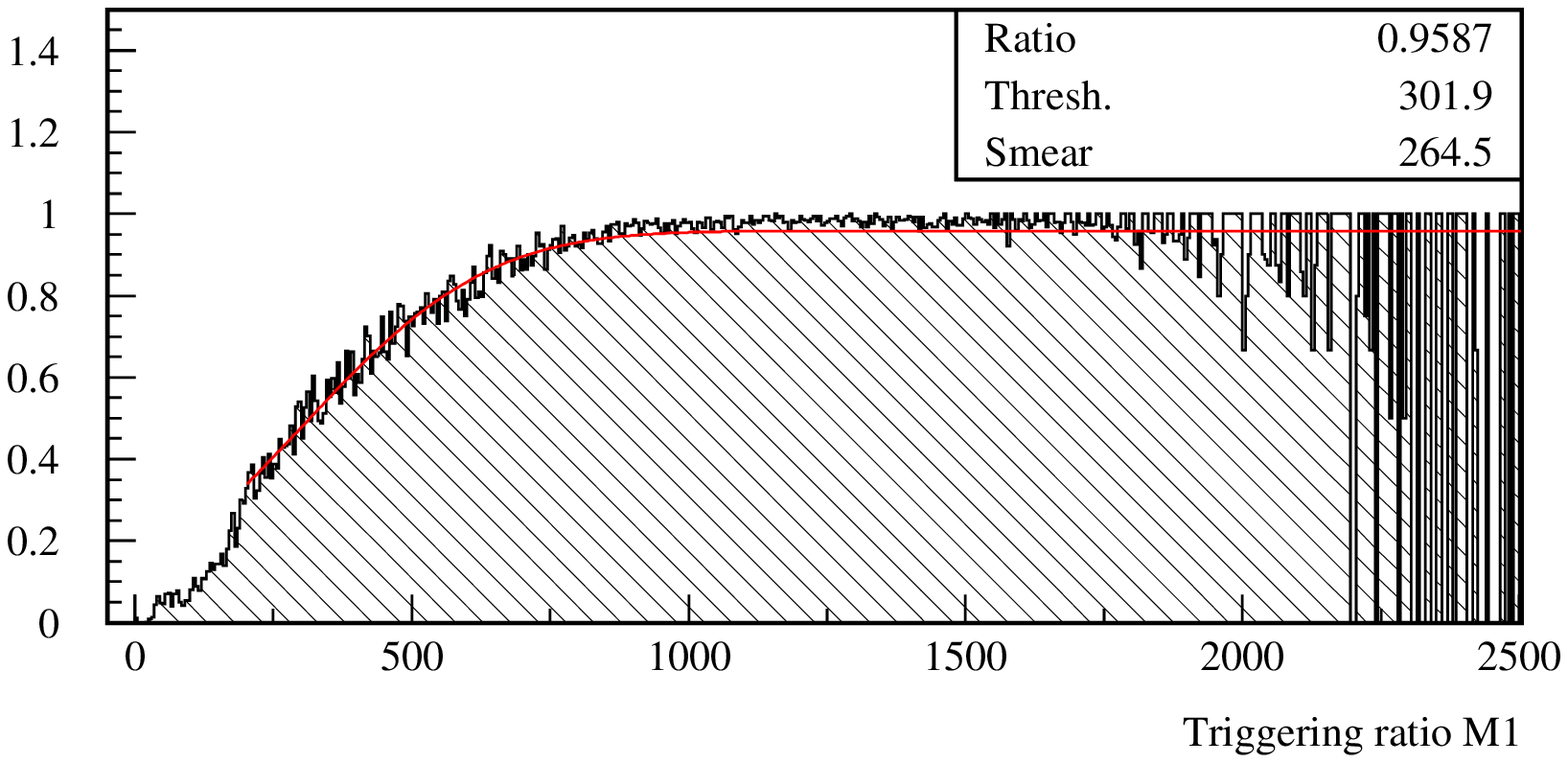}
\includegraphics[width=6cm]{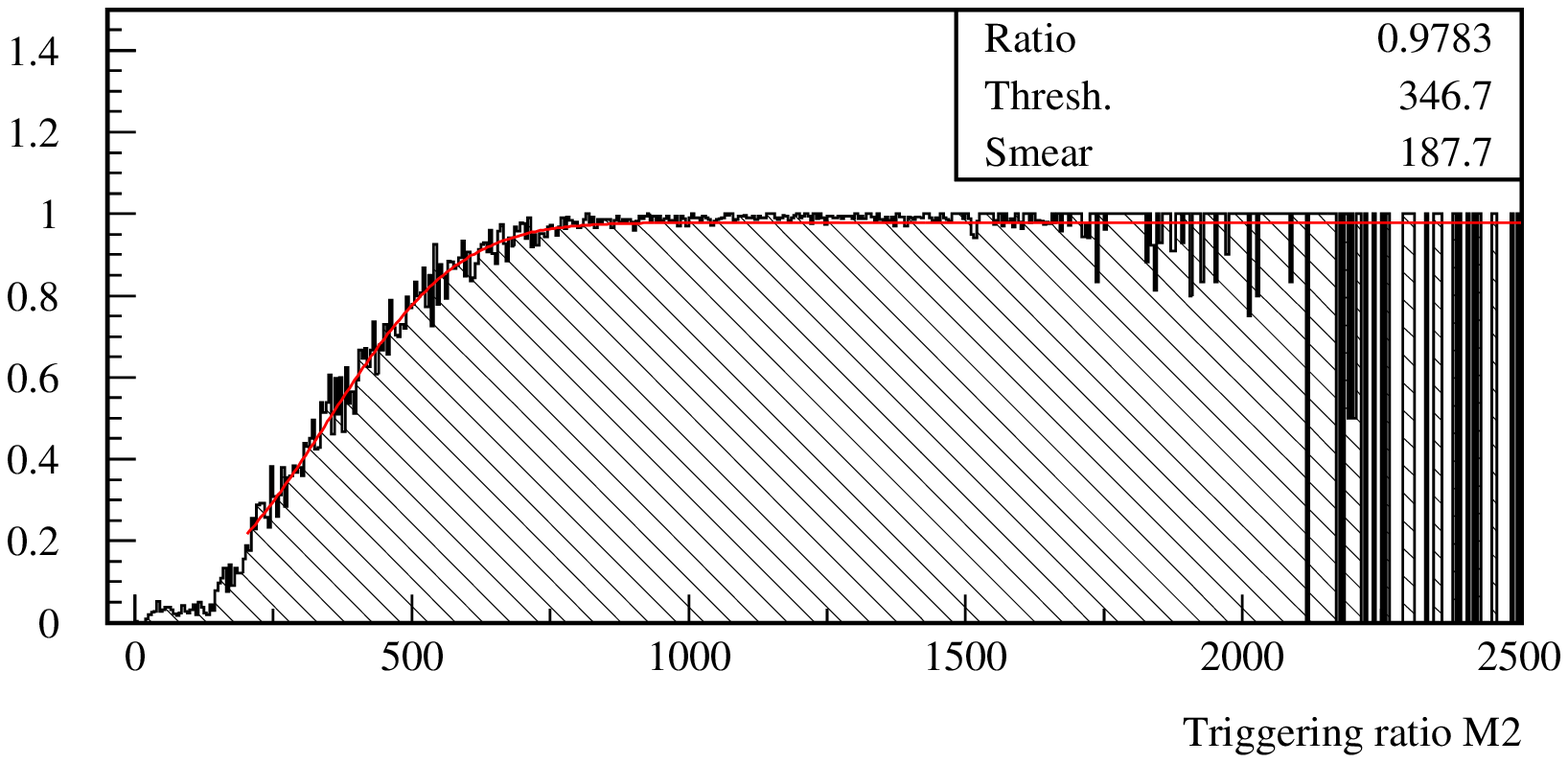} \\
\includegraphics[width=6cm]{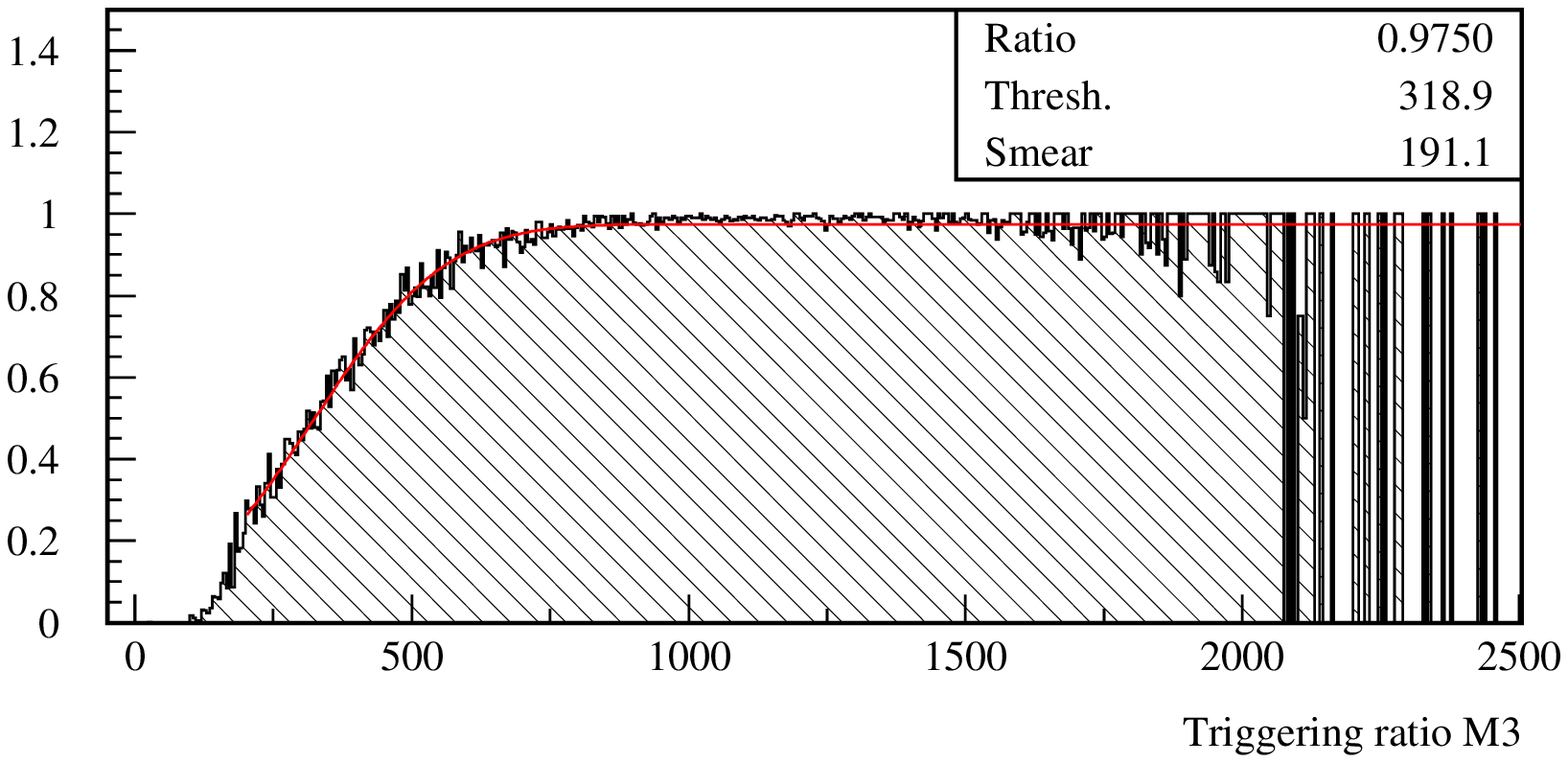}
\includegraphics[width=6cm]{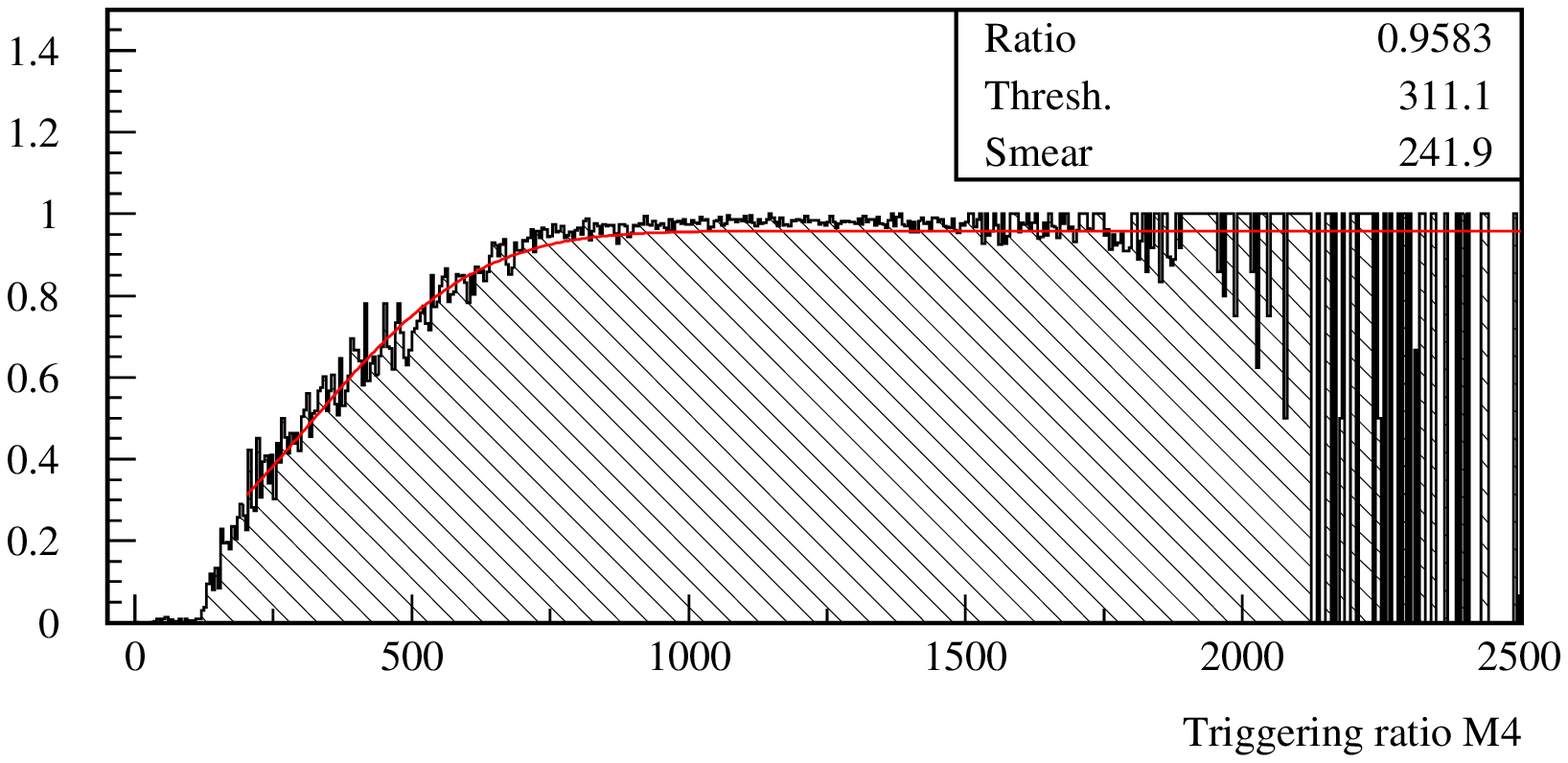} \\
\end{center}
\caption{ \it { Fits of the fractions of triggering events vs. ADC cuts for the four calorimeter modules. The captions show the fit results, as described in the text. } }
\label{fig:module_hresholds}
\end{figure}

Fig.~\ref{fig:module_hresholds} show the fit results for the four calorimeter modules, using as input the data from a representative \DAF\ run. The uncertainty in the threshold determination is mainly related to the background contribution which distorts the left tail of the fitted function. The systematics effect, estimated by varying the low boundary of the fit range, is around 5\%.

Finally, the ADC thresholds need to be converted to energy values. As genuine Bhabha events deposit in average 510~MeV in triggering modules, the peak of the ADC distribution obtained by only keeping events for which the module triggered provides the required conversion factor, with a systematic uncertainty of the order of 1\% (again background-driven). Table~\ref{tab:thres} summarizes the threshold values of the four calorimeter modules, using as ADC-to-energy conversion factors the fit results shown in Fig.~\ref{fig:bhabha3}.

\begin{table*}[h]
\begin{center}
\begin{tabular}{lll}
\hline\hline
         Module   & ADC Threshold value $\tau$  &  Energy   \\ \hline
           1      & 301.9                       &  135 MeV  \\
           2      & 346.7                       &  151 MeV  \\
           3      & 318.9                       &  146 MeV  \\
           4      & 311.1                       &  145 MeV  \\ \hline
\end{tabular}
\caption {\it { ADC count and energy thresholds for the four calorimeter modules. } }
\label{tab:thres}
\end{center}
\end{table*}

\subsection{Expected Bhabha rates} 

Finally, Table~\ref{tab:rate_MC} summarizes the predictions of the \MC\ simulation for the event rate in the Bhabha calorimeter for various shielding configurations. These rates correspond to a benchmark luminosity of $10^{32} \cms$. The systematics uncertainties which are quoted here are computed in Section~\ref{sec:valid}. These rates are high enough to allow an online measurement refreshed every few seconds with a small statistical uncertainty.

\begin{table}[!h]
\begin{center}
\begin{tabular}{|c|c|c|c|} \hline 
   &   Soyuz only    & Soyuz $+$ Sputnik  & Final Setup \\ 
\hline 
Bhabha Calorimeter & 470 $\pm$  65  Hz &  280 $\pm$  40 Hz  &  230 $\pm$  30  Hz \\
\hline 
\end{tabular} 
\caption{ \it { Event rates predicted by the simulation in the Bhabha calorimeter, assuming a luminosity of $10^{32} \cms$. } }
\label{tab:rate_MC} 
\end{center} 
\end{table} 

\section{Validation studies}
\label{sec:valid}

This section deals with the evaluation of the systematic uncertainties for the online luminosity measurements performed with the Bhabha calorimeter. It also presents the data-\MC\ comparison studies which have been carried out to assess their validity.

\subsection{Systematic uncertainties on the calorimeter-based luminosity measurements}
\label{sec:syst_calo}

The uncertainty on the Bhabha luminosity measurements stems from a variety of sources. Most of them are related to inaccuracy or uncertainties in the \MC\ simulation. 

\subsubsection{Bhabha generator uncertainties}

The systematic uncertainty due to potential inaccuracies in the \MC\ kinematic variable distributions and differential cross-sections is given by the variation in the Bhabha rate predicted by the simulation when using the BABAYAGA event generator instead of \BHWIDE. It is found to be small.

\subsubsection{Uncertainty on the interaction region phase space}

The distributions of the position of the Bhabha vertices and of their CM momenta depend on the phase space distribution of the colliding bunches, on the hourglass effect~\cite{ref:witold} and on the beam crossing angle. A precise prediction of the shape of those distributions is not an obvious task, made even more delicate by the crab waist scheme. Writing a realistic IP phase space simulation would be very time-consuming. However, this is not necessary as the crab waist scheme reduces the impact of the hourglass effect by making the length of the overlap between two colliding bunches of the same order as the vertical betatron function. Consequently, the hourglass effect is not simulated: the IP distribution used in the Monte Carlo simulation, is a mere three dimensional Gaussian, whose standard deviations are taken as the quadratic sum of the colliding bunch sizes. The longitudinal sizes are reduced to account for the crossing angle. The uncertainty associated with this approximate description is estimated in a conservative way by quoting the variation observed in the Bhabha rate when the length of the longitudinal component of the Gaussian is set to the value corresponding to a null crossing angle. 

\subsubsection{Geometrical uncertainties}

The location of the various hardware elements (detectors, shields, etc.) defining or impacting on the Bhabha acceptance is known with a precision of the order of $\sigma_{pos}=2$~mm. We vary these positions by $\pm \sigma_{pos}$ and rerun the \MC\ simulation using these slightly modified setups. The corresponding variation in the acceptance is taken as the associated systematic uncertainty. It is the dominant contribution to this error and its main contribution comes from the uncertainties in the \SID\ shield position.

\subsubsection{Uncertainty on the simulated energy reconstruction}

The procedure followed in the \MC\ simulation to derive the reconstructed energy from the deposits associated with the \GEANTTHREE\ raw hits is described in Section~\ref{sec:montecarlo}. It is the source of several systematic uncertainties which are reviewed in the following. The photon yield in the scintillator follows a Poissonian statistics whose parameter depends on the nominal yield per unit energy which is not perfectly known. Consequently, the resolution on the reconstructed energy~-- and thus the simulated acceptance of the trigger~-- could be wrong. The energy resolution is also impacted by the noise level in the PMT amplification. Another contribution to this uncertainty comes from the absorption of the scintillation photons while they are traveling through the tiles or along the WLS fibers. For instance, two equivalent energy deposits can yield different reconstructed energy values if one is closer to a fiber than the other. This makes the average reconstructed energy dependent on the track polar and azimuthal angles, as shown in Fig.~\ref{npho_fig}. This dependence broadens the overall reconstructed energy distribution in a way that has to be correctly taken into account if one wants to reach an accurate simulation of the energy resolution. Since the attenuation length of the materials the scintillator tiles and WLS fibers are made of is not precisely known, it was not the case in the first place. This problem has been solved thanks to the test beam data: plots similar to those shown on Fig.~\ref{npho_fig} have been produced and used as reference to tune the \MC\ attenuation length.

Finally, in Section~\ref{sec:comparison}, a quite good agreement between the actual Bhabha event energy distribution and the corresponding \MC\ prediction is shown. This explains why we eventually quote no systematics associated with the energy reconstruction.

\subsubsection{Uncertainties due to the electronic threshold determination}

The treatment of the trigger by the simulation is described in section~\ref{section:Reco}. The energy thresholds are varied within ranges determined from the data. The associated variation of the Bhabha rates is taken as the corresponding systematic uncertainty.

\subsection{Background-related uncertainties}

The background treatment is described in Section~\ref{sec:performances}. Two sources of systematic uncertainties associated with the real-time background subtraction procedure have been identified. First, the number of events in the sideband of the time difference distribution between two back-to-back triggering module exhibits statistical fluctuations which are directly translated into a systematic error on the corrected event rate. Secondly, the procedure also assumes that the peaking background is negligible. The validity of this assumption has been verified by checking that the calorimeter luminosity variations are in agreement with the expectations when beam currents and sizes vary. Moreover, the genuine Bhabha timing peak disappears completely when the beams are put out of collision~-- see Fig.~\ref{fig:r1520-r1522-ttr}.

For completeness, we also run the \MC\ simulation on two-photon events generated by \BHWIDE\ as their back-to-back topology makes them an irreducible background. The low rate predicted by the simulation (about 0.1\%) is subtracted from the number of signal events obtained after the online subtraction procedure. The same number is conservatively quoted as systematics.

\subsection{Results}

Table~\ref{tab:systema} summarizes the various contributions to the systematic uncertainties presented in the previous paragraphs. The total systematic uncertainty on the measured luminosity, taken as the quadratic sum of all these contributions, is of the order of 15$\%$ and is dominated by the geometrical uncertainty. 

\begin{table}[!h]
\begin{center}
\begin{tabular}{|c|c|} \hline 
\textbf{Source} & \textbf{Value} $\%$ \\ 
\hline 
Bhabha Generation          & 2\% \\
IP Phase Space             & 4\% \\
Geometry and Alignment     & 11\% \\
Threshold Determination    & 5\% \\
Background Treatment       & 3\%  \\
\hline 
Total                      & 13\% \\
\hline
\end{tabular} 
\caption{\label{tab:systema} {\it Systematic uncertainties on the calorimeter luminosity measurements. } }
\end{center} 
\end{table} 

\subsection{Data -- \MC\ comparison}
\label{sec:comparison}

\begin{figure}[!h]
\centering
 \rotatebox{0}{\includegraphics[width=10.0cm]{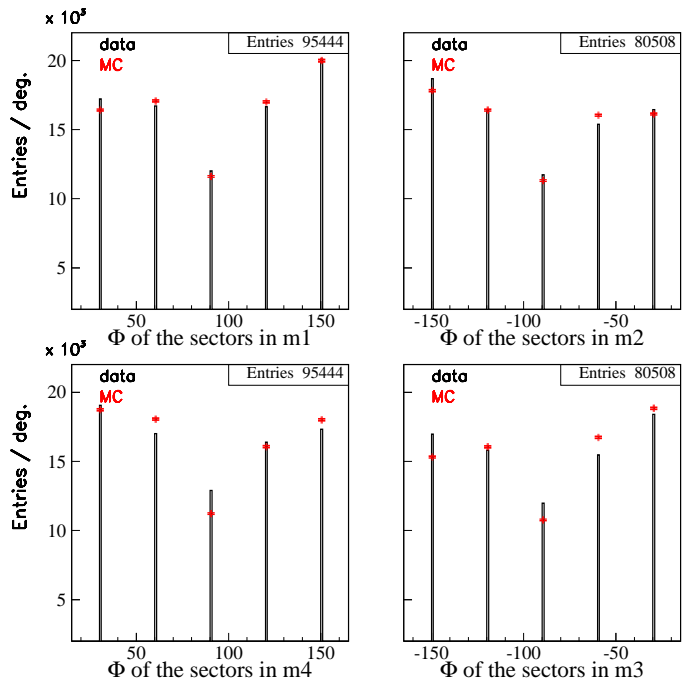}}
\caption{{\it Trigger rates per sector for the four calorimeter modules, with the final setup. Red points are the \MC\ expectations while black histograms are from real data. The rate variations between sectors are due to the boost and to the angular coverage of the various shields.} }
\label{fig:RperSec}
\end{figure}

The predictions of the \MC\ simulation for a number of quantities have been compared with the data to verify that no effect larger than the systematic errors described in the previous section has been overlooked. 

\begin{figure}[!h]
\centering
 \rotatebox{0}{\includegraphics[width=10.0cm]{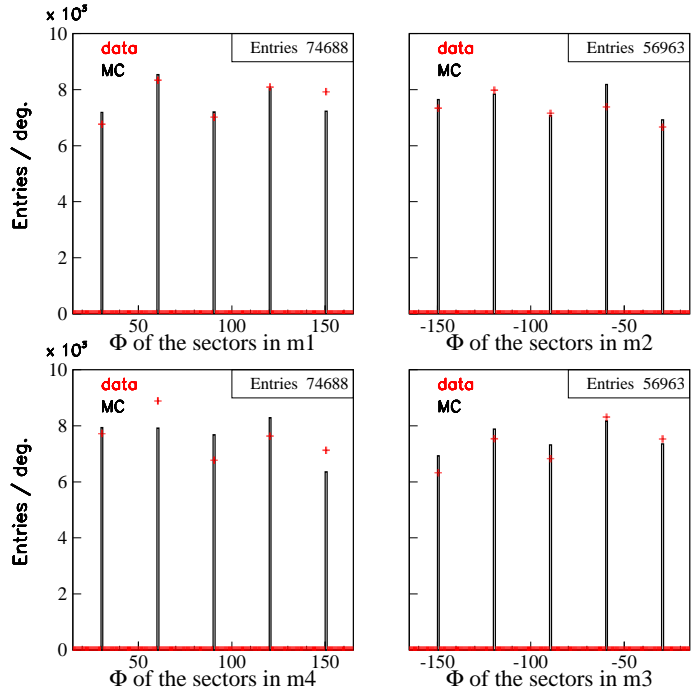}}
\caption{{\it Trigger rates per sector for the four calorimeter modules before the installation of the \SID\ final shield. Red points are the \MC\ expectations while black histograms real data. The rate variations between sectors are due to the boost, and to the angular coverage of the various shields.} }
\label{fig:RperSec2}
\end{figure}

The Bhabha rate per calorimeter sector is shown on Fig.~\ref{fig:RperSec}, both for data and \MC. This distribution is not flat because of the boost and of the the various IR shieldings. The \MC-data agreement is satisfactory although the various sectors are not equally affected by simulation inaccuracies. This figure also shows that all the sectors work properly in the real detector (no broken fiber, no module assembly issue, no problems in the PMT connection quality or isolation from outside light, etc.). It is interesting to notice that a similar agreement was observed between data and \MC\ during the period preceding the \SID\ shield installation (see Fig.~\ref{fig:RperSec2}). Obtaining this result for two different setups is an indication that the overall acceptance is correctly described. 


\begin{figure}[!h]
\centering
 \rotatebox{0}{\includegraphics[width=8.0cm]{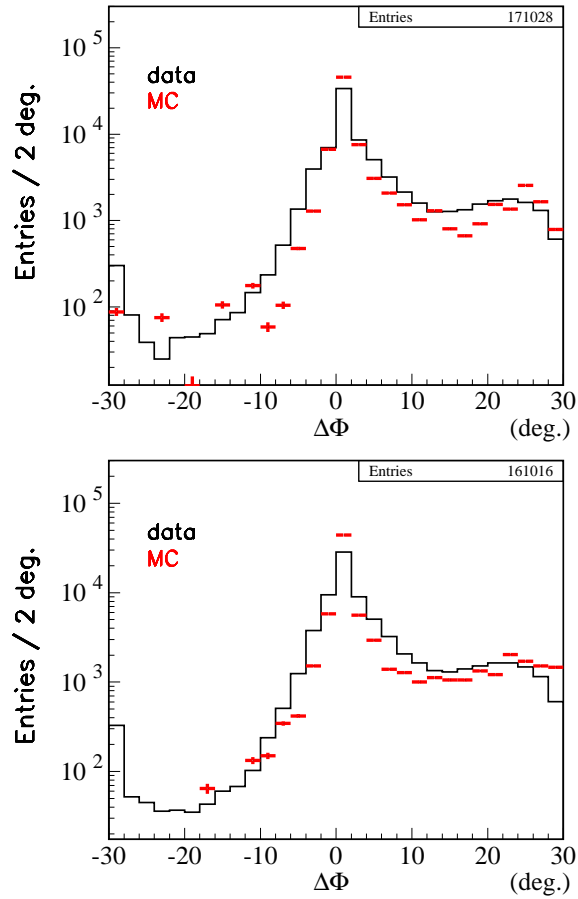}}
\caption{{\it Resolution on the azimuth angle, reconstructed as the weighted average of the $\phi$ positions of the calorimeter sector centers, the weights being the values of the energy deposited in each sector~-- this method is called 'barycenter' in the text. The resolution is taken as the difference between the electron and positron $\phi$ angles. The shape asymmetry seen in both plots (top:M1-M4 coincidences; bottom: M2-M3) is due to the boost caused by the beam crossing angle. The missing Monte-Carlo points in the left part of these plots is due to the logarithmic scale and to the limited statistic of the Monte-Carlo samples which have been rescaled to the data sample statistics to ease the comparison.} }
\label{fig:ResolPhi}
\end{figure}

\begin{figure}[!h]
\centering
 \rotatebox{0}{\includegraphics[width=10.0cm]{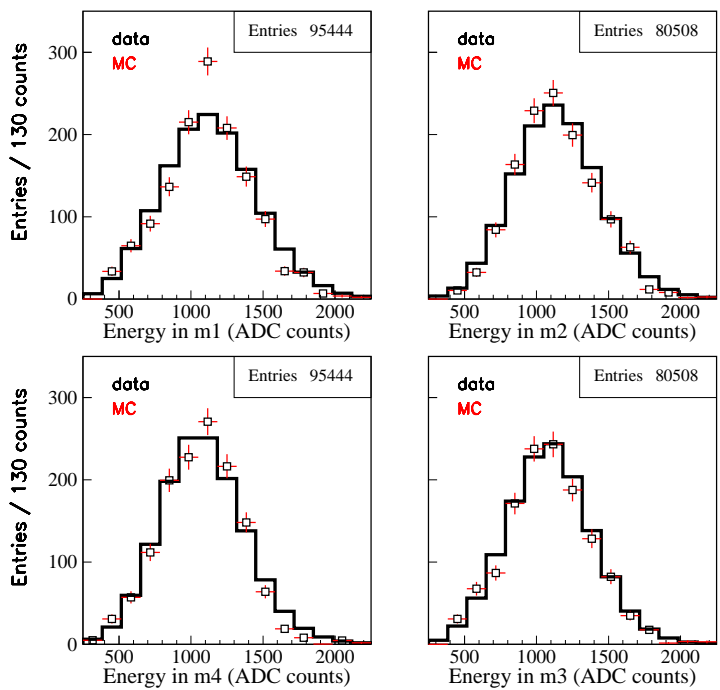}}
\caption{ {\it Comparison of the reconstructed energy distributions for data (black histogram) and the \MC\ simulation (red dots) in the four calorimeter modules: M1 (top left), M2 (top right), M3 (bottom left) and M4 (bottom right). } }
\label{fig:ErecDataMC}
\end{figure}

The effects mentioned above, as well as the imperfect knowledge of the attenuation length in the tiles and WLS fibers also affect the energy reconstruction, which can be tested in detail by studying the resolution on the electron or positron azimuthal angle, reconstructed via a 'barycenter' method (average of the hit azimuth angles using the corresponding energy deposits as weights). The resolution can be estimated by subtracting the electron reconstructed azimuth $\phi$ from that of the positron. In the absence of boost, this quantity only differs from zero because of resolution effects. The distortion due to the boost makes this variable also useful to check that it is correctly implemented in the simulation. Fig.~\ref{fig:ResolPhi} compares the azimuthal resolution in \MC\ with the one extracted from data. The agreement, although not perfect, is reasonable. Comparing the shape of the Bhabha energy peaks in data and \MC\ (Fig.~\ref{fig:ErecDataMC}) show that the discrepancies observed on Fig.~\ref{fig:ResolPhi} do not impact enough the energy distribution to bias the trigger acceptance. It also validates the simplified description of the trigger thresholds in the simulation. 

\section{Results}
\label{sec:results}

The Bhabha calorimeter and the gamma monitors have been designed, built and operated to monitor the expected improvements of the \DAF\ machine after its 2007 upgrade and to provide a fast and useful feedback to the control room operators. These goals have been achieved demonstrating that the accelerator operations are successful. The last section of this paper reviews the main results of the 2008-2009 \DAF\ runs which reflect the performances of both the machine and the associated detectors.

\subsection{Luminosity}

\begin{figure}
\begin{center}
\includegraphics[width=4in]{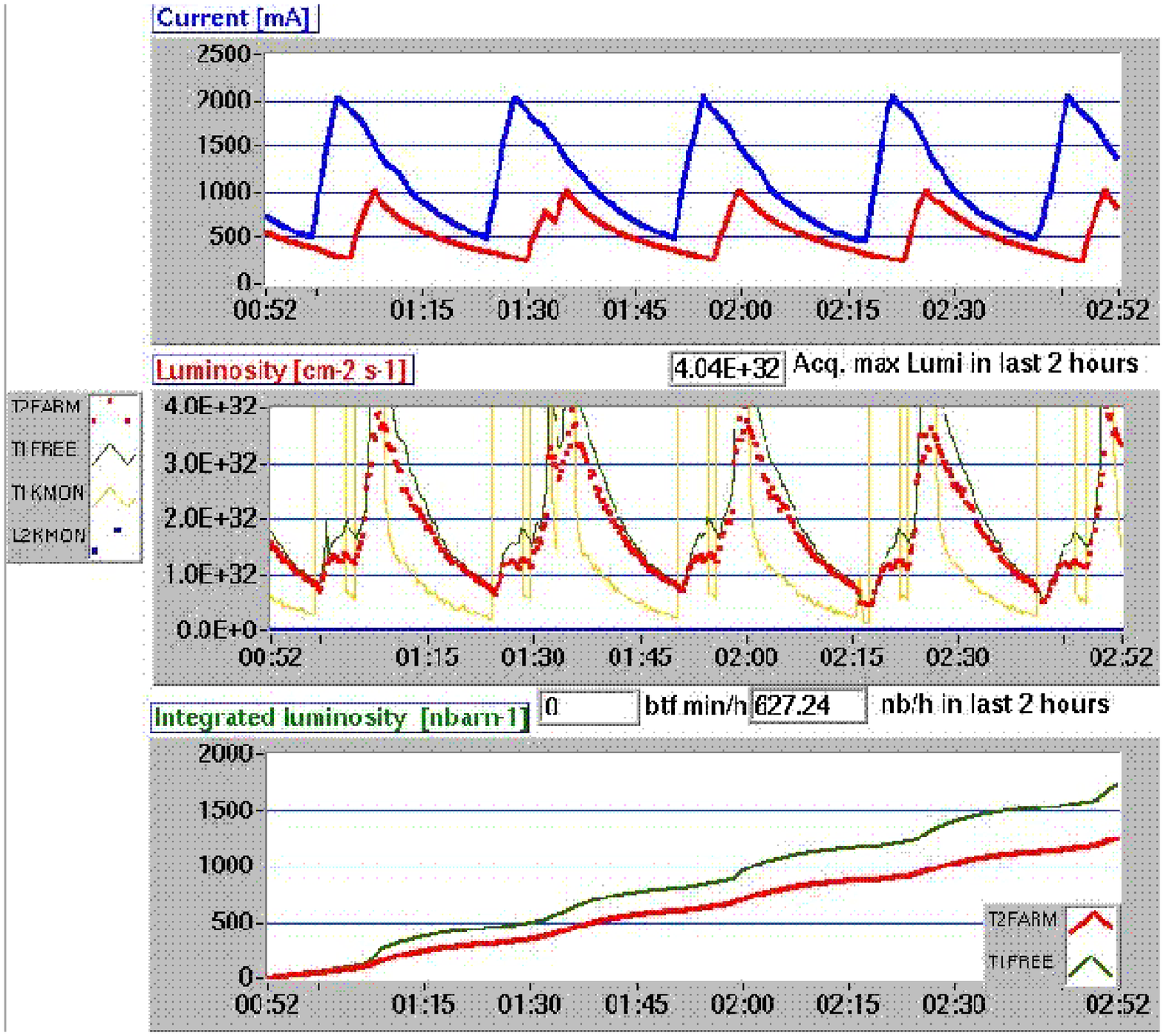}
\end{center}
\caption{ { \it Typical example of a two hours running period at the upgraded \DAF\ accelerator; these data have been taken between 00:52 and 02:52 on April $28^{\mathrm{th}}$~2009. The time evolution of the electron (blue) and positron (red) beam currents are shown on the top plot; the fill and cost pattern used to operate the machine is clearly visible. The middle plot stripcharts, among other quantities, the luminosities measured by the Bhabha calorimeter. The 'T1FREE' (green curve) is the raw luminosity directly computed from the detector trigger rate while the 'T2FARM' quantity (red dots) is background-subtracted. As expected, the latter is much less sensitive to backgrounds than the former: 'T1FREE' exceeds significantly 'T2FARM' when either beam gets injected while the two curves agree almost perfectly when the beam currents are both very low. With currents around 2~Amp for electrons and 1~Amp for positrons, the peak luminosity is routinely between 3.5 and $4.0 \, 10^{32} \cms$. Finally, the bottom plot shows the integrated 'T1FREE' and 'T2FARM' luminosities. Each step in the green curve corresponds to an injection period; after two hours the disagreement can reach 30$\%$ which explains why correcting the calorimeter rate was mandatory to get an accurate luminosity measurement. } }
\label{fig:2hoursRunning}
\end{figure} 

\begin{figure}
\begin{center}
\includegraphics[width=4in]{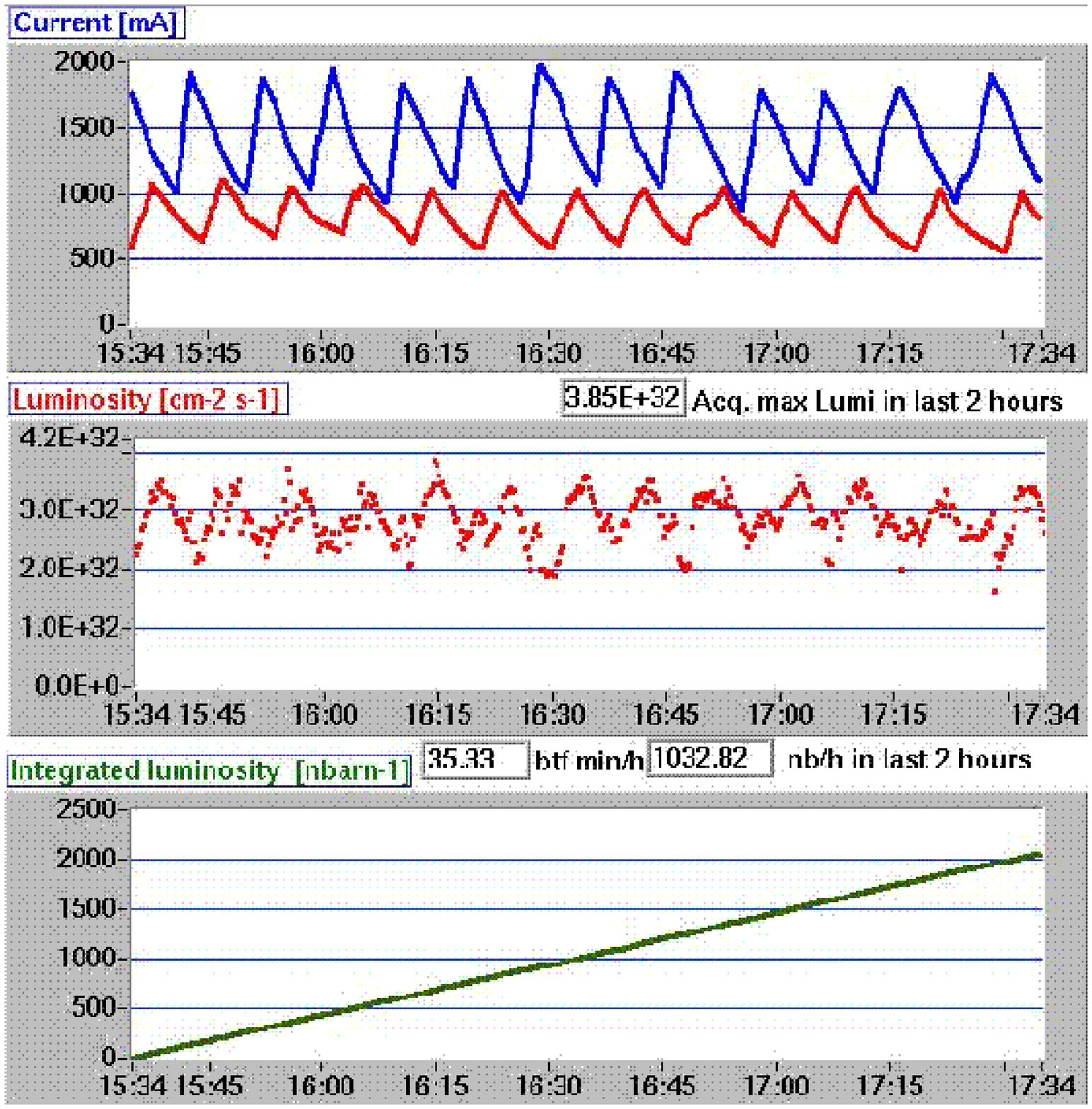}
\end{center}
\caption{ { \it Two hours running period at high injection regime on December $21^{\mathrm{rst}}$ 2008; the integrated luminosity exceeds $1 \invpb$ per hour while its instantaneous value oscillates around $3 \, 10^{32} \cms$. This running mode cannot be maintained over long periods as it is not compatible with the \SID\ operations. } }
\label{fig:highInjectionRegime}
\end{figure}

Fig.~\ref{fig:2hoursRunning} shows a typical two-hours running period at \DAF. The peak currents are around 2~Amp for electrons (top plot, blue curve) and 1~Amp for positrons (top plot, red curve) while the instantaneous corrected luminosity (middle plot, red dots) peaks between 3.5 and $4.0 \, 10^{32}\cms$. As expected, backgrounds (in particular injection-related) only affect the uncorrected luminosity (middle plot, green curve). Although an high-injection regime is not compatible with the operations of the background-sensitive \SID\ experiment, this running mode has been tested for few hours; integration in excess of $1 \invpb$ per hour have been achieved, as shown on Fig.~\ref{fig:highInjectionRegime}.

\begin{figure}[!h]
\centering
 \rotatebox{0}{\includegraphics[width=6.0cm]{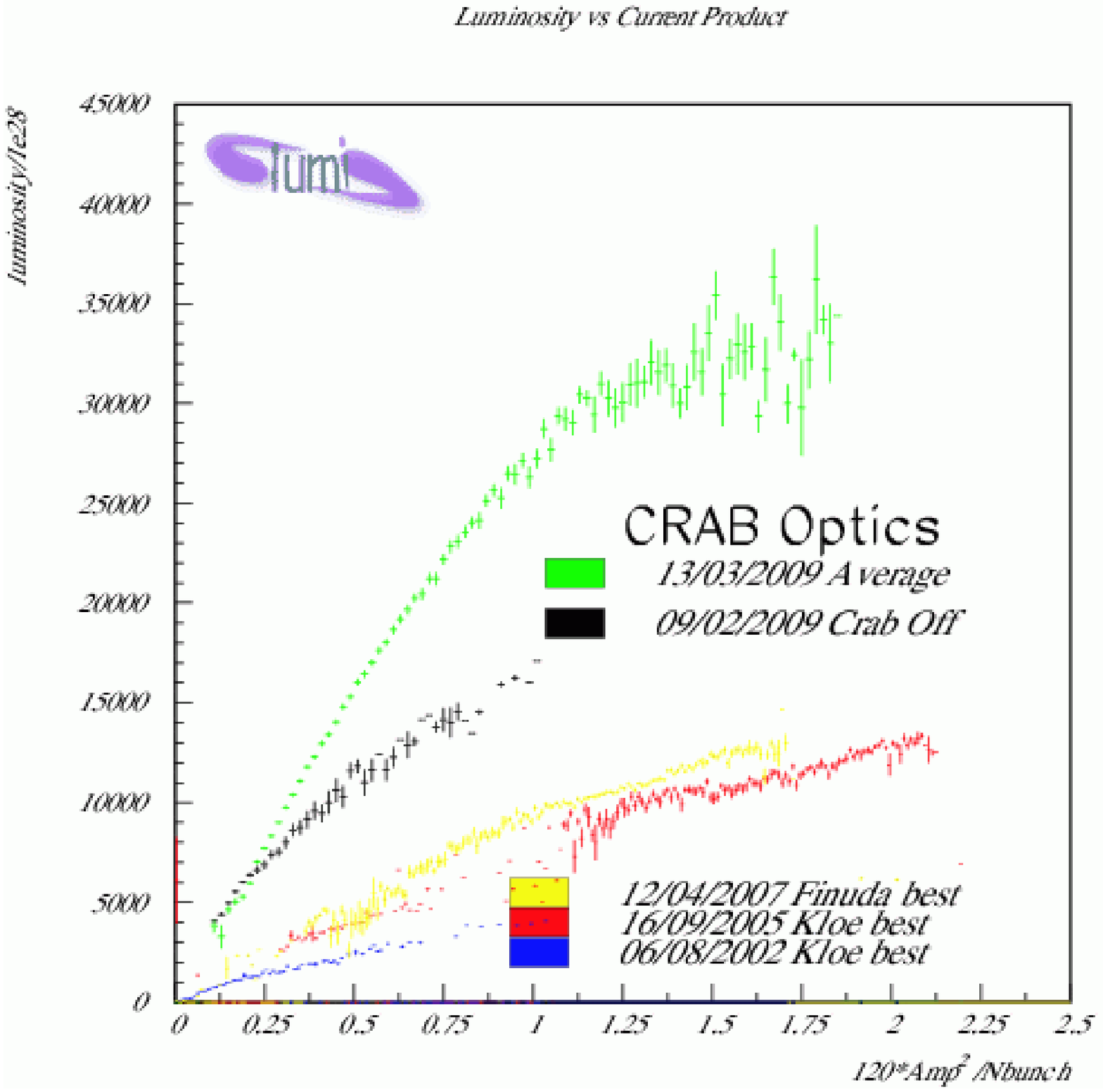}}
 \rotatebox{0}{\includegraphics[width=6.0cm]{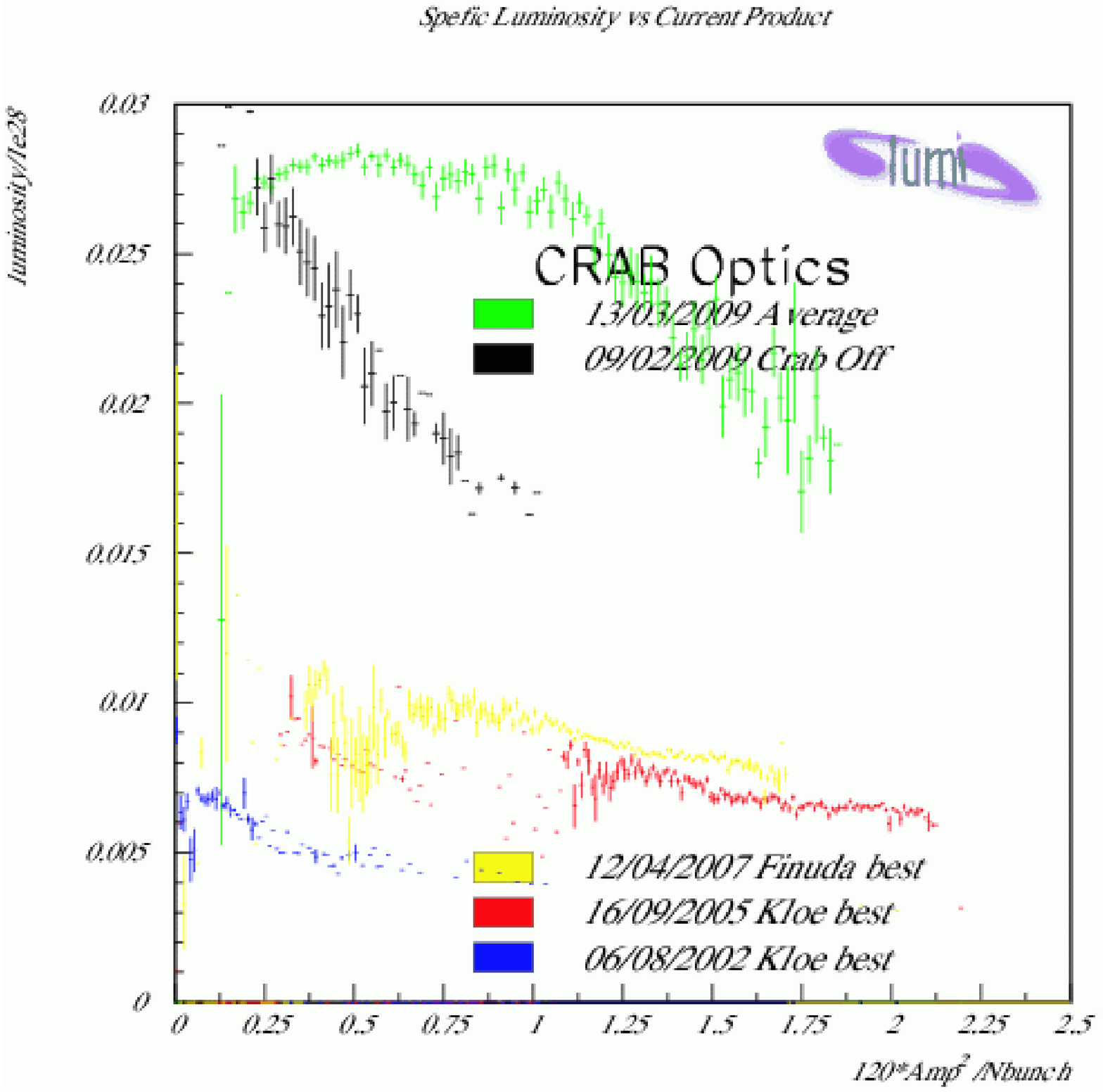}}
\caption{ {\it Comparison of the upgraded \DAF\ performances (green dots) w.r.t. those achieved during the KLOE (blue and red dots) and FINUDA (yellow) runs. The left (right) plot shows the daily average luminosity (daily average specific luminosity) versus the product of the two beam currents. The improvement shown by the data taken on March $13^{\mathrm{th}}$ 2009 is at least a factor 3 over the full range plotted on the $x$-axis. } }
\label{fig:dafne_performances_1}
\end{figure}

\begin{figure}[!h]
\centering
 \rotatebox{0}{\includegraphics[width=10.0cm]{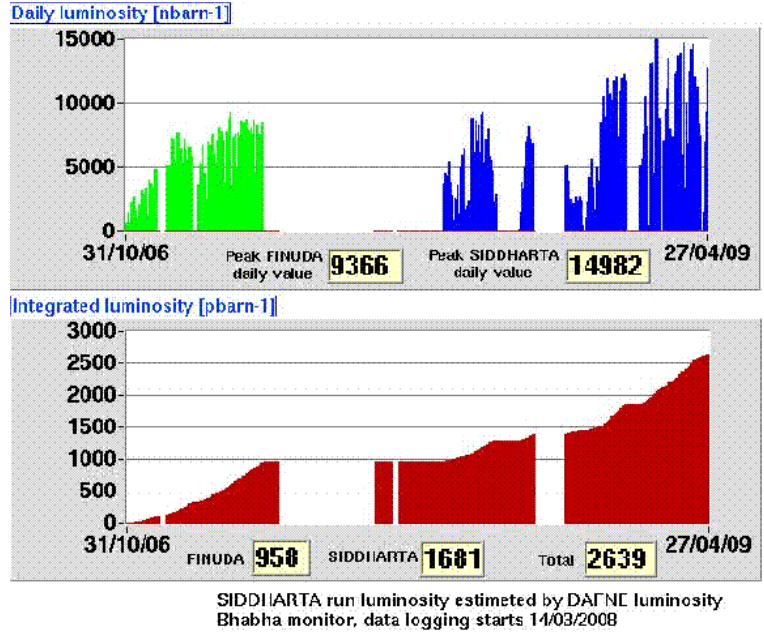}}
\caption{ {\it Evolution of the daily and integrated luminosities for the FINUDA and \SID\ runs of the \DAF\ accelerator. The improvement provided by the crab waist compensation scheme is clear, both for the absolute performances and the rate of improvement. } }
\label{fig:dafne_performances_2}
\end{figure}

A detailed discussion of the machine performances after the implementation of the new interaction scheme can be found elsewhere~\cite{ref:icfa1,ref:icfa2}. The gain provided by the new IR gets higher with the current products and the difference with respect to collisions with the crab sextupoles off can reach 50\%. Figs.~\ref{fig:dafne_performances_1} and~\ref{fig:dafne_performances_2} provide comparisons of the upgraded \DAF\ with respect to past experiments (KLOE and FINUDA). The improvement is striking for all quantities measured: luminosity, specific luminosity and integrated luminosity. The saturation at high currents seen by the Bhabha calorimeter is also present in other monitoring data which indicates that the limitation does stem from a beam size blow up and not from some background-induced PMT saturation preventing accurate luminosity measurements.

\subsection{Bunch-by-bunch luminosity}

\begin{figure}[!h]
\centering
 \rotatebox{0}{\includegraphics[width=6.5cm]{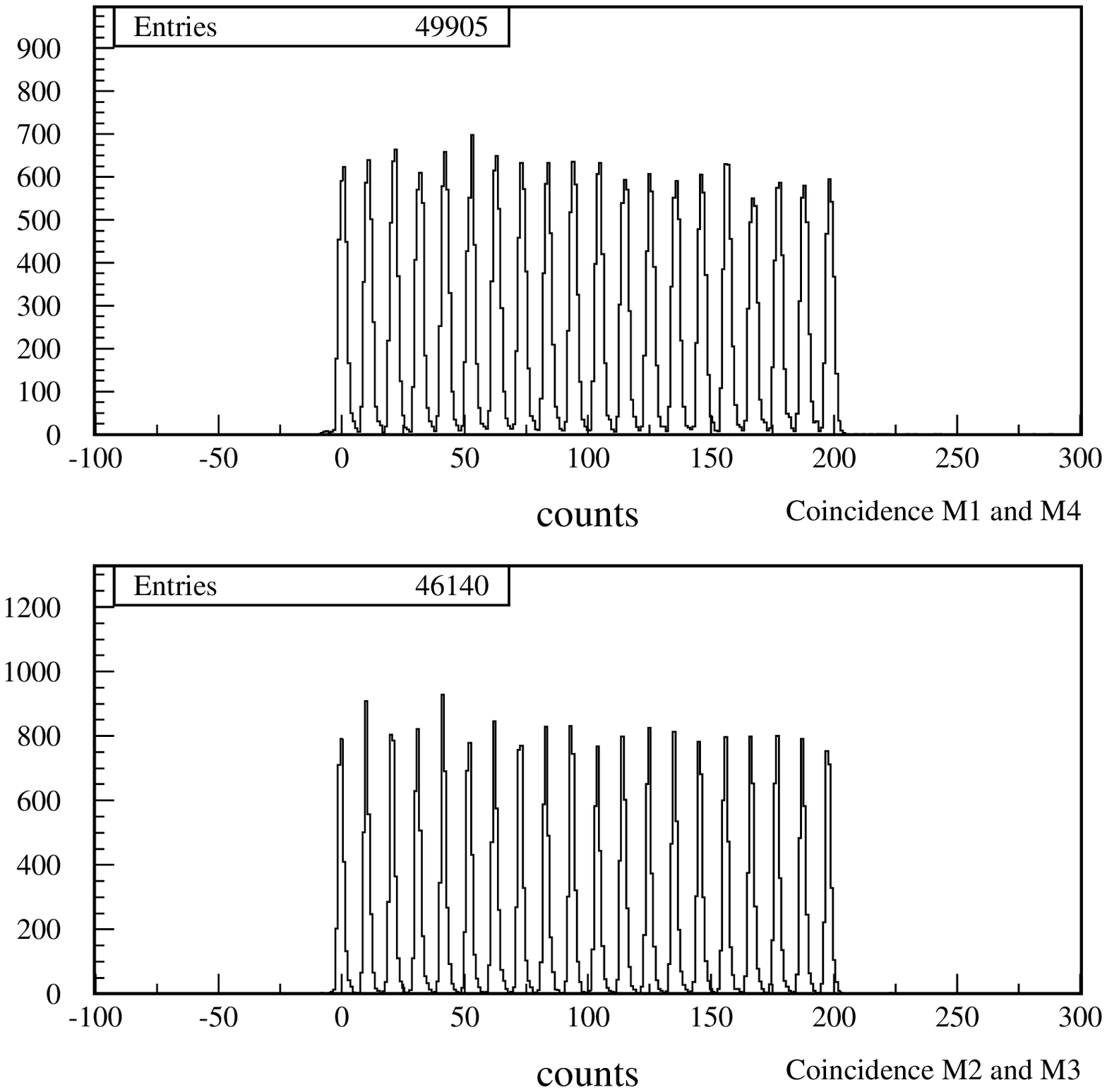}}
 \rotatebox{0}{\includegraphics[width=6.5cm]{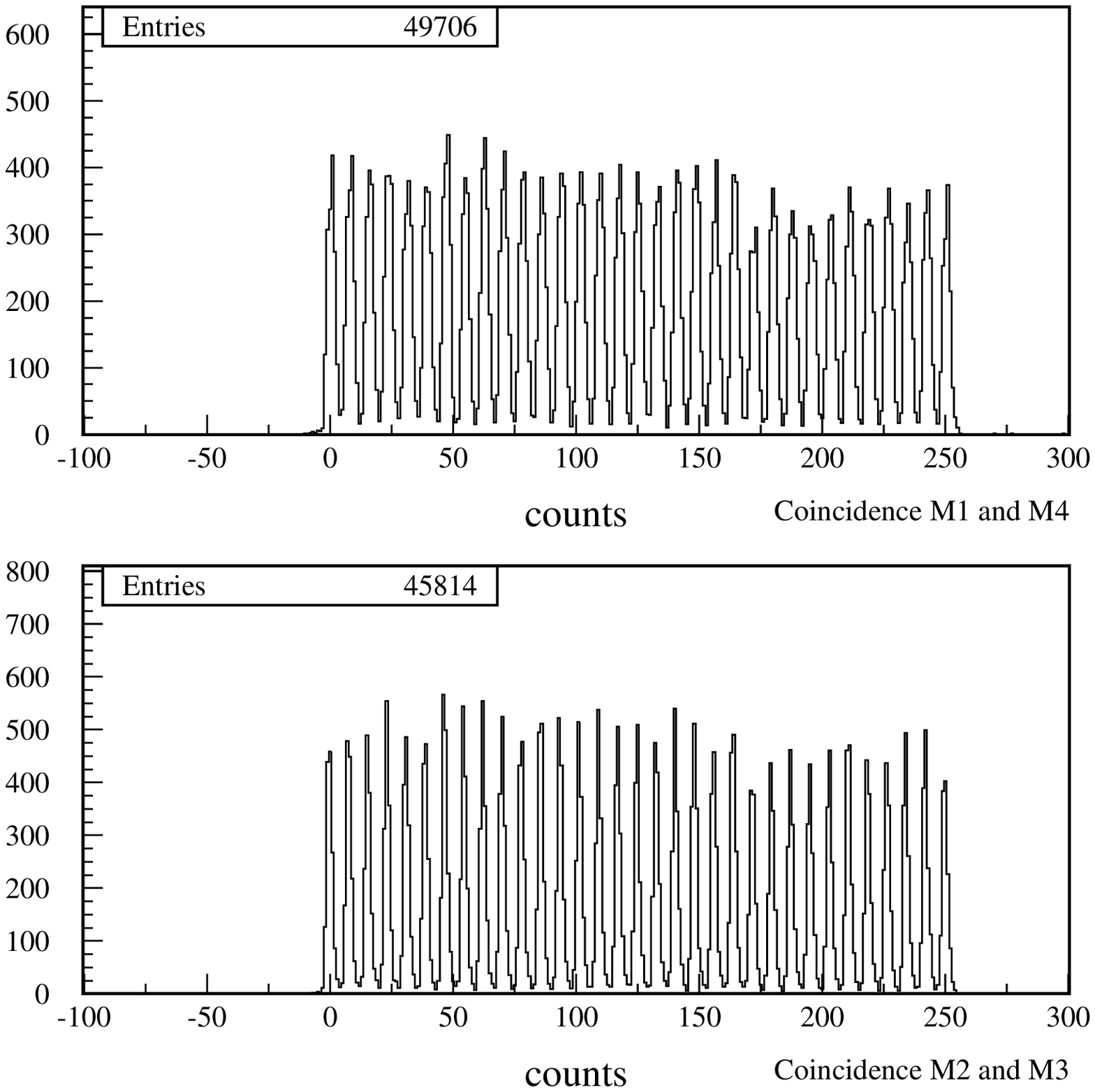}}
 \rotatebox{0}{\includegraphics[width=6.5cm]{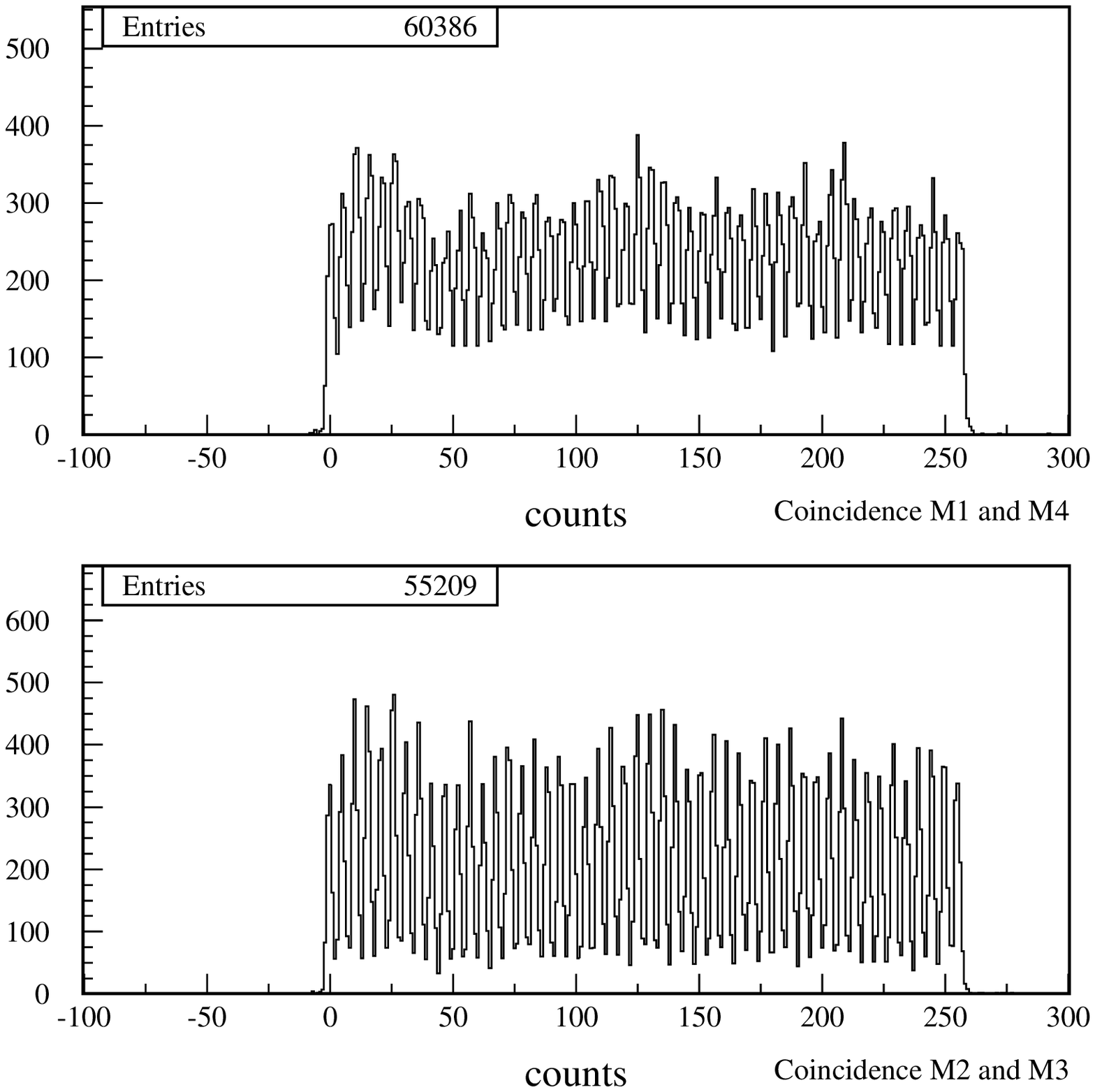}}
 \rotatebox{0}{\includegraphics[width=6.5cm]{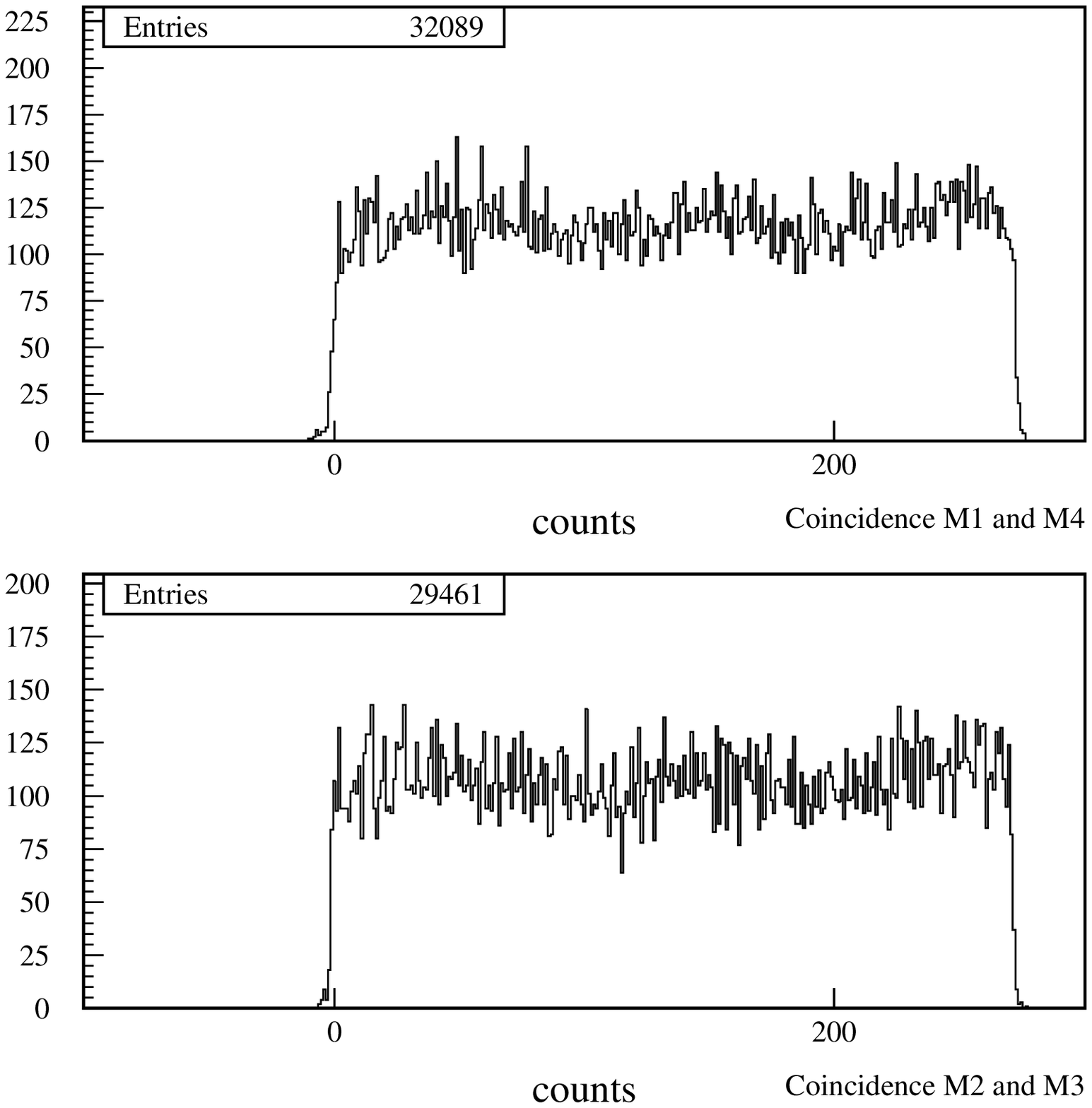}}
\caption{ {\it Histograms of trigger TDC counts (1 count corresponds to 1.04 ns) for the two pairs of back-to-back calorimeter modules (M1-M4 and M2-M3) and different bunch patterns. Top left plots: by-5 pattern (20+20 bunches); top right plots: by-3 pattern (33+33 bunches); bottom left plots: by-2 pattern (50+50 bunches); bottom right plots: full fill (100+100 bunches). These trigger TDC data are corrected by the RF timing signal to give the time elapsed since the last crossing of the first $\en$ and $\ep$ bunches; this allows one to see the contributions from individual bunches separately. } }
\label{fig:bunchbybunch}
\end{figure}

The Bhabha calorimeter timing resolution is good enough to separate the contributions from individual bunches, which is useful for machine studies (bunch filling pattern, etc.). Histograms of trigger TDC counts for the two pairs of back-to-back modules are shown in Fig.~\ref{fig:bunchbybunch} for different accelerator configurations. 
\begin{itemize}
\item Top left plots: by-5 pattern (20+20 bunches).
\item Top right plots: by-3 pattern (33+33 bunches).
\item Bottom left plots: by-2 pattern (50+50 bunches).
\item Bottom right plots: full fill (100+100 bunches).
\end{itemize}
The TDC data are corrected by the RF timing signal: using the first bunch crossing as reference, one only keeps the time elapsed since this event which allows one to separate the contributions of each individual bunch to the trigger. The bunch spacing is found consistent for the different patterns: 2.6~ns in average. One can also note that the luminosity per bunch is quite flat for all bunch patterns and does not really depend on the bunch position in the train, as expected for a well-behaving machine. Obviously all fills are not as good as those exemplified in this section and so having the capability of making such diagnostics quickly helps a lot investigating issues which may involve the beam bunch patterns.

\subsection{Touschek background}

Two machine runs have been dedicated to the measurement of the gamma monitor rates in order to compare the actual numbers with those predicted by a full tracking simulation. The latter simulates the Touschek scattering particles from their generation to their loss in the beam pipe; secondary particles are then simulated and propagated through the IR up to the gamma monitors. The first run (February $3^{\mathrm{rd}}$ 2009) was taken with two beams out of collision while the accelerator was operated in single beam mode for the second run (March $12^{\mathrm{th}}$ 2009), during which the vertical beam sizes (and hence the coupling values) were varied using a skew quadrupole. For this second experiment a ten-bunch pattern was used for a total beam current of about 100~mAmp and the optics were the same as for the high luminosity runs. In addition to the gamma monitor spectra, the beam lifetimes were also measured during the March run. This allows a more complete test of the simulation as its predictions regarding both the background rates and the beam lifetimes can be simultaneously compared with the \DAF\ data.

\subsubsection{Gamma monitor rate}

\begin{figure}[!h]
\centering
\begin{center}
\rotatebox{0}{\includegraphics[width=10cm]{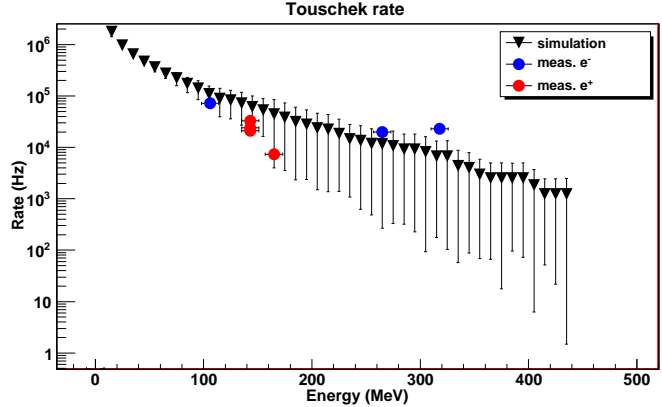}}
\caption{{\it Comparison between simulated (black triangles) and measured (blue dots for the electron beam, red ones for positrons) rates in the gamma monitors. The rates extracted from data have been corrected to take into account the beams coupling in order to allow a direct comparison with the simulation. } }
\label{fig:rateGammaMonitor}
\end{center}
\end{figure}

The black triangles in Fig.~\ref{fig:rateGammaMonitor} show the variation of the gamma monitor rate versus threshold (in MeV), as obtained in simulation. The experimental setup makes the spectrum end around 440~MeV which is below the beam energy (510~MeV). Blue (red) bullets show the actual rates measured by the gamma monitor sensitive to photons emitted by positrons (electrons) for a few thresholds. In order to allow a direct comparison with the simulation, these numbers have been corrected to take into account the coupling between the beams. Although the agreement between simulation and measurements is not perfect, both sets of numbers appear to be in the right ballpark~-- actual rates also depend on various machine conditions which are not easy to reproduce nor to quantify.

\subsubsection{Beam lifetimes}

\begin{figure}[!h]\centering
\begin{center}
\rotatebox{0}{\includegraphics[width=8.0cm]{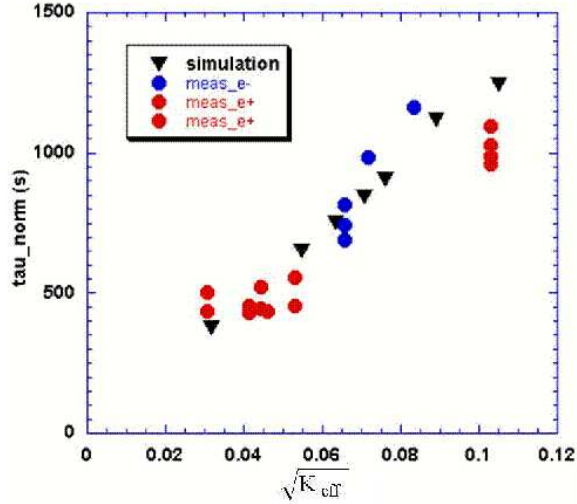}}
\caption{{\it Comparison between predicted (black triangles) and measured (bullets, blue for the electron beam and red for the positron beam) beam lifetimes versus the effective beam coupling $K_{eff}$~-- see text for details. The agreement between simulation and data is quite good. } }
\label{fig:lifetime}
\end{center}
\end{figure}

The resulting simulated Touschek lifetime is as short as 840 s for a 0.5\% beam coupling. This result is in agreement with measurements, as shown in Fig.~\ref{fig:lifetime}. This plot compares the measured normalized lifetime of both beams with the computed one as a function of the square root of the effective beam coupling $K_{eff}$. Black markers refer to simulation while blue and red dots correspond to the measured electron and positron lifetimes, respectively. The plotted lifetime is normalized to the total current

\begin{equation}
\tau_{\mathrm{ normalized } } = \tau_{\mathrm{ measured } } \left( \frac{ I }{ 100 \, \mathrm{mAmp} } \right)^{2/3}
\end{equation}
according to the scaling law $\tau \propto \sigma_{l} \sigma_{x} \sigma_{y} /I$ where  $\sigma_{l} \propto I^{1/3}$ is the current-dependent bunch length.

Larger and smaller markers refer to larger and smaller couplings respectively. The beam coupling 

\begin{equation}
K=\epsilon_{y}/\epsilon_{x} = (\sigma_{y} /\sigma_{x})^{2} \beta_{x} /\beta_{y}\end{equation}

is evaluated at the SLM using the measured transverse beam sizes and the ratio $\beta_{y} /\beta_{x}=2.25$, as indicated by the MAD optical model~\cite{ref:mad}. The effective vertical beam size used for the evaluation of the effective coupling $K_{eff}$ takes into account its measurement resolution: $80~\mu m$~\cite{ref:dafnelogbook}.

\section{Conclusions}
\label{sec:conclusions}

Various diagnostic detectors have been installed around the upgraded \DAF\ IR to monitor and quantify the improvement of performances brought by the new crab waist collision scheme. After the design and building phases, the data taking started in February 2008 and is still going on in spite of a few shutdowns, in particular during the Summer 2008 break. 

The background-subtracted luminosity measured by the Bhabha calorimeter shows a significant improvement with respect to the previous IP design, valid for all currents and still increasing as the machine gets better tuned. This absolute measurement, based on a accurate \GEANT\ simulation of the IR, has a systematics of about 13\% and a negligible statistical error. In addition, gamma monitors allow online measurements of the background which has been found compatible with the predictions of simulation of the Touschek effect. Finally, the qualitative agreement between simulations and measurements is enforcing the validity of the studies currently ongoing to design the SuperB machine~\cite{ref:superb}.

\vspace{0.5 cm}
{\bf Acknowledgments}
\vspace{0.3 cm}
We would like to thank L.~Iannotti, V.~Romano and the LNF SSCR service for their mechanical engineering support; all the electronic staff of the LNF Accelerator Division; G.~Corradi, M.~Pistilli, D.~Tagnani for the GEM electronics and detector construction; M.~Anelli, M.~Iannarelli, E.~Turri for their help during the calorimeter operations; W.~Placzek for his support on the BHWIDE code. Finally, one of us (B.~Viaud) would like to thank the scientific interest group ``Physique des deux infinis'' for its financial support during this work.


\begin{thebibliography}{00}
\bibitem{ref:tyvek} \TYVEK\ is a flash-spun high-density polyethylene fiber synthetic texture and a trademark of the DuPont company.
\bibitem{ref:speclumi} The specific luminosity is the ratio between the instantaneous luminosity and the product of the beam currents times the number of colliding bunches in the accelerator.
\end{thebibliography}
\end{document}